\documentclass[apjl,twocolumn]{openjournal}
\usepackage{natbib} 
\usepackage[dvipsnames]{xcolor}
\usepackage{aas_macros} 
\usepackage{amssymb}
\usepackage{amsmath}
\usepackage{threeparttable}
\usepackage{multirow}
\usepackage[title]{appendix}
\usepackage{hyperref}	
\hypersetup{colorlinks=true,linkcolor=blue,citecolor=blue,filecolor=blue,urlcolor=blue}
\usepackage[caption=false]{subfig}
\usepackage{soul}

\newcommand{\cii}{[C\,{\sc ii}]}
\newcommand{\oiii}{[O\,{\sc iii}]$_{88}$}
\newcommand{\oiiinowave}{[O\,{\sc iii}]}
\newcommand{\oiiidopt}{[O\,{\sc iii}]$_{4959,5007}$}
\newcommand{\oiiiopt}{[O\,{\sc iii}]$_{5007}$}

\newcommand{\ciii}{C\,{\sc iii}]}
\newcommand{\civ}{C\,{\sc iv}}
\newcommand{\oii}{[O\,{\sc ii}]}
\newcommand{\oiidopt}{[O\,{\sc ii}]$_{3727,29}$}
\newcommand{\oi}{[O\,{\sc i}]}
\newcommand{\feii}{[Fe\,{\sc ii}]}
\newcommand{\nii}{[N\,{\sc ii}]}
\newcommand{\niii}{[N\,{\sc iii}]}

\newcommand{\hii}{H\,{\sc ii}}

\newcommand{\oiiitocii}{[O\,{\sc iii}]/[C\,{\sc ii}]}
\newcommand{\oiiil}{[O\,{\sc iii}] 88\,$\mu{\rm m}$}
\newcommand{\ciil}{[C\,{\sc ii}] 158\,$\mu{\rm m}$}

\def\EW {EW([O\,{\sc iii}]+H$\beta$)}

\begin{document}

\title[O$^{++}$ at Redshift 7]{REBELS-IFU: on the origin of the elevated \oiiil{}/\ciil{} ratios in the early Universe \vspace{-15mm}}

\author{Hiddo Algera$^{1}$\thanks{$^*$E-mail: \href{mailto:hsbalgera@asiaa.sinica.edu}}}
\author{Lucie Rowland$^{2}$}
\author{Renske Smit$^{3}$}
\author{Rebecca Fisher$^{4}$}
\author{Lise Ramambason$^{5}$}
\author{Nimisha Kumari$^{6}$}
\author{Livia Vallini$^{7}$}
\author{Hanae Inami$^{8}$}
\author{Themiya Nanayakkara$^{9,10}$}
\author{Mauro Stefanon$^{11,12}$}
\author{Manuel Aravena$^{13,14}$}
\author{Tom Bakx$^{15}$}
\author{Rychard Bouwens$^{2}$}
\author{Rebecca A.\ A.\ Bowler$^{4}$}
\author{Karin Cescon$^{2}$}
\author{Chian-Chou Chen$^{1}$}
\author{Pratika Dayal$^{16,17}$}
\author{Ilse De Looze$^{18}$}
\author{Andrea Ferrara$^{19}$}
\author{Yoshinobu Fudamoto$^{20,21}$}
\author{Lena Komarova$^{11}$}
\author{Ivana van Leeuwen$^{2}$}
\author{Katherine Ormerod$^{3}$}
\author{Sander Schouws$^{2}$}
\author{Laura Sommovigo$^{22}$}
\author{Aswin Vijayan$^{23}$}
\author{Wei-Hao Wang$^{1}$}
\author{Paul van der Werf$^{2}$}
\author{Joris Witstok$^{24,25}$}

\affiliation{$^{1}$Institute of Astronomy and Astrophysics, Academia Sinica, 11F of Astronomy-Mathematics Building, No.1, Sec. 4, Roosevelt Rd, Taipei 106319, Taiwan, R.O.C.}
\affiliation{$^{2}$Leiden Observatory, Leiden University, P.O. Box 9513, 2300 RA Leiden, The Netherlands}
\affiliation{$^{3}$Astrophysics Research Institute, Liverpool John Moores University, 146 Brownlow Hill, Liverpool L3 5RF, UK}
\affiliation{$^{4}$Jodrell Bank Centre for Astrophysics, Department of Physics and Astronomy, School of Natural Sciences, The University of Manchester, Manchester M13 9PL, UK}
\affiliation{$^{5}$Institut fur Theoretische Astrophysik, Zentrum f\"{u}r Astronomie, Universit\"{a}t Heidelberg, Albert-Ueberle-Str. 2, 69120 Heidelberg, Germany}
\affiliation{$^{6}$AURA for European Space Agency (ESA), ESA Office, Space Telescope Science Institute, 3700 San Matin Drive, Baltimore, MD, 21218, USA}
\affiliation{$^{7}$INAF, Osservatorio di Astrofisica e Scienza dello Spazio, Via P. Gobetti 93/3, I-40129, Bologna, Italy}
\affiliation{$^{8}$Hiroshima Astrophysical Science Center, Hiroshima University, 1-3-1 Kagamiyama, Higashi-Hiroshima, Hiroshima 739-8526, Japan}
\affiliation{$^{9}$Centre for Astrophysics and Supercomputing, Swinburne University of Technology, PO Box 218, Hawthorn, VIC 3122, Australia}
\affiliation{$^{10}$JWST Australian Data Centre (JADC), Swinburne Advanced Manufacturing and Design Centre (AMDC), John Street, Hawthorn, VIC 3122, Australia}
\affiliation{$^{11}$Departament d’Astronomia i Astrofìsica, Universitat de València, C. Dr. Moliner 50, E-46100 Burjassot, València, Spain}
\affiliation{$^{12}$Unidad Asociada CSIC ”Grupo de Astrofísica Extragaláctica y Cosmología” (Instituto de Física de Cantabria - Universitat de València), Spain}
\affiliation{$^{13}$Instituto de Estudios Astrof\'{\i}cos, Facultad de Ingenier\'{\i}a y Ciencias, Universidad Diego Portales, Av. Ej\'ercito 441, Santiago, Chile}
\affiliation{$^{14}$Millenium Nucleus for Galaxies (MINGAL)}
\affiliation{$^{15}$Department of Space, Earth, \& Environment, Chalmers University of Technology, Chalmersplatsen, SE-4 412 96 Gothenburg, Sweden}
\affiliation{$^{16}$Canadian Institute for Theoretical Astrophysics, 60 St George St, University of Toronto, Toronto, ON M5S 3H8, Canada}
\affiliation{$^{17}$Kapteyn Astronomical Institute, University of Groningen, P.O. Box 800, NL-9700 AV Groningen, the Netherlands}
\affiliation{$^{18}$Sterrenkundig Observatorium, Ghent University, Krijgslaan 281 - S9, 9000 Gent, Belgium}
\affiliation{$^{19}$Scuola Normale Superiore, Piazza dei Cavalieri 7, I-56126 Pisa, Italy}
\affiliation{$^{20}$Center for Frontier Science, Chiba University, 1-33 Yayoi-cho, Inage-ku, Chiba 263-8522, Japan}
\affiliation{$^{21}$Steward Observatory, University of Arizona, 933 N Cherry Avenue, Tucson, AZ 85721, USA}
\affiliation{$^{22}$Center for Computational Astrophysics, Flatiron Institute, 162 5th Avenue, New York, NY 10010, USA}
\affiliation{$^{23}$Astronomy Centre, University of Sussex, Falmer, Brighton BN1 9QH, UK}
\affiliation{$^{24}$Cosmic Dawn Center (DAWN), Copenhagen, Denmark}
\affiliation{$^{25}$Niels Bohr Institute, University of Copenhagen, Jagtvej 128, DK-2200, Copenhagen, Denmark}

\begin{abstract}

We present new ALMA \oiii{} observations of eight previously \cii{}$_{158}$-detected galaxies at $6.8 \lesssim z \lesssim 7.7$. Six of our targets -- the primary sample -- are massive, UV-luminous galaxies drawn from the REBELS survey, while the remaining two are UV-fainter galaxies that were previously serendipitously detected through their luminous \cii{} lines in the REBELS fields. We detect \oiii{} emission in all eight galaxies at $6.2 - 17.7\sigma$ significance, and find them to be consistent with the local dwarf galaxy relation between $L_\text{\oiii{}}$ and star formation rate. Our sample spans $\text{\oiiitocii{}}\approx 1.9 - 9.6$, which is typical for the high-redshift galaxy population. Five of the primary targets benefit from \textit{JWST}/NIRSpec observations, enabling a direct comparison of the \oiiitocii{} ratio against rest-optical ISM diagnostics. We supplement our high-redshift sample with eleven $z\approx6-14$ galaxies in the literature for which similar ALMA and \textit{JWST} observations are available, and furthermore compare to the \oiiitocii{} ratios measured for local dwarf galaxies. We find that, at fixed metallicity and ionization parameter, $z>6$ galaxies show elevated \oiiitocii{} ratios compared to local dwarfs. Instead, we find that a large $\text{\oiiidopt{}}+\mathrm{H}\beta$ equivalent width -- a proxy for burstiness -- is the main driver of the high \oiiitocii{} ratios seen in the early Universe, which is primarily due to \cii{} being suppressed in bursty galaxies. Given the apparent validity of the \oiii{}-SFR relation across most of cosmic time, as well as the abundance of young, bursty galaxies at high redshift, \oiii{} is set to remain a powerful ISM tracer at the cosmic dawn.

\end{abstract}

\section{Introduction}
\label{sec:introduction}

With the formation of the first stars and galaxies, the rapid enrichment of the early Universe with elements heavier than helium -- known as metals -- commenced \citep{dayal2018,maiolino2019}. In atomic or ionic form, these metals are typically characterized by a large number of electronic transitions that give rise to emission lines at specific rest-frame wavelengths. Observing these lines in high-redshift galaxies provides key insights into the nature and physical conditions of their interstellar medium (ISM), and is thus vital for our understanding of early galaxy evolution.

Among the metals produced in the first core-collapse supernovae, oxygen is of particular importance. Not only is oxygen expected to be produced in large quantities \citep{woosley_heger2002,takahashi2018,kobayashi2020}, it also gives rise to a wealth of bright cooling lines at rest-frame optical (e.g., \oii{}$_{3727,29}$ and \oiiiopt{}) and far-infrared (e.g., \oiiinowave{}$_{52}$, \oiii{}, \oi{}$_{63}$ and \oi{}$_{145}$) wavelengths.\footnote{We denote a given transition by writing its rest-frame wavelength -- either in Angstrom for optical or micron for far-infrared lines --- in subscript, unless only a single relevant transition exists, in which case the subscript is omitted.} Making use of the Mid-Infrared Instrument (MIRI) aboard the \textit{James Webb Space Telescope (JWST)}, \citet{zavala2025_miri} have detected rest-optical oxygen emission as far back as $z=12.33$ (see also \citealt{calabro2024,castellano2024}). Moreover, with the Atacama Large Millimeter/submillimeter Array (ALMA), the \oiii{} line was recently detected in one of the most distant galaxies currently known, at $z=14.18$ \citep{carniani2024_oiii,schouws2024}. 

The production of the present-day second most abundant heavy element, carbon, is expected to lag behind that of oxygen (e.g., \citealt{berg2019}) as it is mainly believed to be produced by Asymptotic Giant Branch (AGB) stars. However, recent \textit{JWST} observations of high-redshift galaxies have in a handful of instances revealed C/O abundances in excess of what is expected from supernova enrichment alone, suggesting the existence of alternative carbon production pathways operating on shorter timescales \citep[e.g.,][]{bunker2023,cameron2023,castellano2024,deugenio2024,hsiao2024}

At rest-frame far-infrared wavelengths, two particularly well-studied transitions of (ionized) carbon and oxygen are the \cii{}$_{158}$ and \oiii{} lines -- both locally with \textit{Herschel} and \textit{SOFIA} \citep[e.g.,][]{delooze2014,cormier2015,diazsantos2017,herrera-camus2018a,herrera-camus2018b,ramambason2022,ura2023,kumari2024,kovacic2025} and at $z\gtrsim 6$ with ALMA \citep[e.g.,][]{inoue2016,hashimoto2019,harikane2020,witstok2022,algera2024}. While in the nearby Universe the \cii{}$_{158}$ line is found to be the brightest far-infrared cooling line across the typical star-forming galaxy population (e.g., \citealt{delooze2014,diazsantos2017}), the \oiii{} line tends to be brighter in low-metallicity ($Z\lesssim0.2\,Z_\odot$) dwarf galaxies, which show an average \oiiitocii{} $\sim2$ \citep{cormier2015,cormier2019}.\footnote{From hereon, we omit the subscript for the \oiii{} line when referring to the \oiiitocii{} ratio.} At $z\gtrsim6$, observations with ALMA have shown that the \oiii{} line is by far the most luminous far-infrared cooling line, exhibiting line ratios of \oiiitocii{}$\approx2-10$ (e.g., \citealt{inoue2016,hashimoto2019,laporte2019,carniani2020,harikane2020,witstok2022}), with only few exceptions (\oiiitocii{}$\lesssim2$; e.g., \citealt{marrone2018,algera2024,bakx2024}). 

There are multiple possible reasons for these elevated \oiiitocii{} ratios in the high-redshift galaxy population. First of all, the physical conditions and resulting radiation fields of high-redshift galaxies are likely to be more extreme than in the local galaxy population, which should boost their ionization parameter $U$, defined as the dimensionless ratio of incident flux of ionizing photons and the number density of hydrogen. Most high-redshift galaxies show enhanced values of $U$ with respect to galaxies in the local Universe (e.g., \citealt{shapley2015,papovich2022,reddy2023}), and such elevated ionization parameters are expected to boost the luminosity of the \oiii{} line, while potentially also ionizing C$^{+}$ into C$^{++}$, thereby reducing the luminosity of \cii{} (e.g., \citealt{arata2020,harikane2020,vallini2020,vallini2021,vallini2024,katz2022}). 

In addition, the typical metallicities of high-redshift galaxies are lower than those of equally massive galaxies at the present day \citep{sanders2021,sanders2024,curti2023,curti2024,chemerynska2024,sarkar2025}. Such low metallicities are expected to suppress \cii{} emission for a variety of reasons. To zeroth order, the \cii{} luminosity scales linearly with the number of C$^+$ atoms, and thus depends on the total carbon abundance which naturally decreases at low metallicities \citep{vallini2015,olsen2017,lagache2018,katz2019,liang2024}. On top of that, the aforementioned time lag in the production of carbon with respect to oxygen should further reduce the overall luminosity of the \cii{} line in metal-poor systems where sub-solar C/O abundances are expected \citep{arata2020,katz2022,nyhagen2024}. The low-metallicity ISM is furthermore thought to be more porous, and the resulting lower PDR covering fraction implies a smaller \cii{}-emitting volume \citep{cormier2019,hagimoto2025}. Finally, the fraction of \cii{} emission emanating from the ionized phase has been suggested to decrease in the low-metallicity ISM \citep[][though see \citealt{peng2025c}]{croxall2017,cormier2019,ramambason2022}.

In contrast, the suppression of \oiii{} is likely less severe at low metallicities due to the typically harder radiation fields in metal-poor galaxies, increasing the overall volume filling fraction of ionized gas and thus (partially) compensating for the lower oxygen abundance \citep[e.g.,][]{cormier2015,arata2020,witstok2025}. Overall, this suggests that the few $z\gtrsim6$ galaxies for which $\text{\oiiitocii{}}\lesssim2$ ratios are observed, are likely to be among the oldest, least bursty, and most metal-enriched at their epoch (e.g., \citealt{algera2024,bakx2024}).

Testing these various predictions on the origin of the elevated \oiiitocii{} ratio in high-redshift galaxies presents an opportunity for a powerful synergy between \textit{JWST} and ALMA, as already noted by several theoretical works (e.g., \citealt{katz2019,nakazato2023,vallini2024}). Specifically, \textit{JWST} can provide the key metallicity, ionization parameter and burstiness measurements, which can then quantitatively be compared against the \oiiitocii{} ratios measured by ALMA. However, the sample size of galaxies with all of these measurements has so far remained rather limited. Indeed, despite the success in probing \oiii{} emission out to $z\approx14$, the line has to date only been detected in just $\sim20$ galaxies at $z\gtrsim6$ \citep[e.g.,][and references therein]{harikane2020,algera2024,bakx2024}. Moreover, \cii{} has not confidently been detected beyond $z=8.3$ \citep{tamura2019,bakx2020}, despite several attempts targeting the line in more distant galaxies \citep{laporte2019,carniani2020,fujimoto2024_s4590,schouws2025}. Finally, most galaxies with an \oiii{} and \cii{} detection do not have the required ancillary \textit{JWST}/NIRSpec data, and in the few cases where such observations are readily available, they have yet to be analyzed in the context of the \oiiitocii{} ratio.

In this work, we present new ALMA \oiii{} observations towards eight $z=6.8 - 7.7$ galaxies first spectroscopically confirmed via \cii{} emission as part of the Reionization Era Bright Emission Line Survey (REBELS; \citealt{bouwens2022}). Five of these have also been observed by \textit{JWST}/NIRSpec as part of the Cycle 1 REBELS-IFU program (GO 1626, PI Stefanon; Stefanon et al.\ in preparation; see also \citealt{algera2025,fisher2025,rowland2025}), yielding measurements of, among other quantities, their metallicity and ionization parameter. Leveraging this powerful dataset, and supplementing it with similar observations in the literature, we explore the synergy between ALMA and \textit{JWST} in the context of the elevated \oiiitocii{} ratios seen in $z\gtrsim6$ galaxies in this work.

In Section \ref{sec:observations}, we present the new ALMA observations, and introduce the \textit{JWST} spectroscopic data. In Section \ref{sec:results}, we present our \oiii{} detections, and we study the \oiii{}-star formation rate relation in Section \ref{sec:oiiiSFR}. We proceed by discussing the \oiiitocii{} ratio, and its dependence on metallicity, ionization parameter and burstiness in Section \ref{sec:oiii2cii_discussion}. Finally, we leave the reader with some caveats in Section \ref{sec:caveats} and summarize our conclusions in Section \ref{sec:conclusions}. We assume a standard $\Lambda$CDM cosmology throughout this work, with $H_0=70\,\text{km\,s}^{-1}\text{\,Mpc}^{-1}$, $\Omega_m=0.30$ and $\Omega_\Lambda=0.70$. Star formation rates (SFRs) assume a \citet{chabrier2003} initial mass function (IMF), and we adopt the solar oxygen abundance of $12 + \log(\mathrm{O/H}) = 8.69$ following \citet{asplund2009}.

\begin{table*}
    \def\arraystretch{1.4}
    \centering
    \caption{Properties of the ALMA Band 8 observations presented in this work.}
    \label{tab:data}
    \begin{threeparttable}
    \begin{tabular}{lccccccc}
    
    \hline\hline
    Field ID & Frequency & $\theta_\mathrm{major}$ & $\theta_\mathrm{minor}$ & $\mathrm{PA}$ & $\mathrm{RMS}^a$ & ALMA PID & PI \\
    \hline 
    & [GHz] & [arcsec] & [arcsec] & [deg] & [mJy/beam] & & \\
    \hline
REBELS-12$^b$ & 407.4 & 0.72 & 0.59 & -62.6 & 0.37 & 2021.1.01297.S, 2024.1.00406.S & Fudamoto, Algera \\
REBELS-12-2$^b$ & 407.3 & 0.61 & 0.45 & 56.8 & 0.53 & 2021.1.01297.S & Fudamoto \\
REBELS-14 & 418.7 & 0.89 & 0.68 & -71.0 & 0.46 & 2024.1.00406.S & Algera \\
REBELS-18 & 392.2 & 0.69 & 0.58 & 43.7 & 0.50 & 2024.1.00406.S & Algera \\
REBELS-25 & 408.5 & 1.09 & 1.03 & -61.9 & 0.64 & 2022.1.01324.S & Algera \\
REBELS-39$^{c}$ & 431.7 & 0.76 & 0.69 & -89.4 & 0.37 & 2021.1.01297.S, 2024.1.00406.S & Fudamoto, Algera \\
REBELS-39-2$^{b,c}$ & 431.7 & 0.76 & 0.69 & -89.4 & 0.37 & 2021.1.01297.S, 2024.1.00406.S & Fudamoto, Algera \\
REBELS-40 & 406.6 & 0.82 & 0.73 & 85.5 & 0.63 & 2021.1.00318.S & Inami \\
\hline\hline 
\end{tabular}
    \begin{tablenotes}
    \item[a] Median root-mean-square noise per $30\,\mathrm{km/s}$ channel, except for REBELS-25 where a $35\,\mathrm{km/s}$ channel (and $1''$ tapered datacube) is used.
    \item[b] The primary beam sensitivities at the position of REBELS-12/REBELS-12-2/REBELS-39-2 are $0.95$/$0.69$/$0.94$. The remaining targets are in the pointing center (primary beam sensitivity of one). 
    \item[c] REBELS-39 and REBELS-39-2 are in the same ALMA Band 8 datacube. \\
    \end{tablenotes}
\end{threeparttable}
\end{table*}

\section{Targets and Data}\label{sec:observations}

\subsection{Targets}
\label{sec:data_targets}

In this work, we present new ALMA Band 8 \citep{shan2005,sekimoto2008,sekimoto2009} observations targeting the \oiii{} line in eight previously \cii{}-detected galaxies at $6.8 \lesssim z \lesssim 7.7$.\footnote{A compilation of recommended references for the ALMA receivers can be found in \citet{bakx2024_alma_receivers}.} Our targets are either part of ($N=6$; henceforth denoted as the `primary sample') or were serendipitously discovered in ($N=2$; the `serendipitous sample'), the REBELS survey, which is a Cycle 7 ALMA Large Program aimed at studying the dust and ISM properties of galaxies within the first $\approx800\,\mathrm{Myr}$ after the Big Bang. Following several successful pilot studies \citep{smit2018,schouws2022,schouws2023}, REBELS targeted 36 UV-luminous galaxies with photometric redshifts $z_\mathrm{phot}\gtrsim6.5$ in \cii{} emission using a spectral scanning technique.\footnote{Four additional galaxies were targeted in \oiii{} as opposed to \cii{} emission, bringing the total to 40 sources, though none were detected in \oiii{} \citep{bouwens2022,vanleeuwen2025}.} In total, the \cii{} line was detected in 27/36 targets (\citealt{bouwens2022,vanleeuwen2025}; Schouws et al.\ in preparation), 15 of which were furthermore detected in underlying dust continuum emission \citep{inami2022}. 

Our six primary targets are all detected in \cii{} and underlying ALMA Band 6 \citep{ediss2004} dust continuum emission. Whilst the \oiii{} observations presented in this work were drawn from a variety of observing programs, each with its distinct selection criteria as detailed in Section \ref{sec:ALMAobservations}, the sample is biased towards relatively luminous \cii{} and continuum emitters within REBELS. Two sources -- REBELS-12 and REBELS-25 -- were previously detected in \oiii{} emission \citep{algera2024}, though we present deeper observations towards these targets in this work (Section \ref{sec:ALMAobservations}). Observations targeting the \oiii{} line in the remaining four primary sources -- REBELS-14, REBELS-18, REBELS-39 and REBELS-40 -- are presented for the first time here.

The shallower \oiii{} observations of REBELS-12 and REBELS-25 revealed a low $\text{\oiiitocii{}}\approx1.3$ for the latter, which was ascribed to its evolved nature and a low burstiness \citep{algera2024}. For REBELS-12, the picture was less clear. Based on its \cii{} morphology and line profile, REBELS-12 was identified as a merger of two \cii{}-emitting components. However, only one of these components -- the redshifted one -- was detected in \oiii{} and rest-frame UV emission. Moreover, even in recent \textit{JWST}/NIRSpec IFU spectroscopy, no emission lines or stellar continuum were detected from the \oiii{}-faint component \citep{rowland2025}, leaving its nature unclear. As the deeper \oiii{} observations of REBELS-12 presented in this work also fail to detect emission from this component (Section \ref{sec:results}), we focus our analysis on the \oiii{}- and UV-emitting component and refer to \citet{algera2024} for an in-depth discussion of the full system.\footnote{We note that two ALMA programs targeting REBELS-12 were recently approved in Cycle 12, with the aim of detecting its \oi{}$_{145}$ emission in Band 6 (2025.1.00418.S, PI Witstok) and resolving its dust continuum in Band 7 (2025.1.01318.S, PI Stefanon). These future observations will hopefully further elucidate the nature of this intriguing galaxy.}

In three of the REBELS fields, a bright emission line was serendipitously discovered within $|\Delta v| \lesssim 250\,\mathrm{km/s}$ of the main target (\citealt{fudamoto2021,vanleeuwen2024}). Unlike the primary targets, these serendipitous line emitters are fainter in the rest-UV, suggesting they are more strongly dust-obscured (Van Leeuwen et al.\ in preparation). The small velocity offset with respect to the emission line detected in the primary target suggests that the serendipitously detected line corresponds to \cii{} emission at the same redshift. In this work, we target two of the serendipitous sources where atmospheric transmission of \oiii{} is favorable (REBELS-12-2 and REBELS-39-2), as described in detail below. While REBELS-39-2 is still relatively UV-luminous ($M_\mathrm{UV} \approx -20.5$; Van Leeuwen et al.\ in preparation), it is $>1$ magnitude fainter than the primary REBELS galaxies, which were selected to have $-23 < M_\mathrm{UV} < -21.5$ \citep{bouwens2022}. REBELS-12-2 is particularly UV-faint, with Van Leeuwen et al.\ (in preparation) inferring $M_\mathrm{UV} > -19.5$ based on deep \textit{JWST}/NIRCam observations. Given their different selection and therefore potentially different nature from the primary REBELS targets, these serendipitous sources form an interesting comparison sample in terms of \oiii{} properties (c.f., \citealt{bakx2024}). Altogether, we thus target eight $z\approx7$ galaxies in \oiii{} emission in this work.

\subsection{ALMA Data}

\subsubsection{Observations}
\label{sec:ALMAobservations}

The ALMA \oiii{} observations of our sample are drawn from several observing programs. In ALMA Cycle 8, REBELS-12, REBELS-39 and their serendipitous neighbors were observed as part of PID 2021.1.01297.S (PI Fudamoto). Besides providing insight into the ISM conditions of these sources through the \oiiitocii{} ratio, an \oiii{} detection would unambiguously confirm the high-redshift nature of the serendipitous companions, whose identification hinges on the detected line indeed being \cii{}. Given the close angular separations of the primary targets and their neighbors, both were observed in a singular Band 8 pointing. The \oiii{} observations of REBELS-12 were previously presented in \citet{algera2024}, although they did not discuss the serendipitous neighbor in the ALMA pointing. The Band 8 observations towards REBELS-39 and its neighbor have not previously been presented.

Two further REBELS targets, REBELS-25 and REBELS-40, were observed in \oiii{} emission as part of a different ALMA Cycle 8 campaign (2021.1.00318.S; PI Inami). These observations had the primary aim of constraining the dust temperatures of REBELS galaxies bright in ALMA Band 6 continuum emission. Where the atmospheric transmission allowed it -- as in the case of REBELS-25 and REBELS-40 -- the \oiii{} line was simultaneously observed. We note that the \oiii{} observations of REBELS-25 were previously presented in \citet{algera2024}, while the observations of REBELS-40 are presented for the first time here.

During Cycle 10, deeper ALMA Band 8 observations of REBELS-25 were carried out as part of program 2022.1.01324.S (PI Algera). These observations were designed to spatially resolve the \oiii{} line and dust continuum emission in REBELS-25, thereby capitalizing on the high-resolution \cii{} and dust maps of this galaxy presented in \citet{rowland2024}. In this work, however, we focus on the integrated properties of the \oiii{} line, while a high-resolution perspective will be presented in Rowland et al.\ (in preparation). As the Cycle 10 observations of REBELS-25 are significantly deeper than the Cycle 8 data previously presented in \citet{algera2024}, we do not use the latter in this work.

More recently, in ALMA Cycle 11, new Band 8 observations targeting the \oiii{} line were taken for REBELS-12, REBELS-14, REBELS-18 and REBELS-39 (2024.1.00406.S; PI Algera). The primary objective of this ALMA program is to provide sensitive dust continuum observations of eight REBELS-IFU galaxies with \textit{JWST}/NIRSpec follow-up (Section \ref{sec:dataJWST}). For these four targets, the \oiii{} line could simultaneously be observed without loss of continuum sensitivity. While REBELS-12 and REBELS-39 were previously observed in \oiii{}, they were not continuum-detected in the Cycle 8 observations, which motivated re-observing these two galaxies in Cycle 11. A full overview of the new Band 8 continuum data of the eight REBELS-IFU targets will be the subject of Algera et al.\ (in preparation). The Cycle 11 observations also cover REBELS-39-2, while REBELS-12-2 falls outside the field-of-view.\footnote{Unlike the Cycle 8 observations, which adopted a pointing center in between REBELS-12 and REBELS-12-2, the Cycle 11 observations are centered on REBELS-12, such that the primary beam sensitivity at the location of the serendipitous neighbor is only $\sim 0.2$. }

\subsubsection{Calibration \& Imaging}
\label{sec:calibrationImaging}

The observations presented in this work constitute eight ALMA Band 8 pointings towards six separate fields (REBELS-12 [$2\times$], REBELS-14, REBELS-18, REBELS-25, REBELS-39 [$2\times$], REBELS-40) drawn from four separate observing programs. We restore the calibrated visibilities of the Band 8 observations in {\sc{casa}} \citep{mcmullin2007,casa2022}, using the standard {\sc{scriptForPI.py}} provided by the ALMA observatory upon retrieving the data from the archive. 

We create datacubes from the eight ALMA observations to search for \oiii{} emission from our primary and serendipitous targets. We use a slightly different imaging approach for REBELS-25, given the high native spatial resolution of its observations, as detailed below. For the other seven pointings, we start by creating naturally-weighted datacubes using the {\sc{tclean}} task in {\sc{casa}}. First, we create dirty cubes with a channel width of $30\,\mathrm{km/s}$, and use these to determine the RMS in the cubes via a $3.5\sigma$ sigma-clipped standard deviation. We subsequently use twice the RMS as our cleaning threshold, making use of the automasking capabilities implemented in {\sc{casa}} \citep{kepley2020}. We present the properties of the ALMA Band 8 datacubes in Table \ref{tab:data}. The typical native resolution of our datacubes is $\sim0.7'' \times 0.6''$, while the $1\sigma$ line sensitivity ranges from $0.37-0.63\,\mathrm{mJy/beam}$ measured in $30\,\mathrm{km/s}$ channels.

The imaging strategy for the high-resolution \oiii{} observations of REBELS-25 closely follows the approach by \citet{rowland2024} and will be presented in Rowland et al.\ (in preparation). For the purposes of this work, however, we focus on the source-integrated emission of REBELS-25, for which we use the \oiii{} datacube tapered to a resolution of $1.0''$ and imaged at a velocity resolution of $35\,\mathrm{km/s}$ that will also be presented in Rowland et al.\ (in preparation). The reason for adopting a tapered cube is that flux measurements in the high-resolution ($\sim0.17''$) cube may be biased due to a mismatch between the areas of the dirty and clean beams, known as the ``\citet{jorsater1995} effect'' \citep[see also][for an extensive discussion]{czekala2021}. For the rest of our targets, with a typical resolution of $\sim0.6-0.8''$, this is not a concern, as demonstrated in Appendix \ref{app:tapering}. The tapered \oiii{} cube of REBELS-25 achieves a sensitivity of $0.64\,\mathrm{mJy/beam}$ per $35\,\mathrm{km/s}$ channel around the line frequency (Table \ref{tab:data}).

\subsubsection{Line Flux Measurements}
\label{sec:fluxMeasurements}

We adopt an iterative procedure to extract the \oiii{} line fluxes of our targets, except for REBELS-25, for which we directly adopt the flux from Rowland et al.\ (in preparation). We start by extracting the flux in each velocity channel, using a circular aperture centered on the \cii{} position of our target. The size of the aperture is chosen manually for each source to ensure the full line flux is captured, and thus depends on the spatial extent of the \oiii{} emission of the galaxy and the resolution of the data. We fit the extracted spectrum with a single Gaussian to obtain a first guess for the center, amplitude and FWHM of the emission line. Where necessary, this fit also includes a constant offset from zero to account for a marginal level of continuum emission.\footnote{This is only the case for REBELS-18; for the other sources the continuum level is found to be consistent with zero at the $<1\sigma$ level, and we therefore repeat the fit without a constant offset. We prefer this method over using {\sc{uvcontsub}} as the latter is formally only applicable to sources in the phase center of the observations, whereas our sample includes off-center serendipitous sources. Moreover, allowing for a constant offset in the Gaussian fit enables propagating the uncertainty on the continuum level into the error on the line flux. Nevertheless, we have confirmed that the \oiii{} fluxes of the primary targets are consistent within $<1\sigma$ when using {\sc{uvcontsub}} instead of fitting a flat continuum in the image domain.} Following this initial Gaussian fit, we create a moment-0 map across $1.2\times$ the FWHM of the emission line, which maximizes its S/N assuming a Gaussian profile (e.g., \citealt{novak2020}). We fit the moment-0 map with a 2D Gaussian, re-center the aperture based on the Gaussian centroid while keeping its size fixed, and re-extract and fit the spectrum. We repeat this procedure until convergence is reached, which is the case in $\leq3$ iterations. We then adopt the area under the 1D Gaussian fit as our line flux measurement. 

We confirm that the \oiii{} line centroids, FWHMs, and fluxes measured for REBELS-12, REBELS-39 and REBELS-39-2 are consistent to within $1\sigma$ between the Cycle 8 and Cycle 11 observations. To improve our sensitivity, we then concatenate these two datasets using {\sc{casa}} task {\sc{concat}} and ensure the visibilities are appropriately weighted via task {\sc{statwt}}. We subsequently re-image the concatenated data following the aforementioned procedure, noting that for REBELS-12 this yields a small mosaic given the different Band 8 pointing centers. We utilize these deeper, concatenated datasets when discussing the properties of REBELS-12, REBELS-39 and REBELS-39-2.

We re-extract the \cii{} fluxes of our targets, which were previously measured by Schouws et al.\ (in preparation; see also \citealt{bouwens2022}) following the same procedure for consistency.\footnote{As discussed in Section \ref{sec:data_targets}, REBELS-12 has a peculiar \cii{} morphology. To extract its full \cii{} flux while simultaneously avoiding picking up noise, we adopt a large aperture of $2.5''$, but extract the spectrum only across pixels with significance $\geq2\sigma$ in the moment-0 map (following \citealt{algera2024}).} In doing so, we also re-image the \cii{} datacubes using a velocity resolution of $30\,\mathrm{km/s}$, to match the channel width of the \oiii{} data. For REBELS-25, we directly adopt the \cii{} flux from Rowland et al.\ (in preparation), which was measured in a consistent manner to the \oiii{} flux (i.e., utilizing a $1''$ tapered \cii{} datacube, albeit with a higher spectral resolution of $15\,\mathrm{km/s}$). We present the \cii{} and \oiii{} flux measurements of our full sample in Table \ref{tab:lineFluxes}.

\begin{figure*}
    \centering
    \includegraphics[width=1.0\textwidth]{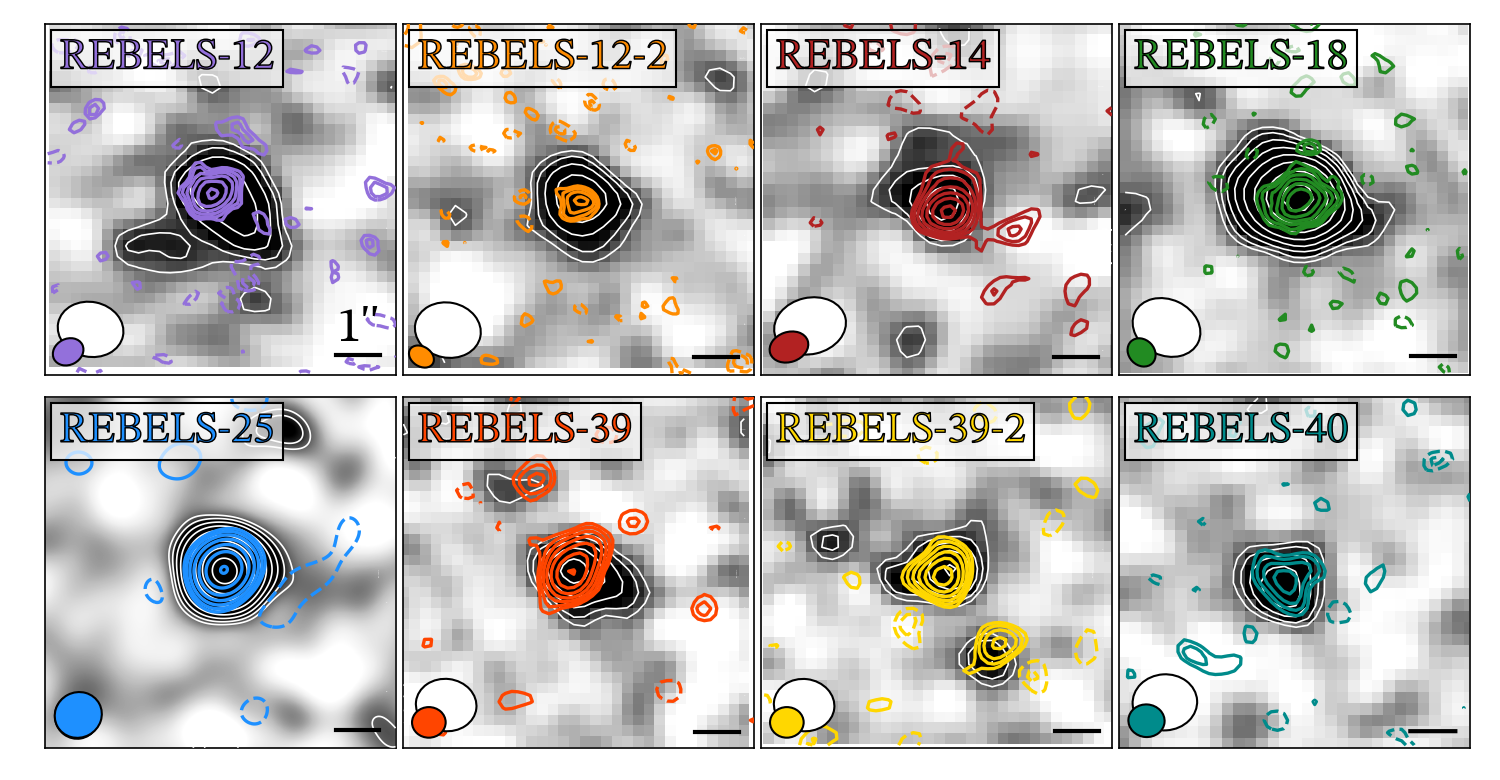}
    \caption{Moment-0 maps of the \cii{} (background image and thin, white contours) and \oiii{} lines (colored contours) of our eight targets. North is up and east is to the left. Contours are spaced as $\pm 2^N \times \sigma$ where $N=1, 1.5, 2, 2.5, \ldots$ and $\sigma$ is the RMS in the moment-0 map. Negative contours are shown as dashed lines, and the background colorscale ranges from $-1.5\sigma$ to $3.5\sigma$. The \cii{} (\oiii{}) beam size is shown via the white (colored) ellipse in the bottom left corner, and the scalebar in the lower right corner indicates a length of $1'' \sim 5.2\,\mathrm{kpc}$ in all panels. All of our eight targets are detected in both \cii{} and \oiii{} emission at a peak S/N of $6.2 - 17.7\sigma$.}
    \label{fig:momentZero}
\end{figure*}

\subsection{JWST Data}
\label{sec:dataJWST}

Among the full \cii{}-detected REBELS sample, twelve galaxies were observed with the \textit{JWST}/NIRSpec Integral Field Unit (IFU) as part of the Cycle 1 `REBELS-IFU' program (GO 1626; PI Stefanon). For details on the science aims, data reduction, and line flux measurements, we refer to Stefanon et al.\ (in preparation) and \citet{rowland2025}, while we provide a brief summary here.

The 12 REBELS galaxies at $6.5 \lesssim z \lesssim 7.7$ were observed with the NIRSpec IFU using the prism mode. The targets were selected with a preference towards galaxies with bright \cii{} lines and dust continuum detections, as discussed in more detail in \citet{rowland2025} and Stefanon et al.\ (in preparation). This selection likely biases the sample towards the most massive and metal-rich galaxies within REBELS, while the parent survey itself already targeted some of the most UV-luminous and therefore likely massive galaxies at $z\gtrsim6.5$ \citep{bouwens2022,topping2022}.

The NIRSpec/IFU observations yield contiguous spectral coverage from $0.6 - 5.3\,\mu\mathrm{m}$ at low spectral resolution ($R\sim100$), and cover a field-of-view of $3''\times3''$ (approximately $15.7\,\mathrm{kpc}\times 15.7\,\mathrm{kpc}$ at $z=7$). This wavelength range provides coverage of a variety of key optical lines such as \oiidopt{}, H$\gamma$, H$\beta$ and \oiiidopt{}, as well as H$\alpha$ below $z\approx7$ (though see \citealt{deugenio2025,pollock2025} for possible extensions of NIRSpec to longer wavelengths). In total, among the eight galaxies observed in \oiii{} emission as part of this work, REBELS-12, REBELS-14, REBELS-18, REBELS-25, and REBELS-39 are targeted in the IFU observations, with only the latter being at $z < 7$ (Table \ref{tab:lineFluxes}). We will refer to this subset of five sources as the `REBELS-IFU sample'.

In each of the REBELS-IFU targets, the \oiidopt{}, H$\beta$ and \oiiidopt{} lines are detected at high significance, while for REBELS-39 the H$\alpha$ line is also detected. Emission line fluxes were measured by \citet{rowland2025}, after first subtracting the continuum using a third-order polynomial. In their fits, the widths of the various emission lines were set according to the NIRSpec prism line spread function. The \oii{} doublet is unresolved at the prism resolution, and is therefore fitted with a single Gaussian. The \oiiidopt{} doublet can be spectrally resolved, and is fitted with two Gaussians with the constraint that their flux ratio is \oiiiopt{}/\oiiinowave{}$_{4959} = 2.98$, as set by atomic physics \citep{storey_zeippen2000}. Moreover, H$\beta$ and H$\gamma$ are both fitted with a single Gaussian, although the latter is likely blended with the \oiiinowave{}$_{4363}$ and \feii{}$_{4360}$ lines, and we therefore follow \citet{rowland2025} by not using the H$\gamma$ fluxes in this work. The H$\alpha$ line, which we cover only for REBELS-39, is moderately blended with the \nii{}$_{6548,84}$ doublet, and \citet{rowland2025} therefore deblend the lines through a triple-Gaussian fit, keeping the flux ratio of the two nitrogen lines fixed to \nii{}$_{6584}$/\nii{}$_{6548} = 3.049$ \citep{osterbrock_ferland2006}.

\begin{table*}
    \def\arraystretch{1.5}
    \centering
    \caption{Emission line parameters for \cii{} and \oiii{}}
    \label{tab:lineFluxes}
    \begin{threeparttable}
    \begin{tabular}{lccccccccc}
    
    \hline\hline 
    REBELS ID & $z_\text{\cii{}}$ & FWHM$_\text{\cii{}}$ & $S_\text{\cii{}}$ & $L_\text{\cii{}}$ & $z_\text{\oiii{}}$ & FWHM$_\text{\oiii{}}$ & $S_\text{\oiii{}}$ & $L_\text{\oiii{}}$ & \oiiitocii{} \\
    \hline 
    &  & [km/s] & [mJy km/s] & [$10^8\,$L$_\odot$] &  & [km/s] & [mJy km/s] & [$10^8\,$L$_\odot$] &  \\
    \hline
REBELS-12$^{a}$ & $7.3517_{-0.0017}^{+0.0017}$ & $377_{-105}^{+292}$ & $437_{-152}^{+183}$ & $5.5_{-1.9}^{+2.3}$ & $7.3495_{-0.0003}^{+0.0003}$ & $179\pm26$ & $753_{-141}^{+151}$ & $16.9_{-3.1}^{+3.5}$ & $3.1_{-1.0}^{+1.9}$ \\
REBELS-12-full$^{a}$ & $7.3448_{-0.0015}^{+0.0016}$ & $1038\pm134$ & $969_{-162}^{+172}$ & $12.2_{-2.0}^{+2.2}$ & - & - & - & - & - \\
REBELS-12-blue$^{a}$ & $7.3358_{-0.0028}^{+0.0028}$ & $502_{-144}^{+298}$ & $449_{-150}^{+187}$ & $5.6_{-1.9}^{+2.4}$ & - & - & $<514$ & $<11.5$ & $<2.0$ \\
REBELS-12-2 & $7.3514_{-0.0005}^{+0.0005}$ & $237\pm45$ & $533_{-125}^{+143}$ & $6.7_{-1.6}^{+1.8}$ & $7.3525_{-0.0012}^{+0.0012}$ & $310\pm98$ & $941_{-369}^{+465}$ & $21.4_{-8.4}^{+9.9}$ & $3.2_{-1.3}^{+2.0}$ \\
REBELS-14 & $7.0833_{-0.0005}^{+0.0005}$ & $156\pm45$ & $282_{-104}^{+125}$ & $3.4_{-1.2}^{+1.5}$ & $7.0842_{-0.0004}^{+0.0004}$ & $311\pm35$ & $1520_{-222}^{+236}$ & $32.5_{-4.7}^{+5.0}$ & $9.6_{-3.1}^{+5.7}$ \\
REBELS-18 & $7.6748_{-0.0002}^{+0.0002}$ & $224\pm20$ & $849_{-97}^{+99}$ & $11.4_{-1.3}^{+1.4}$ & $7.6759_{-0.0003}^{+0.0003}$ & $213\pm28$ & $1391_{-231}^{+250}$ & $33.3_{-5.6}^{+6.0}$ & $2.9_{-0.6}^{+0.7}$ \\
REBELS-25$^{b}$ & $7.3075_{-0.0002}^{+0.0002}$ & $345\pm15$ & $1410_{-150}^{+150}$ & $17.6_{-1.9}^{+1.8}$ & $7.3076_{-0.0004}^{+0.0004}$ & $295\pm34$ & $1460_{-220}^{+220}$ & $32.6_{-4.8}^{+4.9}$ & $1.9_{-0.3}^{+0.4}$ \\
REBELS-39 & $6.8475_{-0.0012}^{+0.0012}$ & $567\pm104$ & $691_{-166}^{+191}$ & $7.8_{-1.9}^{+2.1}$ & $6.8471_{-0.0006}^{+0.0006}$ & $476\pm55$ & $1808_{-260}^{+283}$ & $36.9_{-5.4}^{+5.9}$ & $4.7_{-1.2}^{+1.7}$ \\
REBELS-39-2 & $6.8384_{-0.0006}^{+0.0006}$ & $224\pm56$ & $367_{-111}^{+134}$ & $4.2_{-1.3}^{+1.5}$ & $6.8379_{-0.0006}^{+0.0006}$ & $268\pm51$ & $852_{-207}^{+231}$ & $17.3_{-4.2}^{+4.7}$ & $4.1_{-1.4}^{+2.2}$ \\
REBELS-40 & $7.3649_{-0.0009}^{+0.0009}$ & $329\pm79$ & $377_{-113}^{+134}$ & $4.8_{-1.4}^{+1.6}$ & $7.3652_{-0.0006}^{+0.0006}$ & $259\pm49$ & $991_{-238}^{+273}$ & $22.4_{-5.5}^{+6.0}$ & $4.7_{-1.6}^{+2.5}$ \\

\hline\hline 
\end{tabular}
\begin{tablenotes}
    \item[a] The \cii{} emission of REBELS-12 suggests it is a merger with only one component being detected in \oiii{} \citep{algera2024}. This component appears responsible for the emission seen in \textit{JWST}/NIRSpec, and is therefore the focus of our work. The full and blue-shifted \cii{} emission are denoted as REBELS-12-full and REBELS-12-blue, respectively.
    \item[b] The emission line parameters for REBELS-25 are taken from Rowland et al.\ (in preparation; see also \citealt{rowland2024}). 
\end{tablenotes}
\end{threeparttable}
\end{table*}

\subsubsection{Dust corrections}
\label{sec:methods_dust_correction}

For REBELS-39, in which both the H$\alpha$ and H$\beta$ lines are detected, \citet{rowland2025} correct the emission line fluxes for dust based on the Balmer decrement, assuming a \citet{calzetti2000} attenuation law. This yields a nebular attenuation for REBELS-39 of $A_{V,\mathrm{neb}} = 0.62 \pm 0.40$. For the $z>7$ sample, however, this is not possible due to the lack of an H$\alpha$ detection and no reliable measurement of the H$\gamma$ line. Instead, we therefore apply dust corrections to the four $z>7$ sources based on the stellar $A_{V,*}$ obtained from SED-fitting of their NIRSpec spectra by \citet{fisher2025}. For these four sources, they measure an $A_{V,*} = 0.17 - 0.27$. However, it is well-known that the nebular attenuation often exceeds the stellar attenuation (e.g., \citealt{reddy2015,buat2018,shivaei2020}) due to the ionized regions from where these emission lines originate typically being more dust-enshrouded. For the eight $z<7$ REBELS-IFU galaxies where the \textit{JWST} spectra cover the H$\alpha$ line, Fisher et al.\ (in preparation) compare the stellar $A_{V,*}$ obtained from their SED fits with the nebular value obtained from the Balmer decrement. They find that, on average, the nebular emission is more attenuated by a multiplicative factor of $2.02 \pm 0.33$. We therefore dust-correct the emission line fluxes of our $z>7$ targets using their stellar $A_{V,*}$ inferred from SED-fitting, corrected by this factor, resulting in a nebular attenuation of $A_{V,\mathrm{neb}} = 0.34 - 0.53$. We discuss the effects of dust in further detail in Section \ref{sec:caveats}.

\subsubsection{Metallicities}
Metallicities of the REBELS-IFU sample were measured by \citet{rowland2025}. As any auroral lines are either not detected or blended with stronger emission lines (e.g., \oiiinowave{}$_{4363}$ and H$\gamma$), we rely on the strong-line calibrations from \citet{sanders2024}. As extensively discussed in \citet{rowland2025}, we use the most robust available metallicity diagnostic on a case-by-case basis. For REBELS-39, where a dust correction based on the Balmer decrement is possible, we measure the metallicity through the $R23 = (\text{\oiiidopt{} + \oiidopt{}})/\mathrm{H}\beta$ diagnostic, which is generally considered the most accurate strong-line calibration (e.g., \citealt{maiolino2019}). However, for our four $z > 7$ targets we instead adopt the $R3 = \text{\oiiiopt{}}/\mathrm{H}\beta$ diagnostic, which relies on two lines closely separated in wavelength, and is thus not significantly affected by dust.

Both the $R23$ and $R3$ diagnostics are double-valued, i.e., there is no one-to-one relation between the line ratio and oxygen abundance. \citet{rowland2025} therefore use two other diagnostics, $O32$ and $Ne3O2$, to select the upper or lower branch (see also e.g., \citealt{kewley2008}). For all targets but REBELS-14, this selects the upper branch, while REBELS-14 is consistent with the peak of the $R3$-metallicity relation, implying it has the lowest metallicity among our sample. Overall, the metallicities of our five REBELS-IFU targets span $7.90 \lesssim 12 + \log(\mathrm{O/H}) \lesssim 8.62$, equivalent to $0.16 \lesssim Z/Z_\odot \lesssim 0.85$. Our most metal-rich target is REBELS-25, which was previously suggested to be a `mature' galaxy based on its disk-like \cii{}-kinematics \citep{hygate2023,rowland2024} and high dust mass \citep{algera2024b}. 

\subsubsection{$O32$ and \EW{} measurements}
In Section \ref{sec:oiii2cii_discussion}, we discuss the \oiiitocii{} ratios of our REBELS-IFU targets in the context of two further observables: the $O32 = \text{\oiiiopt{}}/\text{\oiidopt{}}$ ratio, a proxy for ionization parameter (e.g., \citealt{kewley_dopita2002}), and the equivalent width of the \oiiidopt{}+H$\beta$ complex, which we use as a proxy for burstiness. For the $O32$ ratio, we use the \oiiiopt{} and \oii{} fluxes of the REBELS-IFU sample measured by \citet{rowland2025}, applying the dust corrections from Section \ref{sec:methods_dust_correction}. \citet{rowland2025} also provide the $\mathrm{EW}(\text{\oiiiopt{}})$ line for all twelve REBELS-IFU galaxies in their work (see also \citealt{endsley2025}), and here we use their approach to compute $\mathrm{EW}(\text{\oiiidopt{}}+\mathrm{H}\beta)$ in an identical manner.

\subsection{The Dwarf Galaxy Sample}
\label{sec:data_DGS}

Throughout this work, we will compare the \oiiitocii{} ratios of our high-redshift targets to local dwarf galaxies drawn from the \textit{Herschel} Dwarf Galaxy Survey \citep[DGS;][]{madden2013,cormier2015}. In total, 38 of the DGS sources have been observed in both the \cii{} and \oiii{} lines. We here focus on the 26 sources with $\mathrm{sSFR} > 10^{-10}\,\mathrm{yr}^{-1}$ to include only dwarfs that are star-forming, enabling a more direct comparison to \oiii{}-detected galaxies at high redshift.

Metallicities of the DGS sample have been determined in \citet{madden2013}, and make use of optical strong lines following \citet{pilyugin_thuan2005}. Fine structure line fluxes for the DGS galaxies are provided in \citet{cormier2015}, and we adopt their measurements directly. We also make use of the dust-corrected \oiiinowave{}$_{5007}$ and \oii{} fluxes provided by \citet{devis2017} to infer $O32$ for the DGS sources, and adopt \EW{} measurements from \citet{kumari2024}.

The comparison between high-redshift and dwarf galaxies is primarily motivated by the fact that both populations show typical line ratios of $\text{\oiiitocii{}} > 1$ \citep[e.g.,][]{harikane2020,algera2024}. Furthermore, the typical metallicities and ionization parameters of local dwarfs are similar to those of \oiii{}-detected high-redshift galaxies (Section \ref{sec:oiii2cii_discussion}), which are two parameters thought to be important in setting the \oiiitocii{} ratio \citep[e.g.,][]{vallini2015,vallini2020,olsen2017,katz2019}. In addition, high-redshift galaxies and dwarf galaxies are typically found to follow the same relation between \oiii{} luminosity and star-formation rate \citep[e.g.,][see also Section \ref{sec:oiiiSFR}]{delooze2014,harikane2020,zavala2024_alma,witstok2025}.

On the other hand -- as is the case with any comparison to purported low-$z$ analogs -- the comparison to local dwarfs is not perfect, given that the DGS sample has much lower SFRs ($\mathrm{SFR} \sim 10^{-2} - 30\,M_\odot\,\mathrm{yr}^{-1}$) than those of typical \oiii{}-detected galaxies at $z\gtrsim6$ ($\mathrm{SFRs}\sim5-300\,M_\odot\,\mathrm{yr}^{-1}$). While important to keep in mind, dwarf galaxies still constitute the best comparison sample to high-redshift \oiii{}-detected galaxies on the basis of their other physical properties.

\section{Results}
\label{sec:results}

We present the results of our new \oiii{} observations in this section. We first conclusively establish through a second emission line detection that the two serendipitous neighbors are, indeed, at $z\approx7$. Subsequently, we collectively discuss the \oiii{} lines of our full sample of eight high-redshift galaxies.

\subsection{\oiii{} emission and redshift confirmation of the serendipitous neighbors}

As discussed in \citet{fudamoto2021} and \citet{vanleeuwen2024}, the ALMA datacubes of both the REBELS-12 and REBELS-39 fields reveal a bright, serendipitous emission line a distance of $11.5''$ and $2.4''$ away from the primary REBELS targets, respectively. The small velocity difference between the \cii{} lines of the two primary targets and these bright serendipitous lines ($|\Delta v| \lesssim250\,\mathrm{km/s}$) suggests they are emitted by galaxies at the same redshift, and that the serendipitous lines thus also correspond to \cii{}. However, when limited to just a single emission line, one cannot definitively rule out that the detected line instead originates from a foreground galaxy not physically associated with the primary REBELS targets (see e.g., \citealt{bakx2024,jin2024}). 

\begin{figure*}
    \centering
    \includegraphics[width=0.45\textwidth]{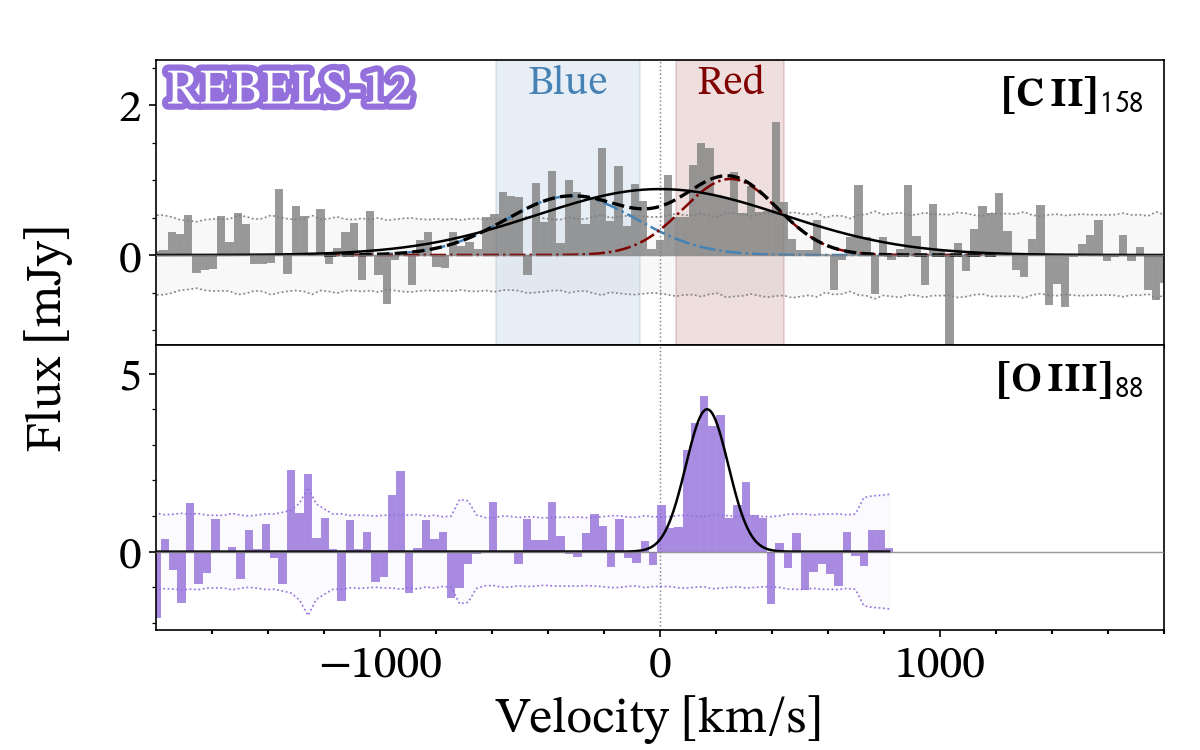}
    \includegraphics[width=0.45\textwidth]{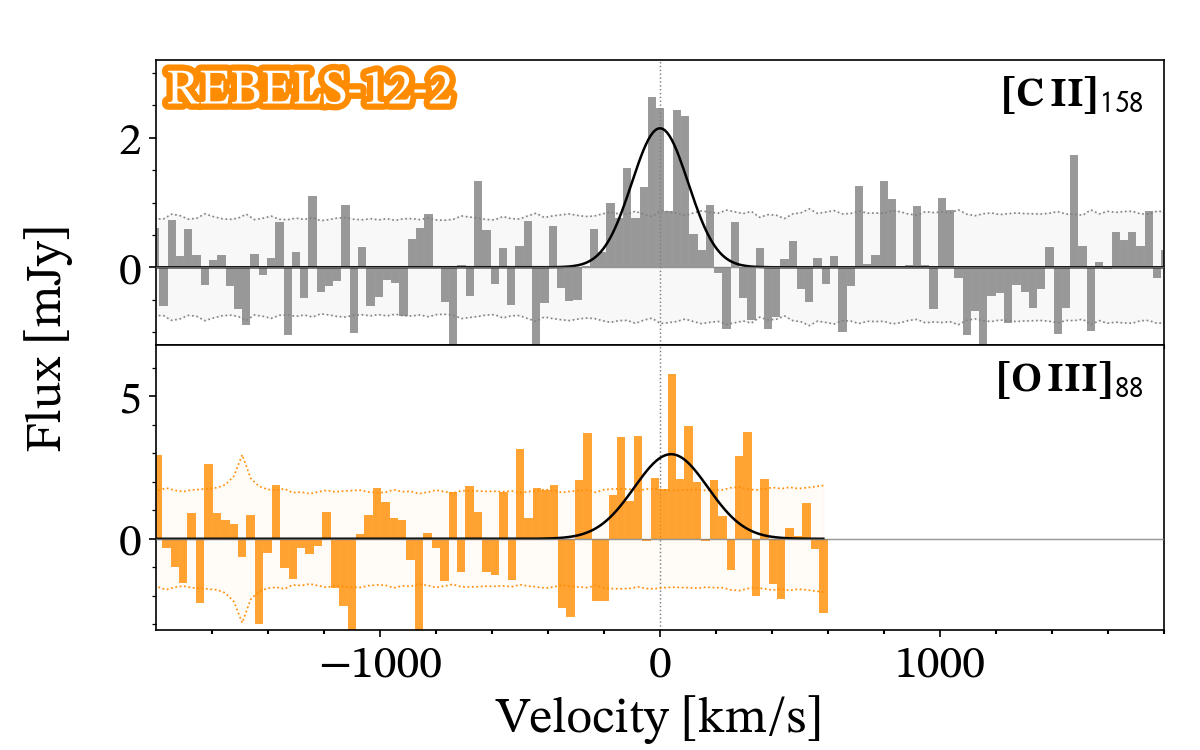} \hfill
    \includegraphics[width=0.45\textwidth]{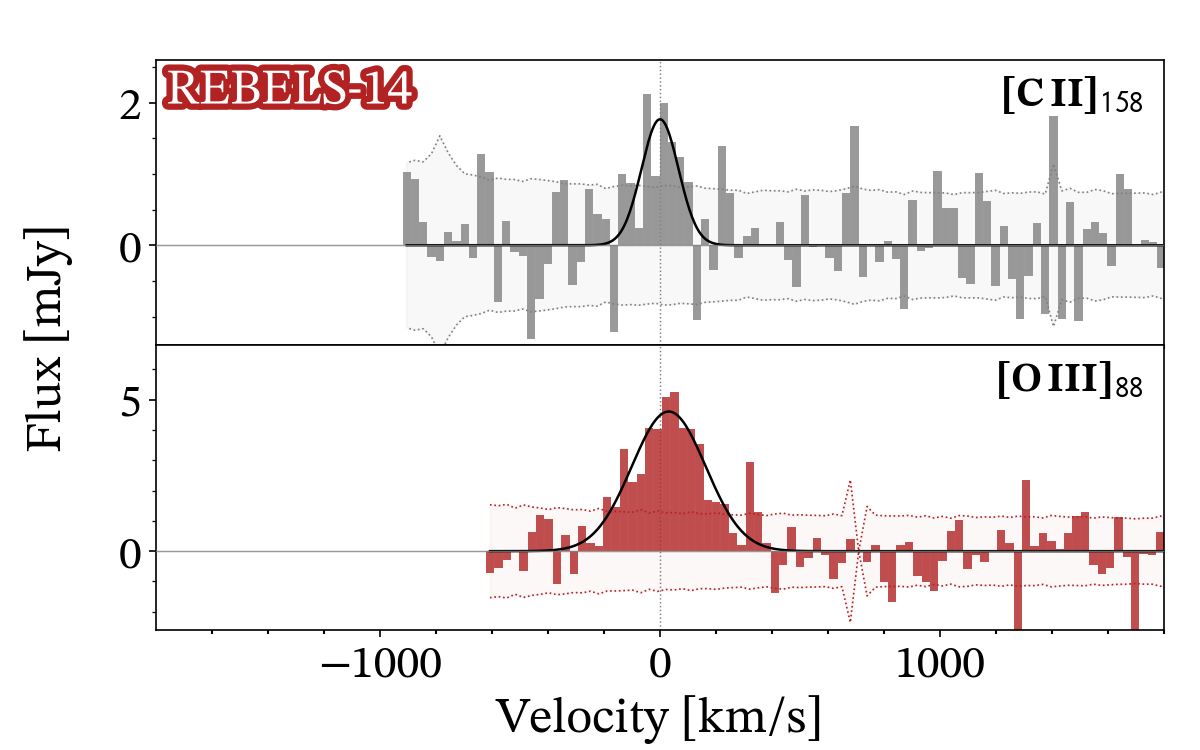}
    \includegraphics[width=0.45\textwidth]{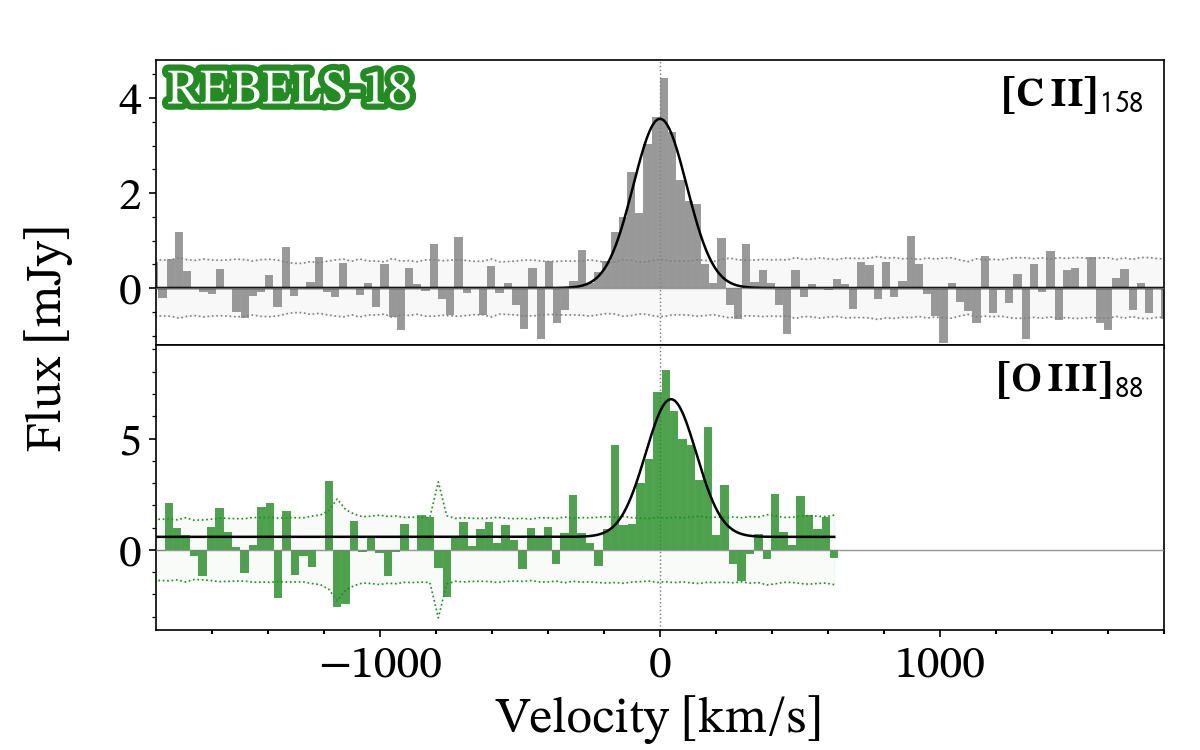} \hfill
    \includegraphics[width=0.45\textwidth]{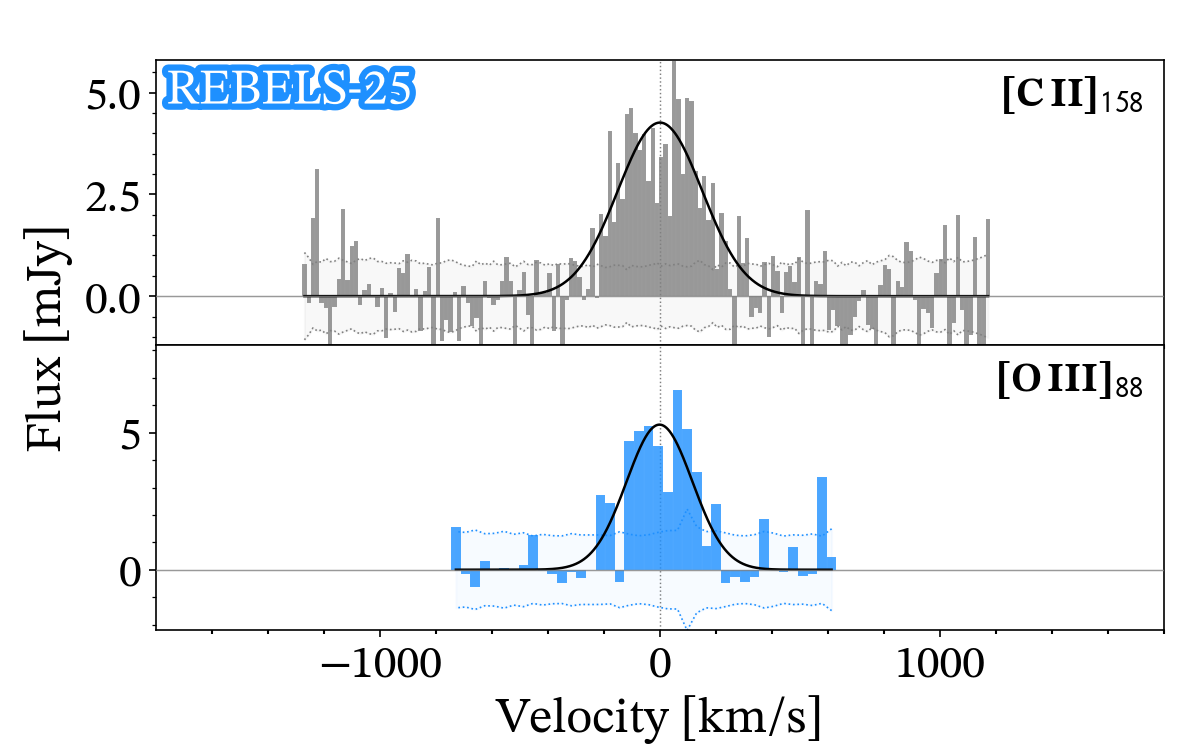}
    \includegraphics[width=0.45\textwidth]{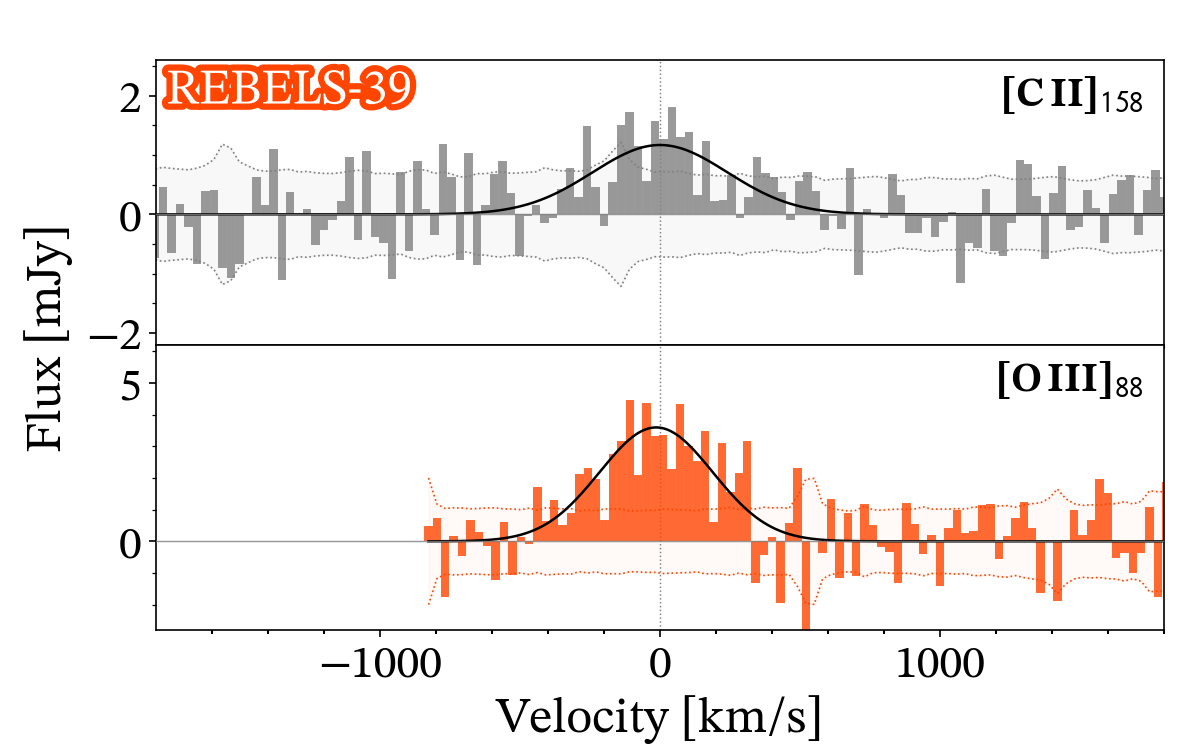} \hfill
    \includegraphics[width=0.45\textwidth]{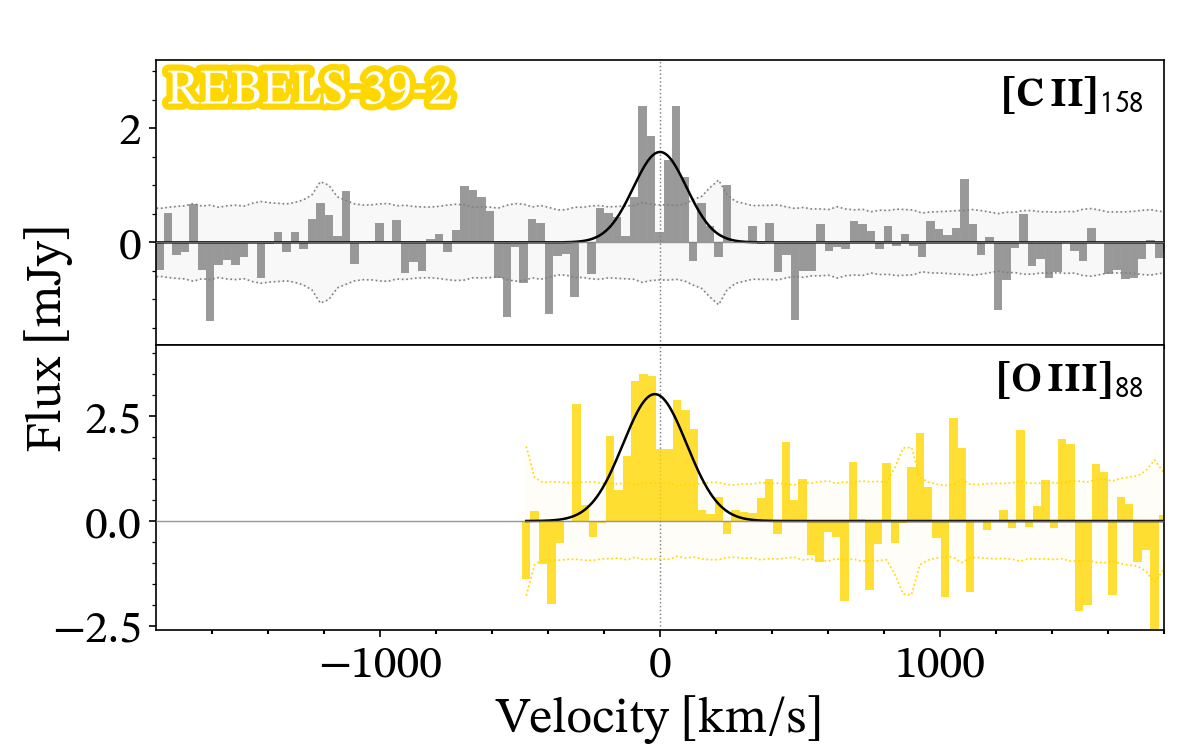}
    \includegraphics[width=0.45\textwidth]{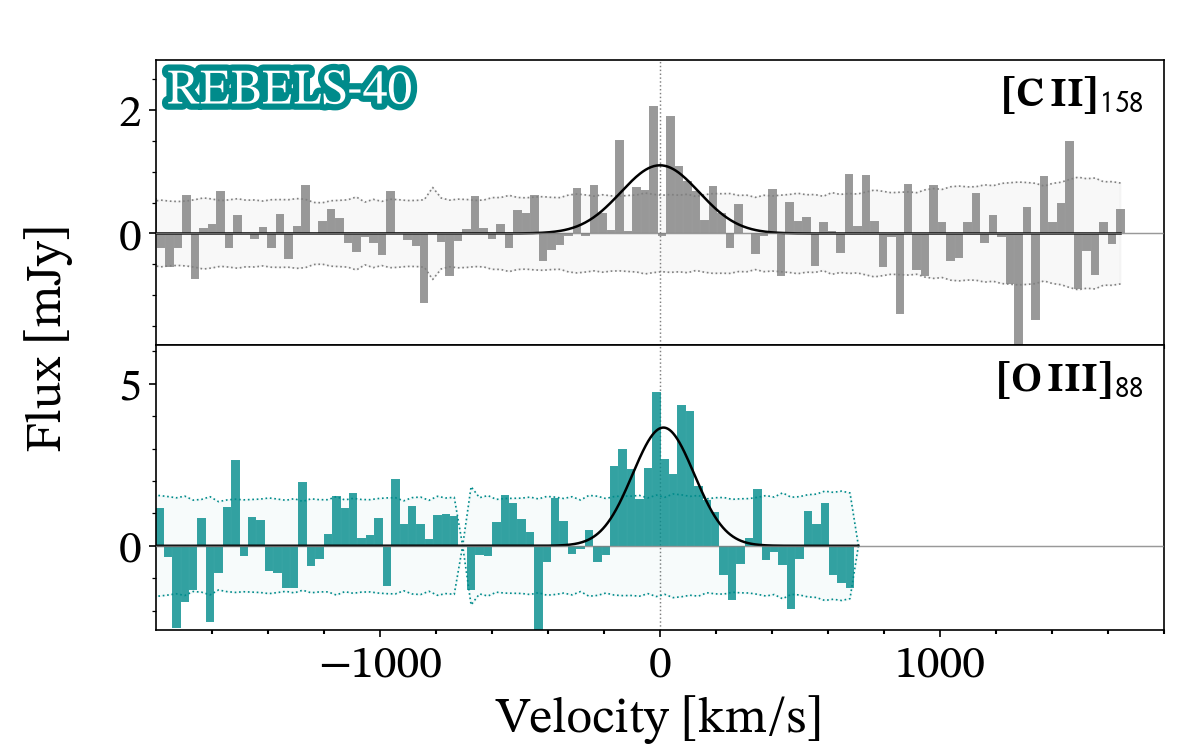} \hfill
    \caption{Extracted \cii{} (\textit{top panels}) and \oiii{} spectra (\textit{bottom panels}) of our six primary and two serendipitous targets (REBELS-12-2 and REBELS-39-2). The velocity zeropoint is defined via the center of the \cii{} line, and is shown through the vertical grey dotted line. The RMS per channel is indicated through the shaded region with dotted outline. We overplot the best Gaussian fit to the spectra as the black line. For REBELS-12 (top left), whose \cii{} and \oiii{} line profiles differ significantly as already noted by \citet{algera2024}, we moreover show a double-Gaussian fit (dashed line), and indicate the FWHM of both components with the blue and red shading. The blueshifted Gaussian component of the \cii{} line is not detected in \oiii{}. The redshifted component, as well as all other \cii{}-detected REBELS galaxies studied here, are all \oiii{}-detected.}
    \label{fig:1Dspectra}
\end{figure*}

Our ALMA Band 8 observations detect a second line in REBELS-12-2 and REBELS-39-2 at $6.2$ and $13.4\sigma$ significance, respectively, measured as the peak pixel S/N in the moment-0 maps. The lines are detected at the exact frequency where \oiii{} emission is expected, if the initial detections by \citet{fudamoto2021} and \citet{vanleeuwen2024} are identified as \cii{}. The detection of this additional emission line thus unambiguously confirms the $z\approx7$ nature of the two serendipitously detected sources, REBELS-12-2 and REBELS-39-2. We show the moment-0 maps and 1D spectra of the lines in Figures \ref{fig:momentZero} and \ref{fig:1Dspectra}, respectively. Moreover, we present the emission line parameters for both the \oiii{} and \cii{} lines in Table \ref{tab:lineFluxes}.

\subsection{\oiii{} emission from the primary targets}
In addition to the two serendipitous galaxies, our ALMA Band 8 observations target six previously \cii{}-detected galaxies from the REBELS survey. We detect the \oiii{} line in all six at $6.6 - 17.7\,\sigma$ significance, recalling that REBELS-12 and REBELS-25 were previously \oiii{}-detected at lower S/N in the shallower observations presented in \citet{algera2024}. 

We show the \oiii{} moment-0 maps and 1D line profiles in Figures \ref{fig:momentZero} and \ref{fig:1Dspectra}, respectively, and refer to Table \ref{tab:lineFluxes} for the emission line parameters. From the Gaussian fits to the 1D line profiles, we find the line luminosities of our targets to span $L_\text{\oiii{}} = (1.7 - 3.7) \times 10^9\,L_\odot$. 

We note that the line luminosity obtained for REBELS-25 of $L_\text{\oiii{}} = (3.3 \pm 0.5) \times 10^9\,L_\odot$ is about $1.6\times$ larger than what was previously measured by \citet{algera2024} based on shallower observations [$L_\text{\oiii{}} = (2.0 \pm 0.5)\times10^9\,L_\odot$]. This is likely due to limitations in the shallower data; the \oiii{} line from REBELS-25 partially coincides with an atmospheric absorption feature, increasing the noise per channel at $+100\,\mathrm{km/s}$ (Fig.\ \ref{fig:1Dspectra}). In this region of elevated noise, no \oiii{} signal was detected in the shallower Cycle 8 data from \citet{algera2024}, which resulted in a Gaussian fit returning a narrower FWHM and a blueshifted line centroid compared to the new Cycle 10 observations. In these newer, high-S/N data, the line centroids and FWHMs are consistent between the \cii{} and \oiii{} lines (Section \ref{sec:results_oiii2cii}).

\subsection{\oiii{} and \cii{} morpho-kinematics}
\label{sec:results_oiii2cii}

\begin{figure*}
    \centering
    \includegraphics[width=0.45\textwidth]{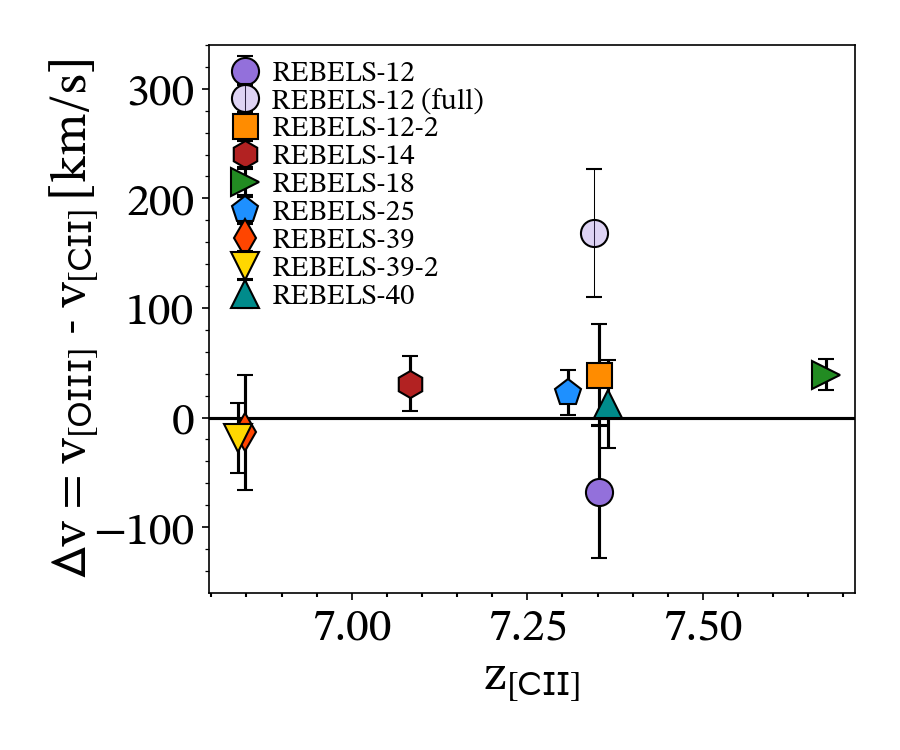}
    \includegraphics[width=0.45\textwidth]{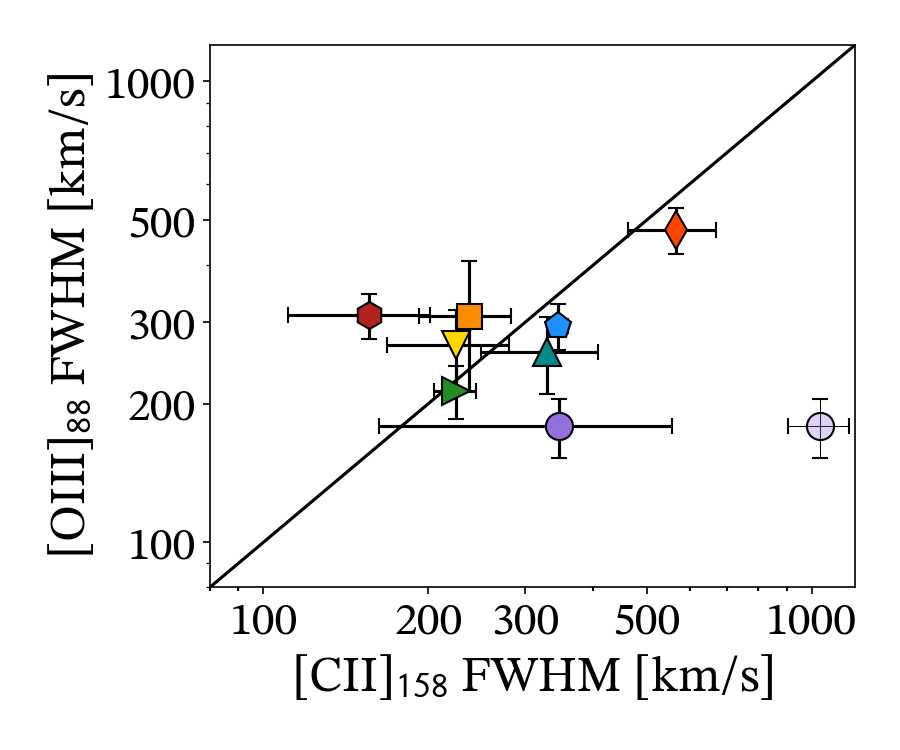}
    \caption{\textit{Left}: the velocity offset between the \oiii{} and \cii{} lines of our sample is consistent with zero within $\lesssim2\sigma$ when only the redshifted \cii{}-emitting component of REBELS-12 is used (see text). \textit{Right}: comparison of the FWHMs for the \cii{} and \oiii{} lines. These too are consistent for 7/8 sources; only in REBELS-14 is the \oiii{} line noticeably broader than its \cii{} emission.}
    \label{fig:CII_OIII_comparison}
\end{figure*}

As both the \cii{} and \oiii{} lines are robustly detected across our entire sample, it is interesting to compare their global line kinematics. In Figure \ref{fig:CII_OIII_comparison} we compare the centroids (left panel) and FWHMs (right panel) of both emission lines. We find that the redshifts implied by the \cii{} and \oiii{} lines are consistent to within $\lesssim40\,\mathrm{km\,s}^{-1}$ for 7/8 sources, with only REBELS-12 showing a clear difference between the two line centers. Moreover, the FWHMs of both lines are consistent within $\lesssim 1\sigma$ for 6/8 sources, suggesting that both lines generally trace the same underlying gravitational potential. For two of our targets -- REBELS-14 and again REBELS-12 -- the line FWHMs are markedly different, with the \oiii{} line (\cii{} line) being broader for REBELS-14 (REBELS-12). 

The nature of REBELS-12, being a merger of an \oiii{}-luminous and \oiii{}-faint component, was previously discussed by \citet{algera2024} and in Section \ref{sec:data_targets}. We therefore perform a double-Gaussian fit to the \cii{} line profile (Fig.\ \ref{fig:1Dspectra}), and associate the redshifted component with the \oiii{} emission observed from REBELS-12 as in \citet[][their Fig.\ 6]{algera2024}. When considering only the redshifted component, the \oiii{} and \cii{} centroids are consistent at the $\sim1\sigma$ level. The \cii{} line appears slightly broader than the \oiii{} line, although the FWHMs are consistent within the uncertainties. From here on, we will use `REBELS-12' to refer to this redshifted \oiii{}- and \cii{}-emitting component, while we reserve `REBELS-12-blue' and `REBELS-12-full' to refer to the (blueshifted) \oiii{}-faint and complete system, respectively (see also Table \ref{tab:lineFluxes}).

For REBELS-14, the \oiii{} line is approximately twice as broad as the \cii{} line. Both lines show a slight elongation towards the north (Figure \ref{fig:CII_OIII_comparison}), where the \textit{JWST}/NIRSpec IFU data reveal an \oiiiopt{}-luminous component (Figure \ref{fig:momentZeroJWST}). The \oiii{} moment-0 map also reveals a $\sim4\sigma$ feature to the west of REBELS-14, connected to the main component via a $\sim2\sigma$ `bridge', without an obvious counterpart in \cii{}. Upon tapering the \oiii{} data to the \cii{} resolution in Appendix \ref{app:tapering}, this manifests as an extended \oiii{} tail towards the south-west, which could hint at the presence of a nearby \oiii{}-emitting companion or an (ionized) outflow. While a detailed investigation of this particular system is beyond the scope of this work, we do note that REBELS-14 shows the highest \oiiitocii{} ratio among our sample (Section \ref{sec:oiii2cii_discussion}), which may be linked to its different morpho-kinematics compared to the rest of our targets. \\

In the subsequent analysis, we are primarily interested in the global \oiiitocii{} luminosity ratios of our targets, and how these depend on galaxy physical properties such as metallicity and ionization parameter. While a resolved analysis is undoubtedly interesting (c.f., \citealt{akins2022,vallini2024}), our goal is to first set an unresolved `benchmark' on what governs the \oiiitocii{} ratio at high redshift, whereupon future works may expand to resolved scales. As such, a further, more detailed comparison of the morpho-kinematics of \cii{} and \oiii{} is beyond the scope of this work. Moreover, such a comparison is hindered by the large difference in angular resolution -- and in some cases relative depth -- of the current emission line data. For our purposes, it suffices to demonstrate that the \cii{} and \oiii{} emission trace each other reasonably well, which we do in Appendix \ref{app:tapering} by tapering the \oiii{} data to the \cii{} resolution. Even if \oiii{} and \cii{} are emitted from different physical regions in the galaxy -- which is ultimately expected given that they trace different gas phases (e.g., \citealt{kohandel2019,akins2022}) -- their global line ratio is still a meaningful quantity tracing the multi-phase nature of the ISM. We furthermore highlight that a detailed spatially resolved analysis of the \oiii{} and \cii{} morpho-kinematics for REBELS-25 will be presented by Rowland et al.\ (in preparation), while high-resolution \cii{} morpho-kinematics for several other REBELS galaxies will be the subject of Phillips et al.\ (in preparation).

\subsection{\oiii{} and \oiiiopt{} morphologies}
\label{sec:results_NIRSpec}

\begin{figure*}[t]
    \centering
    \includegraphics[width=0.95\textwidth]{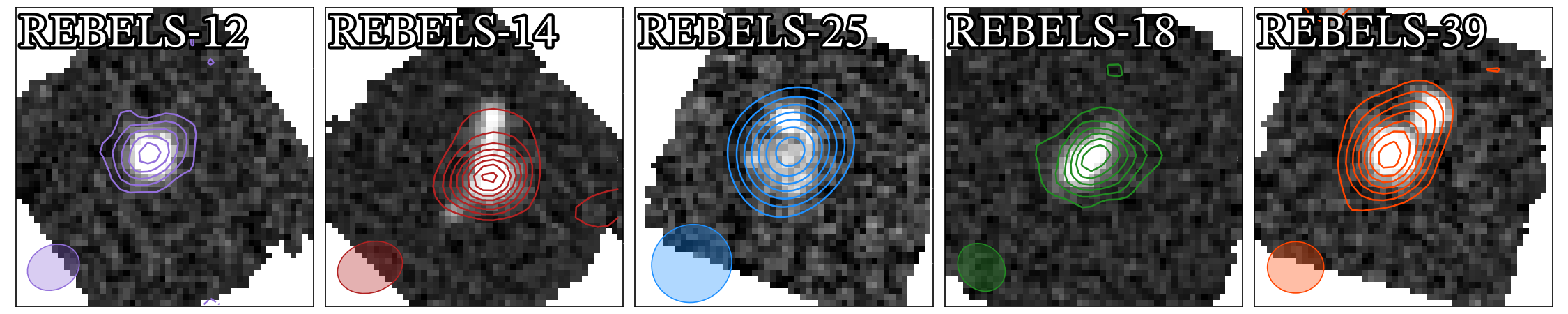}
    \caption{ALMA \oiii{} emission line maps (contours) on top of \textit{JWST}/NIRSpec \oiiiopt{} narrow-band images for our five REBELS-IFU targets (greyscale background). North is up, east is to the left, and the size of the panels equals $4''\times4''$. The ALMA contours start at $3\sigma$, and increase in steps of $2\sigma$. Overall, the two oxygen lines trace each other well on spatial scales of several $\mathrm{kpc}$. A similar comparison using NIRSpec/IFU images convolved to the \oiii{} resolution is shown in Figure \ref{fig:momentZeroConvolvedJWST} in Appendix \ref{app:jwst_convolved}.}
    \label{fig:momentZeroJWST}
\end{figure*}

As discussed in Section \ref{sec:observations}, five of our targets benefit from \textit{JWST}/NIRSpec observations as part of the REBELS-IFU program. For each, the \oiiiopt{} line is detected at high significance \citep{rowland2025}, enabling a comparison of the \oiii{} and \oiiiopt{} morphologies (Figure \ref{fig:momentZeroJWST}). While the resolution of the \oiii{} data is $\sim4-5\times$ coarser than that of the NIRSpec IFU ($\sim0.6-0.8''$ vs.\ $\sim0.15''$), the two oxygen lines generally show similar morphologies; in REBELS-14 and REBELS-39, the \oiiiopt{} emission is clearly clumpy and spatially extended, and despite its lower resolution the ALMA \oiii{} emission traces the emission seen in NIRSpec. Similarly, REBELS-12 and REBELS-18 appear compact in both rest-optical and far-infrared \oiiinowave{} emission. Furthermore, we verify in Appendix \ref{app:jwst_convolved} that the \oiiiopt{} and \oiii{} morphologies are qualitatively consistent when convolving the \oiiiopt{} narrow-band images to the \oiii{} resolution (Figure \ref{fig:momentZeroConvolvedJWST}). Altogether, this suggests that the emission originates from similar regions, at least on spatial scales of a few kpc. 

Only REBELS-25 -- currently the only source in our sample benefiting from high-resolution ALMA \oiii{} observations -- shows a difference in morphology between its rest-optical and FIR \oiiinowave{} lines. This is due to its \oiii{} emission emanating from a highly dust-obscured center, while the \oiiiopt{} emission is dominated by a northern, unobscured clump (see also \citealt{rowland2024}; Rowland et al.\ in preparation). We further discuss the effects of dust obscuration on the direct comparison between far-infrared and rest-optical properties of our sample in Section \ref{sec:caveats}.

\subsection{\oiiitocii{} line ratios}

We proceed by measuring the \oiiitocii{} ratios of our galaxy sample. We find a wide range of $L_\text{\oiii{}} / L_\text{\cii{}} = 1.9 - 9.6$, with REBELS-14 showing the highest line ratio of $L_\text{\oiii{}} / L_\text{\cii{}} = 9.6_{-3.1}^{+5.7}$. Consistent with \citet{algera2024}, REBELS-12-blue shows a low line ratio of $L_\text{\oiii{}} / L_\text{\cii{}} < 2.0$, although we will not consider this component further in this work. The redshifted component instead shows a line ratio of $L_\text{\oiii{}} / L_\text{\cii{}} = 3.1_{-1.0}^{+1.9}$, similar to that of the rest of our sample. For REBELS-25 we also measure a low line ratio of $L_\text{\oiii{}} / L_\text{\cii{}} = 1.9_{-0.3}^{+0.4}$, albeit slightly higher than what was previously inferred by \citet[][c.f., $L_\text{\oiii{}} / L_\text{\cii{}} = 1.3 \pm 0.3$]{algera2024}. We discuss the \oiiitocii{} ratios of our sample in the context of local dwarf galaxies and other high-redshift galaxies in Section \ref{sec:oiii2cii_discussion}.

\section{The \oiii{}-SFR Relation}
\label{sec:oiiiSFR}

\begin{figure*}[t]
    \includegraphics[width=0.48\textwidth]{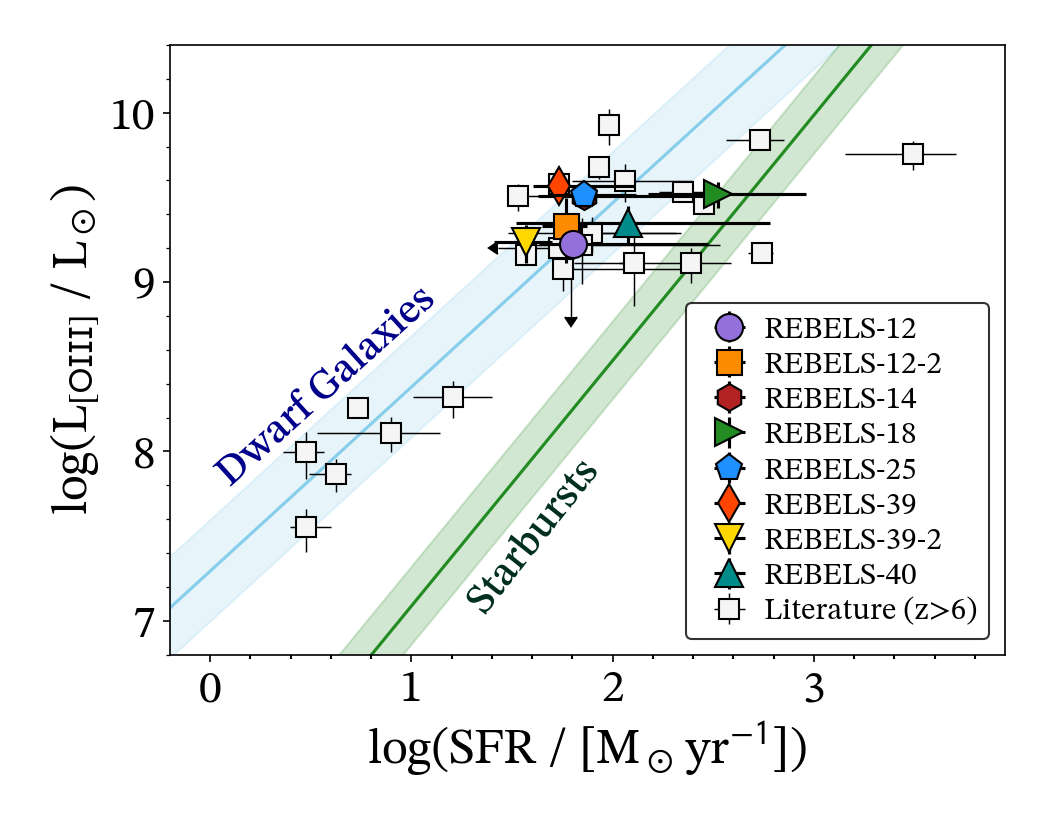}
    \includegraphics[width=0.48\textwidth]{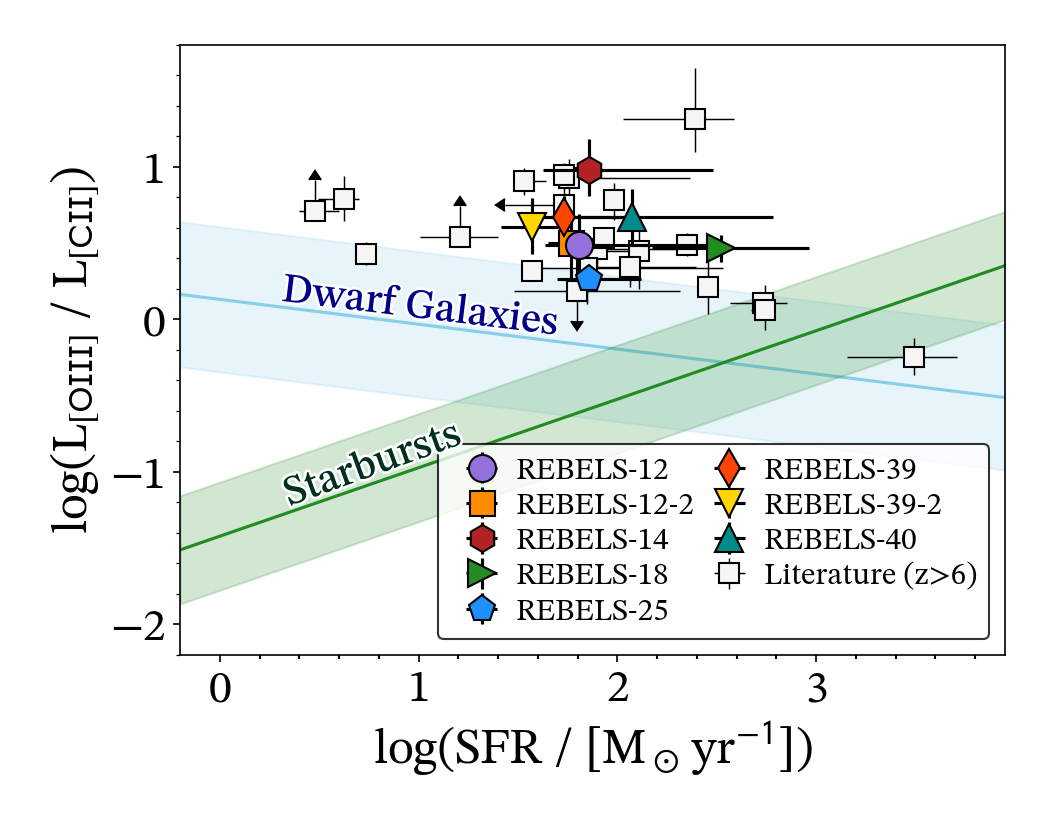}
    \caption{\textit{Left}: \oiii{} luminosity versus star formation rate for our eight targets, as well as a compilation of $z>6$ galaxies from the literature (\citealt{inoue2016,carniani2017,hashimoto2018,hashimoto2019,marrone2018,walter2018,zavala2018,laporte2019,tamura2019,bakx2020,bakx2024,carniani2020,carniani2024_oiii,harikane2020,akins2022,tadaki2022,witstok2022,witstok2025,wong2022,ren2023,algera2024,fujimoto2024_s4590,schouws2024,schouws2025,zavala2024_alma}). We adopt UV+IR-based SFRs where possible, while using \cii{}-based SFRs for our two serendipitous targets (see text). Both our sample and most $z\approx6-14$ galaxies in the literature are consistent with the \oiii{}-SFR relation for local dwarf galaxies from \citet[][blue shaded region; c.f., the relation for local starbursts shown in green]{delooze2014}. \textit{Right}: the \oiiitocii{} ratio as a function of total SFR. Nearly all high-redshift galaxies show a line ratio in excess of what is observed in local galaxies at similar SFR.}
    \label{fig:oiii}
\end{figure*}

The \oiii{} line, with an ionization potential of $35.1\,\mathrm{eV}$ and a critical density of $n_\mathrm{crit} \approx 510\,\mathrm{cm}^{-3}$ (e.g., \citealt{madden2013}) is primarily emitted by ionized gas at low-to-intermediate densities, such as that found in the \hii{}-regions surrounding newly-formed massive stars. This, in turn, makes it a potentially useful tracer of galaxy star formation rates in the early Universe (e.g., \citealt{delooze2014,katz2022,nakazato2023,vallini2024}).

We compare the \oiii{} luminosity and SFR of our sample in Figure \ref{fig:oiii}a. For our primary targets, we determine the total star formation rate as the sum of the UV-based SFR ($\mathrm{SFR}_\mathrm{UV}$) and the infrared-based SFR ($\mathrm{SFR}_\mathrm{IR}$). For the sources observed with \textit{JWST}/NIRSpec as part of REBELS-IFU, we adopt the latest IFU-based measurements of $\mathrm{SFR}_\mathrm{UV}$ from Fisher et al.\ (in preparation), while for REBELS-40 we use the pre-\textit{JWST} measurement from \citet{bouwens2022}. We note that a detailed comparison of the various SFR tracers available for the REBELS-IFU sample (UV+IR, Balmer lines, and \cii{}) will be presented in Fisher et al.\ (in preparation).

For REBELS-25, the IR-based SFR was recently determined by \citet{algera2024b}, making use of six-band ALMA continuum observations that capture both the peak and the Rayleigh-Jeans tail of its dust emission. For the other five primary targets, we make use of the dust continuum emission underlying the \oiii{} line in the ALMA Band 8 observations presented in this work, which in combination with the $160\,\mu\mathrm{m}$ continuum observations presented by \citet{inami2022} enables a measurement of the IR luminosity (across $8-1000\,\mu\mathrm{m}$) based on a dual-band Modified Blackbody fit (e.g., \citealt{algera2024}). We convert $L_\mathrm{IR}$ to a star formation rate adopting the conversion in \citet{inami2022} of $\mathrm{SFR}_\mathrm{IR}/(M_\odot\,\mathrm{yr}^{-1}) = 1.2 \times 10^{-10}\times (L_\mathrm{IR}/L_\odot)$. 

The full set of Band 8 continuum observations -- which also includes an additional four REBELS galaxies for which \oiii{} could not simultaneously be observed due to low atmospheric transmission -- will be presented in an upcoming work (Algera et al.\ in preparation). Briefly, however, we find that REBELS-18 is robustly ($>5\sigma$) detected in Band 8 continuum emission, facilitating a direct measurement of its obscured SFR$_\mathrm{IR}$. The rest of our primary targets (REBELS-12, REBELS-14, REBELS-39 and REBELS-40) are only marginally ($<5\sigma$) detected in Band 8, preventing robust constraints on their total infrared SFRs. We therefore adopt the average dust temperature that Algera et al.\ (in preparation) obtain from a simultaneous Band 6 and Band 8 stack of nine REBELS galaxies, and use this to determine their $L_\mathrm{IR}$. This average temperature of $T_\mathrm{dust} \approx 50\,\mathrm{K}$ is consistent with that predicted at $z\approx7$ by the theoretical model from \citet{sommovigo2022}, as well that derived from multi-band ALMA observations of UV-luminous galaxies at similar or slightly lower redshifts ($z\approx4-7$; \citealt{witstok2023_beta,mitsuhashi2024}). 

For the two serendipitous neighbors no robust far-IR detections exist, and we therefore adopt \cii{}-based SFRs using the \citet{delooze2014} relation for starbursts (following e.g., \citealt{fudamoto2021,bakx2024}). This relation is found to also be valid for \cii{}-luminous galaxies out to $z\sim7$ (e.g., \citealt{schaerer2020}; Schouws et al.\ in preparation). We opt to use UV+IR based SFRs for the rest of the sample, as we explicitly investigate the $L_\text{\cii{}}/\mathrm{SFR}$ ratio for these sources in Section \ref{sec:oiii2cii_discussion}, which naturally requires a SFR tracer independent of \cii{}. We do not include the serendipitous galaxies in this analysis due to a lack of ancillary \textit{JWST}/NIRSpec data, and hence we use their \cii{}-based SFRs in Figure \ref{fig:oiii} for visualization purposes only.

In Figure \ref{fig:oiii}a, we furthermore compare the \oiii{}-SFR relation of our sample to a variety of literature studies comprising -- to the best of our knowledge -- all published, spectroscopically confirmed galaxies at $z > 6$ targeted in \oiii{} emission. This sample builds upon the compilations provided in \citet{harikane2020} and \citet{algera2024}, and includes more recent studies and updated measurements where available. A full suite of references is given in the caption of Figure \ref{fig:oiii}. We note that, except for our two serendipitous targets, we omit galaxies that do not have a SFR measurement from UV and/or IR emission. This excludes the bulk of the \cii{}-selected sample from \citet{bakx2024}, though we explicitly compare to their work in Section \ref{sec:lineRatioCiiSelected}. \\

Overall, we find that the REBELS sample is consistent with the local relation for dwarf galaxies from \citet{delooze2014}, with 7/8 galaxies falling within the $\sim0.3\,\mathrm{dex}$ scatter of the relation. Among our sample, only REBELS-18 is more consistent with the relation for local starbursts, although it still agrees with the dwarf galaxy relation within the errors. While the \oiii{} luminosity of REBELS-18 is similar to that of the other REBELS galaxies, it is characterized by a high obscured SFR owing to its high dust temperature (Algera et al.\ in preparation). 

The recent detection of the \oiii{} line in three galaxies at $z > 10$, GS-z11-0 at $z=11.12$ \citep{witstok2025}, GHZ2 at $z=12.33$ \citep{zavala2024_alma} and GS-z14 at $z=14.18$ \citep{carniani2024_oiii,schouws2024}, has highlighted the potential of detecting this line even in the earliest galaxies. All three galaxies were also found to be consistent with the local \oiii{}-SFR relation for dwarf galaxies, suggesting that this relation may be near-universal across redshift. Focusing on the full $z>6$ sample shown in Fig.\ \ref{fig:oiii}a, there is however an indication that the relation breaks down at the high-$\mathrm{SFR}$ end. These galaxies tend to be more dusty (e.g., \citealt{marrone2018,tadaki2022}), such that the \oiii{} line could potentially be subject to optical depth effects, or be collisionally de-excited if gas densities are large. Alternatively, the suppression of \oiii{} at high SFRs may be a manifestation of the `fine structure line deficit' typically seen in \cii{} \citep[e.g.,][]{herrera-camus2018b,herrera-camus2025,villanueva2024}, but which is also expected to affect other FIR lines \citep{diazsantos2017,peng2025c}. At the low-SFR end, there appears to be a general agreement with the local dwarf galaxy relation (e.g., \citealt{carniani2017,hashimoto2018,fujimoto2024_s4590,witstok2025}), although larger galaxy samples and more uniform SFR measurements are necessary to confirm this. Nevertheless, the apparent validity of the \oiii{}-SFR relation across cosmic time highlights the power of \oiiil{} as a key ISM diagnostic in the distant galaxy population.

\section{The \oiiitocii{} Ratios of high-redshift and local dwarf galaxies}
\label{sec:oiii2cii_discussion}

In recent years, ALMA observations of high-redshift galaxies ($z\gtrsim6$) have demonstrated that their \oiiitocii{} ratios are significantly larger than what is commonly observed in the local Universe (e.g., \citealt{hashimoto2019,laporte2019,bakx2020,carniani2020,harikane2020,akins2022,witstok2022,ren2023,fujimoto2024_s4590}). A likely explanation is that these elevated line ratios are driven by the typically bursty conditions of high-redshift galaxies generating strong ionizing radiation fields (e.g., \citealt{arata2020,harikane2020,katz2022,algera2024,vallini2024}), in combination with \cii{} likely being suppressed at low metallicity (e.g., \citealt{vallini2015,ferrara2019,liang2024,gurman2024}). However, several other -- potentially interdependent -- factors may also play a role, such as selection effects \citep{algera2024,bakx2024}, different morpho-kinematics of \cii{} and \oiii{} \citep{carniani2017,carniani2020,kohandel2019,schimek2024}, time-variability of the relative \cii{} and \oiii{} intensities due to photo-evaporation \citep{vallini2017}, a low PDR covering fraction or otherwise porous ISM suppressing the \cii{} emission \citep{chevance2016,cormier2019,harikane2020,ramambason2022,ura2023,hagimoto2025}, a high PDR density leading to collisional de-excitation of \cii{} emission \citep{harikane2020,vanleeuwen2025}, a top-heavy IMF, and/or a sub-solar C/O abundance \citep{katz2022,nyhagen2024}.

Although limited to a small sample size, we can investigate several of these hypotheses. For five of our targets, as well as several other high-$z$ galaxies in the literature (Section \ref{sec:lineRatioHighzDwarfsComparison}), we have metallicity and ionization parameter measurements from \textit{JWST}/NIRSpec observations. Moreover, the two serendipitous sources (REBELS-12-2 and REBELS-39-2) benefit from a different selection than the typical UV-based selection of high-redshift galaxies, similar to the quasar companion galaxies in \citet[][see also \citealt{walter2018}]{bakx2024}. \\

We show the \oiiitocii{} ratios of our sample as a function of SFR in Fig.\ \ref{fig:oiii}b, and once again compare to $z>6$ galaxies targeted in \oiii{} drawn from a variety of literature studies. We find that the line ratios of our sample, which span \oiiitocii{}$ \approx 1.9 - 9.6$, are typical for the high-redshift galaxy population. REBELS-14, in particular, is confidently detected in both lines, and shows a ratio of $\text{\oiiitocii{}} \approx 10$, among the highest observed at $z > 6$ (c.f., \citealt{ren2023}). The observed line ratios are well above what is observed in starburst galaxies in the local Universe, indicated via the green line and shading in Fig.\ \ref{fig:oiii} adopted from \citet{delooze2014}. Local dwarf galaxies show larger line ratios, with \oiii{} typically being more luminous than \cii{} (c.f., blue line and shading in Fig.\ \ref{fig:oiii}). However, barring some exceptions \citep{kumari2024}, the typical line ratio for dwarf galaxies of \oiiitocii{}$\approx2$ is still below that seen in the high-redshift population \citep{cormier2015,cormier2019}.

In what follows, we quantitatively investigate the origin of the elevated \oiiitocii{} ratios of high-redshift galaxies. Specifically, we first investigate the impact of selection effects on the line ratio (Section \ref{sec:lineRatioCiiSelected}), and then explore scaling relations with metallicity (Section \ref{sec:lineRatioMetallicity}), ionization parameter (Section \ref{sec:lineRatioIonizationParameter}) and burstiness (Section \ref{sec:lineRatioBurstiness}).

\subsection{The \oiiitocii{} Ratios of \cii{}-selected Galaxies}
\label{sec:lineRatioCiiSelected}

Recently, \citet{bakx2024} presented ALMA Band 8 observations targeting the \oiii{} line in 13 \cii{}-selected galaxies at $6 \lesssim z \lesssim 7.5$. These galaxies were serendipitously detected as nearby neighbors around high-redshift quasars by \citet{venemans2020} (see also \citealt{decarli2017}). Given that \cii{} is generally considered to be a reliable star-formation rate tracer (e.g., \citealt{delooze2014,schaerer2020}), mostly independent of obscuration, these galaxies are essentially SFR-selected, distinguishing them from the typically UV-selected galaxy population at high redshift identified based on its level of unobscured star formation. Intriguingly, \citet{bakx2024} detect only one of their targets in \oiii{} emission, and use stacking to infer the typical line ratio of this population to be \oiiitocii{}$<1.2$. This is in reasonable agreement with \citet{walter2018}, who measure $\text{\oiiitocii{}}=1.58 \pm 0.24$ in a single \cii{}-selected quasar companion galaxy, somewhat below the typical line ratio of $\text{\oiiitocii{}}\gtrsim2-10$ of the UV-selected $z\gtrsim6$ population.

Our two serendipitous targets are selected in a similar manner to the \citet{bakx2024} sample, albeit around massive UV-selected galaxies, instead of around quasars. We infer line ratios of \oiiitocii{}$=3.1_{-1.0}^{+1.9}$ and \oiiitocii{}$=4.1_{-1.4}^{+2.2}$ for REBELS-12-2 and REBELS-39-2, respectively, which is comparable to those of the UV-selected population at $z\gtrsim6$. While limited to just two sources, this suggests that the elevated \oiiitocii{} ratios of high-redshift galaxies may not be fully due to a UV- versus \cii{}-based selection. 

Alternatively, the low line ratios for quasar companion galaxies compared to what we find for massive galaxy companions could suggest that the central quasar affects the ISM of galaxies in its immediate environment through feedback and/or outflows (e.g., \citealt{zana2022,ferrara2023_quasar}). Direct evidence of the quasar's effect on the ISM of a nearby companion was recently presented by \citet{decarli2024}, although this represents a system in the process of merging. It is unclear if the quasar is also able to affect the ISM on $\gtrsim50\,\mathrm{kpc}$ scales, which would be necessary to explain the difference in the observed line ratios between our serendipitous targets and the \citet{bakx2024} sample. Moreover, it is possible that a few of the companion galaxies in the \citet{venemans2020} sample are spurious, as blind line searches have yielded seemingly robust emission lines detections that were later found to be due to noise \citep{hayatsu2017,hayatsu2019,uzgil2021}. Moreover, a small fraction of the emission lines detected in the quasar fields -- particularly those at larger angular separations from the central quasar (\citealt{bakx2024}; their Fig.\ 9) -- could correspond to CO lines of foreground emitters. As discussed in detail by \citet{bakx2024}, any such contamination by low-redshift galaxies would also cause the average \oiiitocii{} ratio of the quasar companion galaxies to be underestimated.

While the difference between the \oiiitocii{} ratios of our two serendipitous targets and the \citet{bakx2024} quasar companion galaxies is therefore intriguing, future studies leveraging larger samples of \cii{}-selected galaxies are needed for more robust comparisons. At the same time, further constraints on the stellar emission of \cii{}-selected galaxies is crucial to better understand their place in the framework of galaxy evolution.

\subsection{The dependence of the \oiiitocii{} ratio on metallicity, ionization parameter and burstiness}
\label{sec:lineRatioHighzDwarfsComparison}

\begin{figure*}
    \centering
    \includegraphics[width=1.0\linewidth]{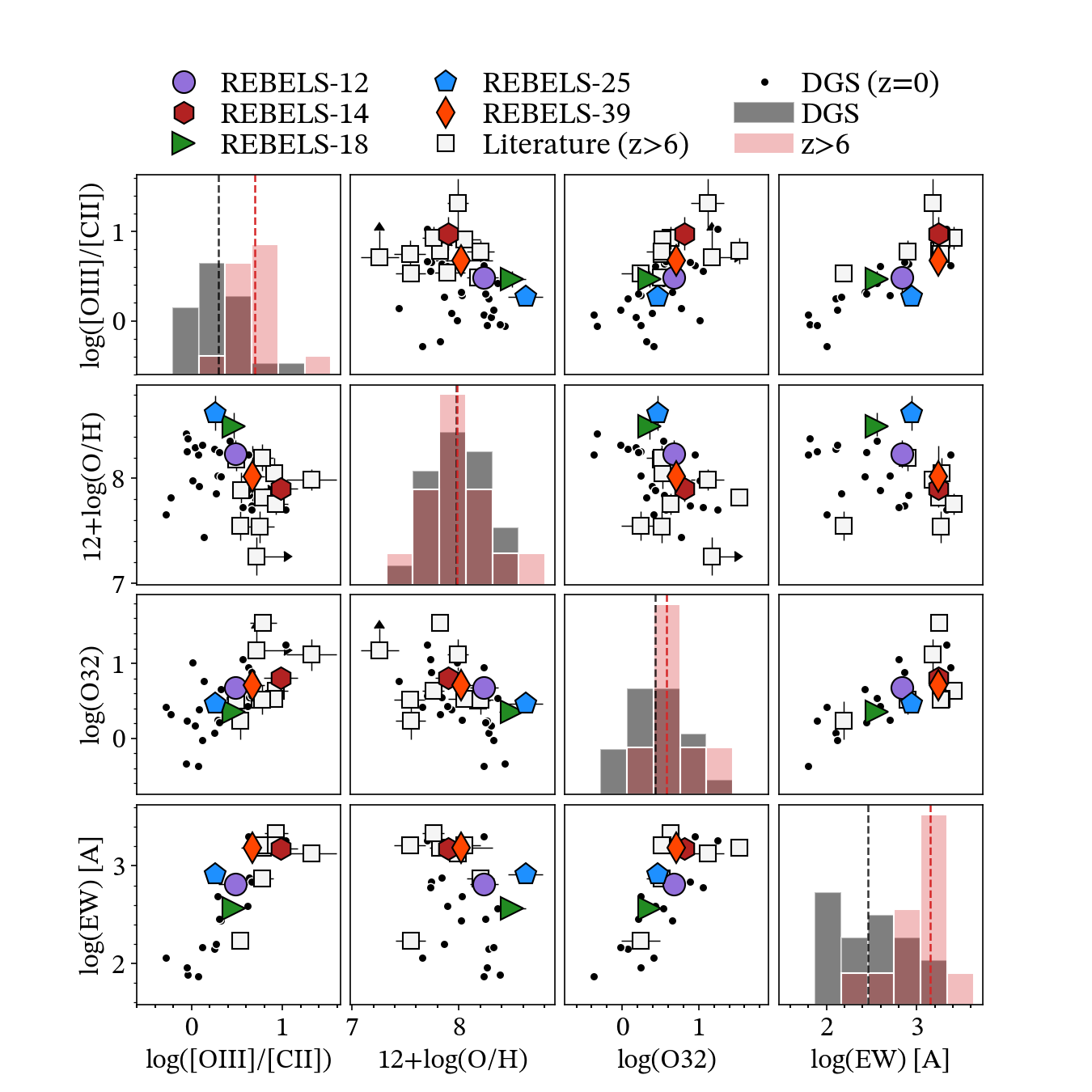}
    \caption{Matrix plot combining the \oiiitocii{} ratio, oxygen abundance, $\log(O32)$, and \EW{} for the REBELS-IFU (colored markers), other high-redshift (white squares) and dwarf galaxy (small points) samples. Errorbars for the DGS are omitted for visual clarity. The red (grey) histograms correspond to $z>6$ (dwarf) galaxies, and the medians are indicated through vertical dashed lines. The high-redshift sample is characterized by similar metallicities and ionization parameters as local dwarfs, but shows higher \EW{} and \oiiitocii{} ratios. It is furthermore clear that various parameters (anti-)correlate not only with \oiiitocii{}, but also with each other. In order to investigate the main drivers of the elevated \oiiitocii{} ratios of high-redshift galaxies, we attempt to disentangle these correlations in Section \ref{sec:oiii2cii_discussion}.}
    \label{fig:matrixPlot}
\end{figure*}

We next explore the dependence of the \oiiitocii{} ratio on three physical parameters -- metallicity, ionization parameter, and burstiness -- for high-redshift and dwarf galaxies alike. We use oxygen abundance, expressed as $12+\log(\mathrm{O/H})$, as a proxy for metallicity -- as is typical in the literature -- but discuss the effects of non-solar abundance ratios in Section \ref{sec:caveats}. As a proxy for the ionization parameter $U$ we use the dust-corrected $O32$ line ratio, defined as $O32 = \text{\oiiiopt{}} / \text{\oii{}}_{3727,29}$. This quantity is known to effectively trace the ionization parameter (e.g., \citealt{strom2018,kewley2019}), and where necessary we adopt the tight linear relation between $O32$ and $U$ obtained by \citet{papovich2022} through photo-ionization modeling of a sample of $z\sim1.1 - 2.3$ galaxies.\footnote{We note that, due to the higher critical density of \oiiiopt{} ($n_\mathrm{crit} = 6.8\times10^5\,\mathrm{cm}^{-3}$) compared to \oii{} ($n_\mathrm{crit} = (3.4 - 15) \times 10^3\,\mathrm{cm}^{-3}$; \citealt{osterbrock_ferland2006}), $O32$ could also be boosted at particularly high densities.} Their best-fit relation is given by $\log(U) = (0.86 \pm 0.07) \log(O32) - (2.84 \pm 0.02)$.\footnote{Note that \citet{papovich2022} use the total \oiiidopt{} flux in their definition of $O32$, whereas we only include the flux in the $5007\,$\AA{} line. We account for this difference when we reproduce their relation, assuming a ratio between \oiiiopt{}/\oiiinowave{}$_{4959}$ = 2.98 as in Section \ref{sec:dataJWST}.}

Finally, we trace galaxy `burstiness' via the equivalent width of the combined \oiiidopt{} and H$\beta$ complex, denoted as \EW{} \citep[e.g.,][]{smit2014,endsley2021,topping2022,witstok2022}. This parameter is known to tightly correlate with specific star formation rate, and is therefore a tracer of starburst age (with a higher EW implying a younger age), with the additional benefit that it is a direct observable and hence does not rely on detailed modeling. In addition, previous works have investigated the \oiiitocii{} ratio as a function of \EW{}: \citet{witstok2022} found a correlation between the line ratio and the photometrically-estimated \EW{} based on \textit{Spitzer}/IRAC imaging for $z\approx6.5$ galaxies, and argue that reproducing the highest EWs requires a young, starbursting population that is potentially already significantly metal-enriched. This is furthermore supported by \citet{kumari2024}, who perform a detailed study of the local high-redshift analog Pox 186 with $\text{\oiiitocii{}}\approx10$ -- the highest such line ratio in the local Universe. By comparing Pox 186 to both low- and high-redshift galaxies, they suggest a possible link between high \oiiitocii{} ratios, young ages, a high \EW{}, and bright high-ionization carbon lines (\ciii{}, \civ{}).

We note, however, that various other definitions of burstiness exist in the literature; for example, \citet{ferrara2019} define it as the upwards deviation from the Kennicutt-Schmidt relation (i.e., an elevated $\Sigma_\mathrm{SFR}$ for a given $\Sigma_\mathrm{gas}$). However, this quantity is hard to measure directly as it requires resolved observations of (molecular) gas. Another definition of burstiness is the ratio of SFRs on short and long timescales, for example $\mathrm{SFR}_\mathrm{10\,Myr} / \mathrm{SFR}_\mathrm{100\,Myr}$ \citep[e.g.][]{endsley2024,mcclymont2025}. Observationally, this is often measured as the ratio of SFRs traced through Balmer lines vs.\ UV emission \citep[e.g.,][]{langeroodi2024,sun2025}. While a high Balmer-to-UV ratio similarly suggests a galaxy is undergoing a starburst and could thus be classified as bursty, we here focus on \EW{} as it is less affected by the presence of dust. As discussed in Section \ref{sec:methods_dust_correction}, 4/5 of our REBELS-IFU targets are detected in only a single Balmer line (H$\beta$), making the ratio of Balmer-to-UV SFR rather uncertain due to limitations in dust corrections.

In what follows, we will use metallicity, ionization parameter and burstiness interchangeably with, respectively, oxygen abundance, $O32$ and \EW{}. We compile these three parameters also for a sample of $z>6$ galaxies with ALMA \oiii{} observations in the literature: three sources from \citet{harikane2025} at $z=6.03-7.21$, COS-2987 at $z=6.81$ \citep{mawatari2025}, COS-3018 at $z=6.85$ \citep{witstok2022,scholtz2024}, B14-65666 at $z=7.15$ \citep{hashimoto2019,jones2024,sugahara2025}, MACS0416-Y1 at $z=8.31$ \citep{tamura2019,bakx2020,harshan2024,ma2024}, ID4590 at $z=8.50$ \citep{nakajima2023,fujimoto2024_s4590}, MACS1149-JD1 at $z=9.11$ \citep{hashimoto2018,laporte2019,carniani2020,tokuoka2022,stiavelli2023,morishita2024}, and GS-z14 at $z=14.18$ \citep{carniani2024,carniani2024_oiii,schouws2024,schouws2025}. In total, we thus have a combined sample of fifteen $z > 6$ galaxies with which we perform a first empirical investigation of how the \oiiitocii{} ratio depends on metallicity in the $z > 6$ Universe. While not (yet) targeted in \cii{} and thus not analyzable in the context of the \oiiitocii{} ratio, we also include GHZ2 at $z=12.33$ \citep{calabro2024,castellano2024,zavala2024_alma,zavala2025_miri} in the analysis in Section \ref{sec:burstiness_on_oiii_and_cii}, where we focus on the $L_\text{\oiii{}}/\mathrm{SFR}$ ratio. A full compilation of the $z>6$ sample with ancillary \textit{JWST}/NIRSpec data used in this work is provided in Table \ref{tab:literatureComplation} in Appendix \ref{app:literatureCompilation}. For the DGS sample, metallicity, ionization parameter and burstiness measurements are also available, as discussed in Section \ref{sec:data_DGS}. \\

We show a `matrix plot' comprising the \oiiitocii{} ratio, metallicity, ionization parameter and burstiness for both the $z>6$ and DGS samples in Figure \ref{fig:matrixPlot}. This provides a compact overview of the possible (anti-)correlations between the various parameters and their effect on the \oiiitocii{} ratio.

Before delving into a more quantitative analysis in the subsequent sections, we highlight a few qualitative features of the high-$z$ and DGS samples. First of all, both span a similar range in $12+\log(\mathrm{O/H})$ and $\log(O32)$, as indicated by the histograms on the diagonal of Figure \ref{fig:matrixPlot}. However, the $z>6$ galaxies, on average, show a higher burstiness and elevated \oiiitocii{} ratios.

Qualitatively, Figure \ref{fig:matrixPlot} also suggests that the \oiiitocii{} ratio may correlate with $O32$ and \EW{}, and anti-correlate with metallicity. Moreover, possible (anti-)correlations between these parameters themselves, such as a correlation between $O32$ and \EW{} and anti-correlation between $O32$ and metallicity, are apparent. Indeed, when investigated across large galaxy samples, metallicity and ionization parameter are often found to anti-correlate \citep[e.g.,][]{nakajima2014,shapley2015,sanders2023}; low-mass, metal-poor galaxies typically have hard radiation fields, and thus large $O32$ values, and vice versa for more massive, metal-enriched galaxies. Similarly, several studies have found correlations between $O32$ and either \EW{} or $\mathrm{EW}(\text{\oiiiopt{}})$ \citep[e.g.,][]{tang2019,tang2023,boyett2024_lineEmitters}, which is similarly unsurprising given that a young starburst is expected to drive powerful radiation fields.

Based on the qualitative discussion above, our approach is the following: we will first, in Sections \ref{sec:lineRatioMetallicity}$-$\ref{sec:lineRatioBurstiness}, explore and quantify the dependence of the \oiiitocii{} ratio on metallicity, ionization parameter and burstiness separately, for both the $z>6$ and DGS samples. Following this, we will explore the combined dependence of the \oiiitocii{} ratio on these three parameters by attempting to disentangle their degeneracies (Section \ref{sec:lineRatioMultivariateFit}). Finally, we will discuss whether any variations in the \oiiitocii{} ratio are driven primarily by variations in \oiii{} or in \cii{} in Section \ref{sec:burstiness_on_oiii_and_cii}.

\begin{table*}
    \def\arraystretch{1.7}
    \centering
    \caption{Results from fitting and correlation analysis}
    \label{tab:correlations}
    \begin{tabular}{ll|ccccc|ccccc}
        \hline \hline 
        & & \multicolumn{5}{c|}{\textbf{\large High-redshift sample}}
        & \multicolumn{5}{c}{\textbf{\large DGS sample}} \\
        \textbf{$\mathbf{x}$-axis} & \textbf{$\mathbf{y}$-axis} & \textbf{Slope} & \textbf{Intercept} & \textbf{Scatter} & \textbf{Corr.} & \textbf{$\mathbf{p}$-value} & \textbf{Slope} & \textbf{Intercept} & \textbf{Scatter} & \textbf{Corr.} & \textbf{$\mathbf{p}$-value} \\
        (1) & (2) & (3) & (4) & (5) & (6) & (7) & (8) & (9) & (10) & (11) & (12) \\ 
        
\hline
12+log(O/H) & log\oiiitocii{} & $-0.50_{-0.26}^{+0.24}$ & $0.71_{-0.07}^{+0.07}$ & $0.20_{-0.07}^{+0.09}$ & $-0.47$ & $0.106$ & $-0.45_{-0.29}^{+0.29}$ & $0.29_{-0.07}^{+0.07}$ & $0.33_{-0.05}^{+0.06}$ & $-0.32$ & $0.114$ \\
log(O32) & log\oiiitocii{} & $0.42_{-0.25}^{+0.27}$ & $0.42_{-0.19}^{+0.19}$ & $0.23_{-0.07}^{+0.09}$ & $0.53$ & $0.064$ & $0.48_{-0.16}^{+0.16}$ & $0.07_{-0.10}^{+0.10}$ & $0.28_{-0.05}^{+0.06}$ & $0.58$ & $0.004$ \\
log(EW) & log\oiiitocii{} & $0.49_{-0.25}^{+0.26}$ & $0.72_{-0.08}^{+0.08}$ & $0.21_{-0.07}^{+0.10}$ & $0.57$ & $0.051$ & $0.67_{-0.10}^{+0.11}$ & $0.69_{-0.07}^{+0.07}$ & $0.12_{-0.04}^{+0.05}$ & $0.89$ & $<0.001$ \\
\hline
12+log(O/H) & log\oiiinowave{}/SFR & $0.14_{-0.40}^{+0.37}$ & $7.48_{-0.13}^{+0.11}$ & $0.37_{-0.09}^{+0.13}$ & $0.17$ & $0.550$ & $0.07_{-0.28}^{+0.28}$ & $7.32_{-0.07}^{+0.07}$ & $0.31_{-0.05}^{+0.06}$ & $0.06$ & $0.763$ \\
log(O32) & log\oiiinowave{}/SFR & $-0.52_{-0.30}^{+0.30}$ & $7.89_{-0.25}^{+0.25}$ & $0.31_{-0.09}^{+0.13}$ & $-0.47$ & $0.090$ & $0.11_{-0.18}^{+0.18}$ & $7.28_{-0.11}^{+0.11}$ & $0.33_{-0.05}^{+0.07}$ & $0.14$ & $0.541$ \\
log(EW) & log\oiiinowave{}/SFR & $-0.07_{-0.45}^{+0.47}$ & $7.60_{-0.15}^{+0.13}$ & $0.38_{-0.13}^{+0.19}$ & $-0.08$ & $0.812$ & $0.21_{-0.17}^{+0.17}$ & $7.54_{-0.11}^{+0.11}$ & $0.24_{-0.05}^{+0.07}$ & $0.29$ & $0.250$ \\
\hline
12+log(O/H) & log\cii{}/SFR & $0.84_{-0.56}^{+0.66}$ & $6.70_{-0.18}^{+0.16}$ & $0.49_{-0.15}^{+0.22}$ & $0.23$ & $0.465$ & $0.58_{-0.30}^{+0.31}$ & $7.02_{-0.08}^{+0.08}$ & $0.36_{-0.05}^{+0.07}$ & $0.39$ & $0.048$ \\
log(O32) & log\cii{}/SFR & $-0.82_{-0.41}^{+0.34}$ & $7.38_{-0.28}^{+0.30}$ & $0.35_{-0.13}^{+0.19}$ & $-0.58$ & $0.046$ & $-0.41_{-0.19}^{+0.18}$ & $7.21_{-0.11}^{+0.11}$ & $0.33_{-0.05}^{+0.07}$ & $-0.47$ & $0.024$ \\
log(EW) & log\cii{}/SFR & $-0.47_{-0.49}^{+0.49}$ & $6.88_{-0.17}^{+0.14}$ & $0.41_{-0.15}^{+0.24}$ & $-0.37$ & $0.257$ & $-0.51_{-0.18}^{+0.18}$ & $6.82_{-0.12}^{+0.12}$ & $0.28_{-0.06}^{+0.08}$ & $-0.63$ & $0.007$ \\
\hline
\hline

    \end{tabular}
    \raggedright \\ 
    \textbf{Notes:} We fit a linear function of the form $y=a(x-x_0) + b$ to both the $z>6$ (Cols.\ 3 - 7) and DGS (Cols.\ 8 - 12) samples. The parameters used for $x$ and $y$ are given in Cols.\ 1 and 2, and we fix $x_0 = (8, 0, 3)$ for the fits against $12+\log(\mathrm{O/H})$, O32 and \EW{}, respectively. The intrinsic scatter about the regression line (Cols.\ 5 \& 10), Pearson correlation coefficient (Cols.\ 6 \& 11) and corresponding $p$-value (Cols.\ 7 \& 12) are also presented.
\end{table*}

\subsubsection{The effect of metallicity on the \oiiitocii{} ratio}
\label{sec:lineRatioMetallicity}

\begin{figure*}
    \centering
    \includegraphics[width=1.0\textwidth]{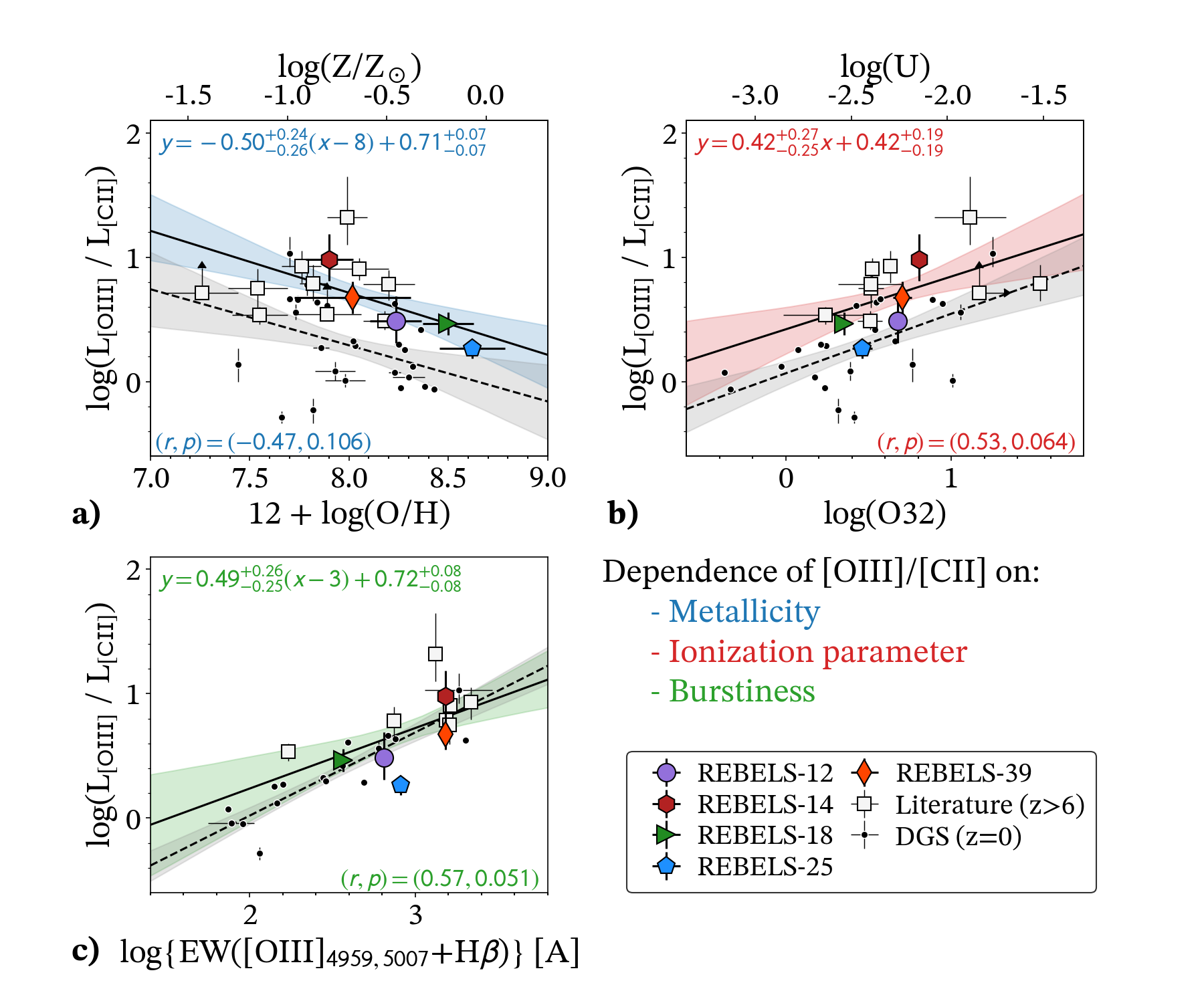}
    \caption{\textbf{a)} the \oiiitocii{} ratio as a function of oxygen abundance, a proxy for metallicity. \textbf{b)} the \oiiitocii{} ratio as a function of $O32 = \mathrm{[OIII]}_{5007} / \mathrm{[OII]}_{3727,29}$, which is directly related to the ionization parameter $U$. \textbf{c)} the \oiiitocii{} ratio as a function of the \oiiidopt{}+H$\beta$ equivalent width, a measure of galaxy burstiness. In addition to our REBELS-IFU targets (colored symbols), we overplot several $z \approx 6 - 14$ galaxies from the literature (white squares). We furthermore show local galaxies from the Dwarf Galaxy Survey \citep{madden2013,cormier2015} as small, black markers. We perform a linear fit to the $z > 6$ sample in each panel, represented by the solid black line and colored shading for the $16-84^\mathrm{th}$ percent confidence interval. The best-fit relation, correlation coefficient $r$ and corresponding $p$-value for the $z>6$ sample are annotated in each panel. A similar fit to the DGS sample is shown in grey and quantified in Table \ref{tab:correlations}. At fixed metallicity and ionization parameter, high-redshift galaxies show higher \oiiitocii{} ratios than local dwarfs, which are likely driven by their higher burstiness (Section \ref{sec:burstiness_on_oiii_and_cii}).}
    \label{fig:oiii_cii_metallicity_ionization}
\end{figure*}

One possible reason for the elevated \oiiitocii{} ratios of high-redshift galaxies is their low metallicity (e.g., \citealt{vallini2015,carniani2017,ferrara2019,ura2023,liang2024}). With the advent of the \textit{JWST}, the first few $z\gtrsim6$ galaxies with both a metallicity and \oiiitocii{} measurement are now being compiled, enabling direct insight into the effects of metallicity on the line ratio. In the local Universe, a possible trend of increasing \oiiitocii{} ratio with decreasing metallicity has been reported for the \textit{Herschel} Dwarf Galaxy Survey (DGS) by \citet{cormier2015}, although the scatter at fixed metallicity was found to be substantial (see also \citealt{cormier2019}).

We use the {\sc{python}} implementation of {\sc{linmix}} \citep{kelly2007} to perform a linear fit between $\log(\text{\oiiitocii{}})$ and oxygen abundance.\footnote{\url{https://github.com/jmeyers314/linmix}} This package provides Bayesian fitting machinery that accounts for uncertainties in both the $x$ and $y$ directions. Moreover, the linear fit includes an intrinsic scatter term, which is useful when the scatter of the data about the regression line is not only driven by the measurement uncertainties, but also by intrinsic source-to-source variation. {\sc{linmix}} also allows for fitting to censored data (i.e., upper limits), which enables us to self-consistently include the \cii{}-non-detected galaxies in the fit (c.f., \citealt{fujimoto2024_s4590,schouws2025}).

Our best fit relation between \oiiitocii{} and metallicity for the combined $z>6$ sample is shown in Figure \ref{fig:oiii_cii_metallicity_ionization}a through the solid black line and blue shaded region. The best-fitting function, Pearson correlation coefficient and corresponding $p$-value can be found in Table \ref{tab:correlations}. The slope of the relation, $\alpha_{z>6} = -0.50_{-0.26}^{+0.24}$ is $\sim2\sigma$ different from zero, suggesting a tentative anti-correlation between \oiiitocii{} and metallicity. The Pearson correlation coefficient of $r=-0.47$ and $p$-value of $0.11$, however, suggest the correlation is modest, and not statistically significant.

We also perform a similar fit to the DGS sources, imposing a minimum uncertainty of $0.1\,\mathrm{dex}$ on the metallicity and line ratio to account for systematic uncertainties in flux measurements and calibrations. The fit, shown as the grey shaded region in Figure \ref{fig:oiii_cii_metallicity_ionization}a, yields a similar slope as that obtained for the high-redshift sample. However, for the DGS too the anti-correlation is not found to be statistically significant (Table \ref{tab:correlations}), which is consistent with \citet{cormier2015}, who found that while some of the highest \oiiitocii{} ratios are seen for low-metallicity dwarfs, the scatter in this regime is large.

\subsubsection{The effect of ionization parameter on the \oiiitocii{} ratio}
\label{sec:lineRatioIonizationParameter}

For our five REBELS-IFU targets, both the \oiiiopt{} and \oii{} lines are detected, and we determine dust-corrected $O32$ ratios as outlined in Section \ref{sec:methods_dust_correction}. We furthermore have $O32$ measurements for nine $z > 6$ galaxies from the high-$z$ literature sample (Table \ref{tab:literatureComplation}). As before, we also compare to dwarf galaxies from the DGS, for which the $O32$ ratio is determined using the dust-corrected \oii{} and \oiiiopt{} fluxes provided in \citet{devis2017}. 

We show the \oiiitocii{} ratio vs.\ $O32$ for the high-redshift and DGS samples in Figure \ref{fig:oiii_cii_metallicity_ionization}b. As before, we fit a linear relation to the full set of $z > 6$ galaxies, which yields a slope $\alpha_{z>6} = 0.42_{-0.25}^{+0.27}$ that is consistent with zero at the $1.7\sigma$ level (Table \ref{tab:correlations}). The $p$-value of $0.064$ moreover suggests no statistically significant correlation. 

For the DGS sample, on the other hand, the correlation is found to be statistically significant with $p=4\times10^{-3}$. The slope of $\alpha_\mathrm{DGS} = 0.48 \pm 0.16$ is different from zero at the $3.0\sigma$ level, and is moreover consistent with the slope obtained for the high-redshift sample, which however has larger uncertainties.

\subsubsection{The effect of burstiness on the \oiiitocii{} ratio}
\label{sec:lineRatioBurstiness}

The high `burstiness' of high-redshift galaxies has previously been suggested as a potential origin of their elevated \oiiitocii{} ratios (e.g., \citealt{pallottini2019,vallini2021,vallini2024,algera2024}). As discussed in Section \ref{sec:lineRatioHighzDwarfsComparison}, we here use \EW{} as a proxy for burstiness. In addition to the five REBELS-IFU sources with \EW{} measurements, we compile equivalent widths for seven additional $z>6$ galaxies with \oiii{} and \cii{} information (Table \ref{tab:literatureComplation}). For a subset of DGS sources, \EW{} was recently measured by \citet{kumari2024}. 

We explore the dependence of \oiiitocii{} on the \oiiidopt{} + H$\beta$ equivalent width in Figure \ref{fig:oiii_cii_metallicity_ionization}c. A linear fit to the $z>6$ sample yields a positive slope of $\alpha_{z>6} = 0.49_{-0.25}^{+0.26}$, deviating from zero at the $2.0\sigma$ level. The Pearson correlation coefficient is $r=0.57$ with a corresponding $p$-value of $p=0.051$, suggesting a possible correlation with modest statistical significance.

For the DGS sample, a consistent linear relation is found, with a slope of $\alpha_\mathrm{DGS} = 0.67_{-0.10}^{+0.11}$. This suggests a particularly tight correlation between \oiiitocii{} and \EW{} for the local dwarfs, with the slope being inconsistent with zero at $6.7\sigma$ significance. The correlation coefficient and $p$-value are $(r,p) = (0.89, <10^{-3})$, respectively, also indicating a strong and statistically significant correlation. 

Interestingly, the fits to the $z>6$ and DGS samples are in good agreement, unlike those for \oiiitocii{} versus metallicity and ionization parameter where the high-redshift sample is offset to higher line ratios at fixed metallicity and $U$. However, as evident from Fig.\ \ref{fig:oiii_cii_metallicity_ionization}c, the \EW{} measurements for the $z>6$ galaxies are generally on the upper end of those measured for the DGS galaxies, suggesting a higher burstiness may be the origin of their elevated line ratios. We discuss this in further detail in what follows.

\subsection{The origin of the elevated \oiiitocii{} ratios at high redshift}
\label{sec:lineRatioMultivariateFit}

In the previous sections, we discussed how the \oiiitocii{} ratio for dwarf and high-redshift galaxies appears to depend on burstiness, and to a lesser extent on ionization parameter, while metallicity does not seem to be a key driver of the line ratio. However, we have also noted that these parameters themselves may (anti-)correlate (c.f., Figure \ref{fig:matrixPlot}), which could complicate this interpretation. We therefore proceed by performing a multivariate fit of the \oiiitocii{} ratio against metallicity, ionization parameter and burstiness simultaneously. By doing so in a Bayesian fashion, we are furthermore able to robustly quantify any degeneracies between the various fitting parameters.

We demonstrated in Section \ref{sec:lineRatioHighzDwarfsComparison} that the dependence of the \oiiitocii{} ratio on metallicity, ionization parameter and burstiness can be well-captured by a power law (i.e., a linear function in log-space), thus motivating a trivariate linear fit of the form

\begin{equation}
\begin{split}
    \log\left(\dfrac{\text{\oiiinowave{}{}}}{\text{\cii{}}}\right) 
   &= \beta_0 + \beta_1 \times \left[12 + \log(\mathrm{O/H}) - 8 \right] \\
   &\quad + \beta_2 \times \log(O32) \\
   &\quad + \beta_3 \times \left[\log\left(\text{\EW{}}\right) - 3\right]
\end{split}
\label{eq:trivariate_fit}
\end{equation}

Here the equivalent width is expressed in Angstrom, and we normalize the metallicity and EW at $12 + \log(\mathrm{O/H}) = 8$ and $10^{3}\,\mathrm{\AA}$, as before. We perform the fit using the {\sc{emcee}} package \citep{foreman-mackey2013}, and simultaneously fit to the high-redshift and DGS samples to improve constraints. This is appropriate given that both populations show similar scaling relations between their \oiiitocii{} ratios and the aforementioned three physical parameters. To account for uncertainties on all of the measurements, we treat the metallicity, ionization parameter and equivalent width as latent variables, adopting a Gaussian prior with mean and standard deviation equal to the measured value and uncertainty, respectively. We moreover impose a minimum error on all quantities of $0.10\,\mathrm{dex}$ and include an intrinsic scatter term to account for additional scatter about the relation not captured by the measurement uncertainties.

\begin{figure*}
    \centering
    \includegraphics[width=1.0\linewidth]{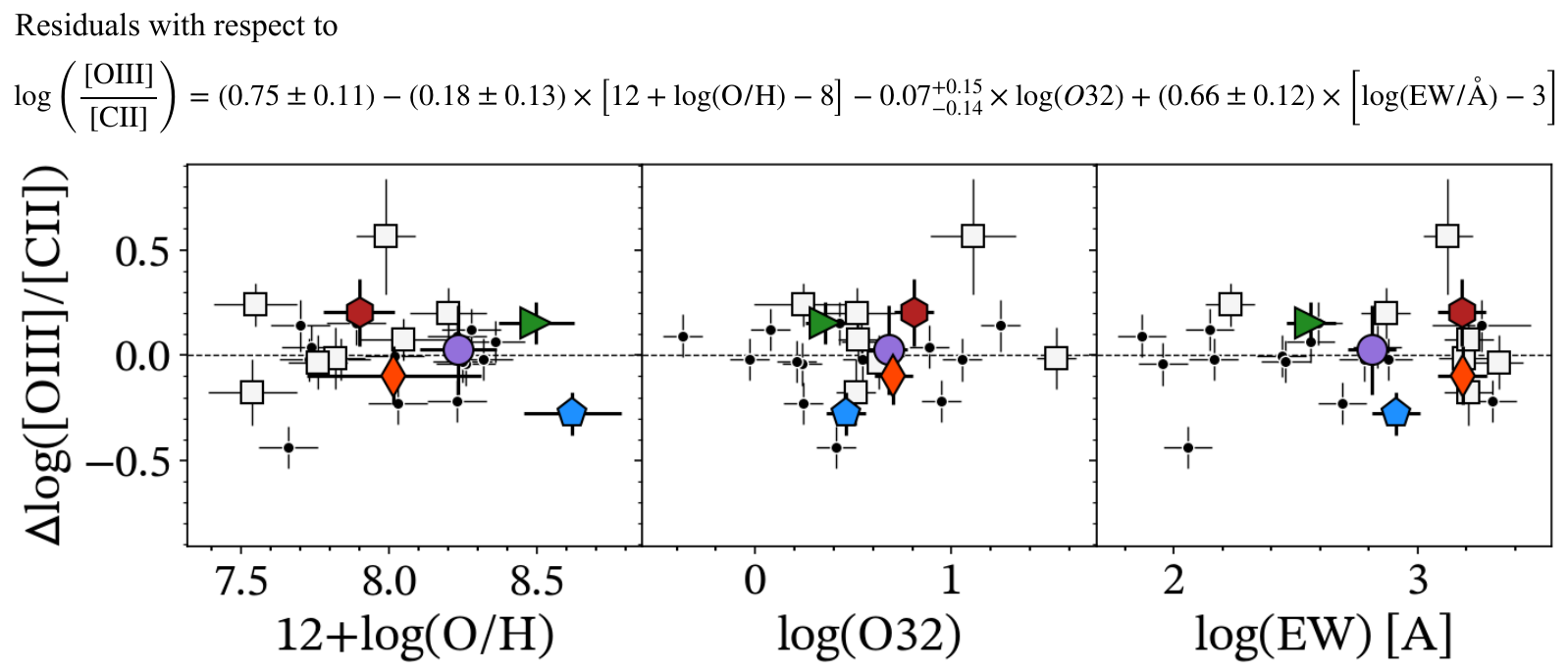}
    \caption{Residuals about the best fit multivariate relation displayed above the panels, defined as $\Delta\log\left(\text{\oiiitocii{}}\right) = \log\left(\text{\oiiitocii{}}\right)_\mathrm{obs}- \log\left(\text{\oiiitocii{}}\right)_\mathrm{fit}$. The residuals do not show any trends with metallicity (\textit{left}), ionization parameter (\textit{middle}), or \EW{} (\textit{right}), suggesting the fit accurately captures the observed variation in \oiiitocii{}. The intrinsic scatter about the relation is found to be $\sigma \approx 0.11\,\mathrm{dex}$. }
    \label{fig:residuals}
\end{figure*}

We show the corner plot (omitting the latent variables) in Figure \ref{fig:corner} in Appendix \ref{app:corner}. The best-fit equation is displayed in full in Figure \ref{fig:residuals}, with the best-fit parameters being $\beta_0 = 0.75 \pm 0.11$, $\beta_1 = -0.18 \pm 0.13$, $\beta_2 = -0.07_{-0.14}^{+0.15}$ and $\beta_3 = 0.66 \pm 0.12$. Figure \ref{fig:residuals} also displays the residuals with respect to the best fit as a function of metallicity, $O32$ and \EW{}. The lack of a clear trend in the residuals suggests that our multivariate fit appropriately captures the behavior of \oiiitocii{} across the full parameter space.

From the fit, we can draw several interesting conclusions. First of all, among the three scaling factors, only the value of $\beta_3 = 0.66 \pm 0.12$ is significantly different from zero (at the $\sim5\sigma$ level). This supports our previous hypothesis that burstiness is indeed the main parameter governing the elevated \oiiitocii{} ratios of high-redshift galaxies. We obtain only a mild anti-correlation with metallicity ($\beta_1 = -0.18 \pm 0.13$) while no correlation with $O32$ is observed ($\beta_2 = -0.07_{-0.14}^{+0.15}$). This suggests that the variation with metallicity and ionization parameter seen in Figure \ref{fig:oiii_cii_metallicity_ionization} is mostly captured by variations in the burstiness. While the corner plot (Figure \ref{fig:corner}) highlights a degeneracy between the coefficients $\beta_2$ and $\beta_3$, scaling with $O32$ and \EW{} respectively, this degeneracy is captured within the quoted fitting uncertainties, and hence does not change our interpretation.

We verify that a bivariate fit, including only metallicity and $O32$, does not yield a good fit to the combined high-$z$ and DGS samples; this consistently underpredicts (overpredicts) the line ratio of the former (latter), as expected given the the offset in \oiiitocii{} between the samples at fixed metallicity and $O32$. Moreover, fitting to the combination of \textit{i)} metallicity+burstiness or \textit{ii)} ionization parameter+burstiness yields similar results as obtained from our trivariate fit; burstiness drives the \oiiitocii{} ratio, whereas only a mild scaling with metallicity and no significance dependence on $O32$ is recovered.

\begin{figure*}
    \centering
    \includegraphics[width=1.0\textwidth]{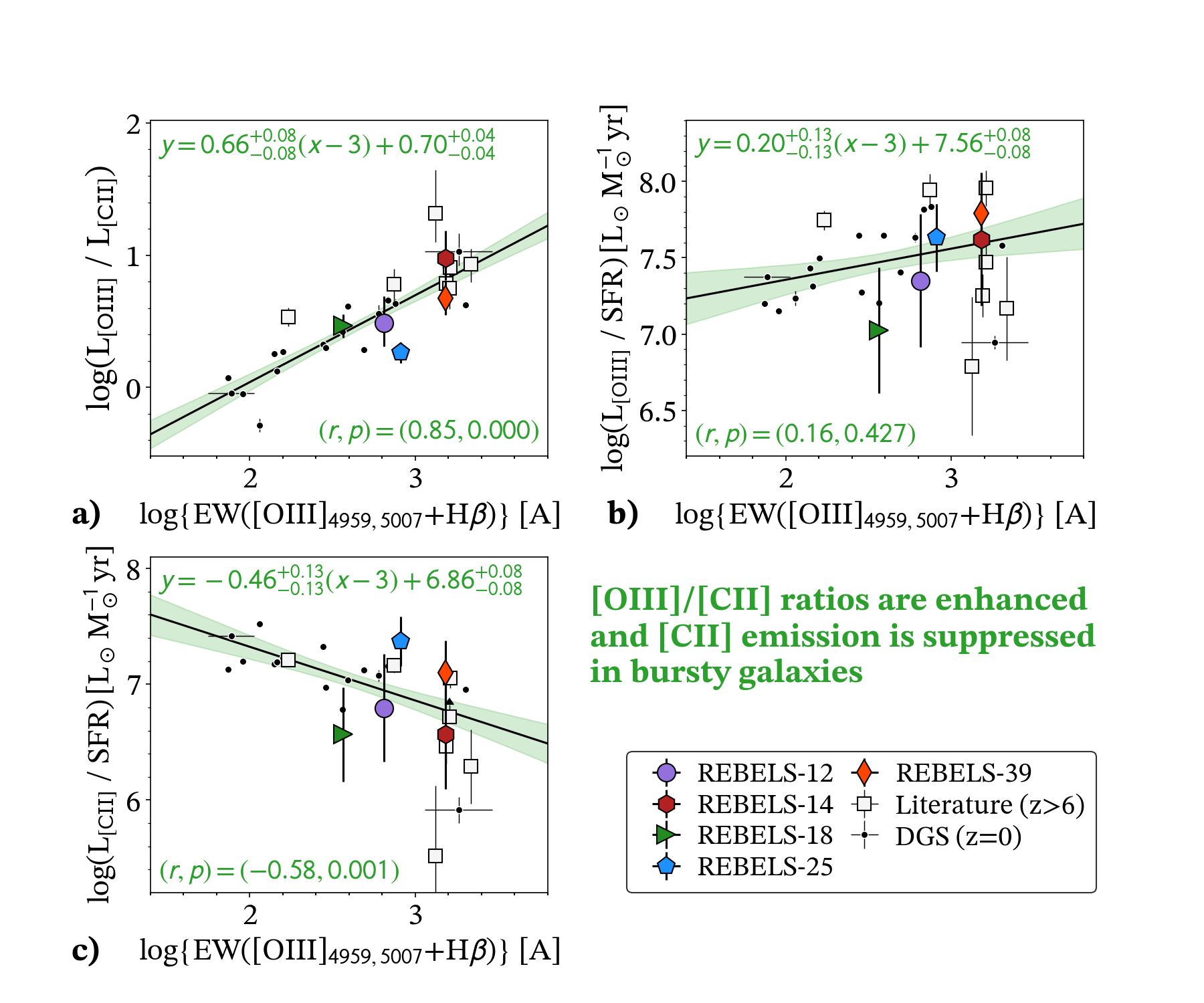}
    \caption{The dependence of \textbf{(a)} \oiiitocii{}, \textbf{(b)} \oiiinowave{}/SFR, and \textbf{(c)} \cii{}/SFR on the \oiiidopt{}+H$\beta$ equivalent width, a measure of galaxy burstiness. We jointly fit the high-$z$ sample (colored symbols and white squares) and local dwarf galaxies (black markers) with a linear function, represented by the solid black line and green shading for the $16-84^\mathrm{th}$ percent confidence interval. The best-fit relation, correlation coefficient $r$ and corresponding $p$-value for the combined high-$z$ and DGS samples are annotated in each panel. These fits suggest that the high \oiiitocii{} ratios observed for both local dwarfs and high-redshift galaxies are driven by a high burstiness, which mainly suppresses their \cii{} emission.}
    \label{fig:EW_masterplot}
\end{figure*}

Given that metallicity and ionization parameter are thus not found to significantly affect the \oiiitocii{} ratios, we perform one additional fit: a simple linear fit between \oiiitocii{} and \EW{} for the combined high-redshift and DGS samples (Figure \ref{fig:EW_masterplot}a). This yields a strong positive correlation (correlation coefficient of $0.85$, and $p$-value $<10^{-3}$). The slope of the relation, $\alpha = 0.66 \pm 0.08$, is different from zero at the $8.3\sigma$ level, and both the high-redshift and dwarf galaxies show a similar scatter about the best-fit trend.

\subsection{The effect of burstiness on the \oiii{} and \cii{} luminosity}
\label{sec:burstiness_on_oiii_and_cii}

In the previous section, we found that the \oiiitocii{} ratio strongly correlates with galaxy burstiness, expressed through the equivalent width of the \oiiidopt{} + H$\beta$ complex. In what follows, we consider the \oiii{} and \cii{} lines separately, in order to disentangle which of the two lines is most strongly impacted by variations in burstiness. To account for the primary dependence of both of these lines on star formation rate, we focus our analysis on the quantities $L_\text{\oiii{}}/\mathrm{SFR}$ and $L_\text{\cii{}}/\mathrm{SFR}$, where the star formation rate is measured as the sum of its UV- and infrared-based components. It should be kept in mind, however, that obscured SFR measurements at $z\gtrsim6$ can be particularly uncertain if one is limited to only one or two bands sampling the far-infrared continuum \citep[e.g.,][]{bakx2021,algera2024b,sommovigo_algera2025}, as is the case for most of the high-redshift sample compiled in this work.

We show the dependence of $L_\text{\oiii{}}/\mathrm{SFR}$ and $L_\text{\cii{}}/\mathrm{SFR}$ on \EW{} in Figure \ref{fig:EW_masterplot}b and \ref{fig:EW_masterplot}c, respectively. As before, we perform a linear fit to the combined high-redshift and dwarf galaxy samples. We do not see any strong correlation between \oiii{}/SFR and $\mathrm{EW}$, with the slope of the linear fit being consistent with zero at the $1.5\sigma$ level. The correlation coefficient ($0.16$) and $p$-value ($0.43$) moreover indicate no correlation. This is qualitatively consistent with the fact that most high-redshift galaxies follow the \oiii{}-SFR relation for local dwarf galaxies, at least for $\mathrm{SFR} \lesssim 100 - 200\,M_\odot\,\mathrm{yr}^{-1}$ (Section \ref{sec:oiiiSFR}).

The $L_\text{\cii{}}/\mathrm{SFR}$ ratio, on the other hand, is found to strongly anti-correlate with \EW{}. A linear fit yields a slope of $\alpha = -0.46 \pm 0.13$, different from zero at the $3.5\sigma$ level. The correlation coefficient ($-0.58$) and $p$-value ($10^{-3}$) moreover suggest the correlation is statistically significant. This demonstrates that the high \oiiitocii{} ratios seen in local dwarfs and high-redshift galaxies are likely driven by their high burstiness suppressing the \cii{} emission.

For completeness, we show the dependence of \oiii{}/SFR and \cii{}/SFR on metallicity and ionization parameter for the high-redshift and DGS samples separately in Appendix \ref{app:oiii2sfr_cii2sfr}. The best-fit parameters, correlation coefficients and $p$-values are provided in Table \ref{tab:correlations}. No joint linear fit is performed given the offset between the two galaxy populations at fixed metallicity and $\log U$, and any correlations found in Appendix \ref{app:oiii2sfr_cii2sfr} are likely manifestations of a secondary correlation mainly driven by burstiness. Regardless, the fits provided in the Appendix may still be useful for estimating the offset with respect to the \oiii{}-SFR or \cii{}-SFR relations when no \EW{} information is available.

\section{Caveats}
\label{sec:caveats}

Our analysis suggests the elevated \oiiitocii{} ratios of typical high-redshift and some dwarf galaxies are likely due to their bursty nature, which suppresses \cii{} relative to \oiii{} emission. However, we discuss some caveats to this interpretation below.

\subsection{Dust and spatial offsets between emission lines}

Throughout this work, we have compared the far-infrared \oiiitocii{} ratio to various rest-optical diagnostics. For these comparisons to be physically meaningful, these different emission lines need to emanate from similar physical regions within the galaxy (c.f., \citealt{vijayan2024,vijayan2025}). However, dust obscuration can affect the observed emission, and indeed spatial offsets between dust, UV, and line emission in high-redshift galaxies have been seen both in observations (e.g., \citealt{inami2022,killi2024,rowland2024}) and simulations (e.g., \citealt{behrens2018,cochrane2019,pallottini2022,esmerian2024,ocvirk2024}). 

Currently, we have no clear evidence for such offsets between rest-optical and far-infrared oxygen emission in our targets on scales of a few kpc, except in the dustiest galaxy among the sample REBELS-25 (Figure \ref{fig:momentZeroJWST}). Similarly, the \oiii{} and \cii{} emission appear mostly co-spatial at the current resolution of the observations (Figure \ref{fig:momentZero}), although high-resolution \cii{} data is needed to verify this. While the currently available data thus suggest that our spatially-integrated analysis of FIR and optical lines is justified, we proceed by discussing how any sub-resolution spatial offsets between the different emission line probes could affect our interpretation. \\

In particularly dust-obscured sources such as ULIRGs, rest-optical metallicity diagnostics have been found to underestimate the true metallicity \citep{chartab2022}. Circumventing this requires far-infrared metallicity diagnostics such as the \oiii{}/\niii{}$_{57}$ ratio (e.g., \citealt{nagao2011}), which are currently not available for our sample. However, the typical dust masses of ULIRGs ($M_\mathrm{dust}\sim10^{8-9}\,M_\odot$; \citealt{dacunha2010,clements2018}) are $\gtrsim10-100\times$ larger than those of REBELS galaxies, which have a typical $M_\mathrm{dust}\sim10^{7-7.5}\,M_\odot$ \citep{sommovigo2022,algera2025}. Moreover, \textit{JWST}/NIRSpec observations reveal REBELS-IFU galaxies to be spatially extended on $>\mathrm{kpc}$ scales \citep{rowland2025}, unlike the typically compact ULIRG population ($<1\,\mathrm{kpc}$). Overall, the dust mass surface densities of ULIRGs thus far exceed those of typical $z\sim7$ galaxies, making it unlikely that the metallicities of moderately dusty high-redshift galaxies are currently significantly underestimated due to a lack of FIR diagnostics. 

For the same reason, it is unlikely that dust obscuration significantly affects the observed trend between \oiiitocii{} ratio and ionization parameter or burstiness. However, if any heavily obscured regions are missed in the NIRSpec observations, these are likely young and star-forming (e.g., \citealt{diazsantos2007,bohn2023}), and thus presumably have a high $U$ and \EW{}. Qualitatively, this would thus drive our dustiest sources to higher values of $O32$ and burstiness. High-resolution \oiii{} as well as ALMA dust continuum observations will be necessary to investigate if this is indeed the case. 

Finally, for 4/5 galaxies within the REBELS-IFU sample we lack robust measurements of the Balmer decrement (Section \ref{sec:methods_dust_correction}). This naturally introduces additional uncertainty into their $O32$ measurements, although metallicities and \oiiidopt{}+H$\beta$ equivalent widths are less likely to be significantly affected (c.f., Section \ref{sec:methods_dust_correction}). For most of the high-redshift comparison sample, multiple Balmer lines are generally detected \citep[e.g.,][]{jones2024,scholtz2024}, making the resulting dust corrections generally more reliable. However, the precise shape of the dust attenuation curve -- which is generally unknown at high redshift and may vary significantly between individual galaxies and/or with redshift \citep[e.g.,][]{fisher2025,reddy2025,shivaei2025} -- still introduces additional uncertainty into any dust-corrected quantities.

\subsection{Other parameters possibly driving the \oiiitocii{} ratio}

We have focused our analysis on the scaling of the \oiii{} and \cii{} emission with metallicity (traced via oxygen abundance), ionization parameter (traced via $O32$), and burstiness (traced via \EW{}). While these parameters have been hypothesized to affect the \oiiitocii{} ratio -- and we indeed find that burstiness plays a major role -- this does not rule out the possibility that different quantities can also, or perhaps even more strongly, affect the line ratio.

\subsubsection{Electron densities}
Indeed, one additional parameter that is likely important in setting the \oiii{} emission from galaxies is the electron density, as a high $n_e$ can lead to collisional de-excitation of the \oiii{} line, and thus suppress the \oiii{} luminosity \citep[e.g.,][]{cormier2015,vanleeuwen2025}. In recent years, the \textit{JWST} has enabled electron density measurements at high redshift primarily through the \oii{} doublet \citep[e.g.,][]{isobe2023_electrondensity,abdurrouf2024}. However, we have not invoked this quantity in our analysis, as most of our high-redshift sample lacks $n_e$ measurements, given that the \oii{} doublet cannot be resolved with NIRSpec prism spectroscopy.

While the ratio of the \oiii{} and \oiiiopt{} lines has also been used to infer electron densities at high redshift \citep[e.g.,][]{fujimoto2024_s4590,abdurrouf2024}, the usefulness of this line ratio is limited in the context of our analysis. For one, the \oiii{}/\oiiiopt{} line ratio is degenerate between both the electron density and temperature ($T_e$), and an auroral line measurement yielding $T_e$ is required to break this degeneracy \citep[e.g.,][]{fujimoto2024_s4590}. Auroral line measurements are, however, not available for the REBELS-IFU sample \citep{rowland2025}. Furthermore, the \oiii{}/\oiiiopt{} line ratio can be affected by dust attenuation of the \oiiiopt{} line, and is thus strongly dependent on the availability and reliability of dust corrections. Finally, recent work by \citet[][see also \citealt{usui2025}]{harikane2025} has suggested that the interpretation of joint ALMA and \textit{JWST} line ratios is complicated in the presence of a multi-phase ISM, resulting in inconsistent electron densities inferred from optical and far-IR oxygen lines. In light of these various uncertainties, we have not attempted to infer electron densities for our targets via the \oiii{}/\oiiiopt{} line diagnostic, nor explicitly investigated its effect on \oiiitocii{} ratios. Future high-resolution ($R \gtrsim 1000$) \textit{JWST}/NIRSpec follow-up of our targets resolving the \oii{} doublet will enable investigating the effect of the electron density in more detail.

\subsubsection{C/O Abundances}
While metallicity was not found to play a major role in setting the \oiiitocii{} ratio in Section \ref{sec:oiii2cii_discussion}, we note that we have assumed solar abundance ratios throughout this work. However, the C/O abundance is known to decrease at low metallicities \citep[e.g.,][]{nicholls2017,berg2019}, and this has been hypothesized to boost the \oiiitocii{} ratio in young, metal-poor galaxies \citep[e.g.,][]{arata2020,katz2022}. While this scenario cannot completely be ruled out as we lack detailed abundance measurements, we did not find a clear trend between \oiiitocii{} and oxygen abundance, which is expected to be enhanced if low C/O abundances further suppress \cii{} at low metallicities (Figure \ref{fig:oiii_cii_metallicity_ionization}). Similarly, we did not find a clear trend between $L_\text{\cii{}}/\mathrm{SFR}$ and oxygen abundance (Figure \ref{fig:cii158sfr_metal_ionization} in Appendix \ref{app:oiii2sfr_cii2sfr}), which suggests that the effects of metallicity or non-solar abundance ratios are subdominant compared to burstiness.

\subsubsection{Time-dependent \oiiitocii{} ratios}
Given the dynamical and bursty conditions expected in high-redshift galaxies, their ISM conditions will undoubtedly vary with time. \citet{vallini2017} show that the \oiiitocii{} ratio of individual giant molecular clouds is likely to be time-dependent, with a short period ($\lesssim5\,\mathrm{Myr}$) after a starburst being characterized by high \oiiitocii{} ratios. On longer timescales, the clouds are photo-evaporated resulting in a sharp drop in \oiiitocii{}. Qualitatively similar results were recently reported by \citet{kohandel2025}, focusing on a simulated high-redshift analog of the $z=14.18$ galaxy GS-z14. They also find that high \oiiitocii{} ratios are transient, being related to temporarily high ionization parameters ($\log U\sim -1$) linked to merger-driven starburst events. Observationally, this time-variation of the high-redshift ISM can of course not be captured. However, these theoretical works qualitatively support our interpretation that high \oiiitocii{} ratios are associated with rapid star formation in young, bursty galaxies, while also suggesting that a biased selection towards such bursty systems may play an additional role.

\section{Summary and conclusions}\label{sec:conclusions}

We have presented new ALMA \oiii{} observations towards eight previously \cii{}-detected galaxies at $6.8 \lesssim z \lesssim 7.7$. Six of these -- our primary targets -- are UV-luminous galaxies drawn from the REBELS survey \citep{bouwens2022}, while the remaining two were serendipitously identified as dust-obscured neighbors to our REBELS targets through their bright \cii{} emission \citep{fudamoto2021,vanleeuwen2024}. Five of our primary targets benefit from \textit{JWST}/NIRSpec IFU prism spectroscopy, which yields strong-line oxygen abundances, a measure of their ionization parameter through the $O32$ diagnostic, and the equivalent width of the \oiiidopt{}+H$\beta$ complex, a proxy for burstiness.

We first focus on the ALMA \oiii{} emission of our eight targets, from which we draw the following conclusions:

\begin{enumerate}
    \item We detect the \oiii{} line in our two serendipitous targets, unambiguously confirming their redshifts to be $z=7.352$ (REBELS-12-2) and $z=6.838$ (REBELS-39-2) through both the \cii{} and \oiii{} lines. We infer $\text{\oiiitocii{}}\approx3-4$ for both, which suggests that the elevated \oiiitocii{} measurements at $z>6$ are unlikely to solely be the result of a biased selection towards UV-luminous galaxies (c.f., \citealt{algera2024,bakx2024}).

    \item We also detect the \oiii{} line in our six primary targets, noting that two were already previously detected in shallower observations \citep{algera2024}. The \oiiitocii{} ratios of our full sample of eight galaxies spans \oiiitocii{}$\approx 1.9 - 9.6$, consistent with what has previously been seen in the $z\gtrsim6$ galaxy population (e.g., \citealt{harikane2020,witstok2022}).

    \item Our full $z\sim7$ sample falls within the scatter of the \oiii{}-SFR relation for local dwarf galaxies \citep{delooze2014}. This is consistent with the bulk of the $z\approx6-14$ \oiii{}-detected galaxies in the literature, suggesting the relation may be near-universal across cosmic time -- at least up to $\mathrm{SFR}\lesssim 100 - 200\,M_\odot\,\mathrm{yr}^{-1}$, beyond which the relation may flatten.
    
\end{enumerate}

We combine our sample with ten $z>6$ galaxies in the literature also benefiting from ALMA \oiii{} and \cii{} observations and \textit{JWST} spectroscopy to explore the various possible origins of the elevated \oiiitocii{} ratios seen in high-redshift galaxies. In particular, we have investigated whether the elevated line ratios are predominantly driven by low metallicities (e.g., \citealt{vallini2017,ferrara2019,katz2019,liang2024}), high ionization parameters (e.g., \citealt{arata2020,harikane2020,katz2022}), and/or a high burstiness (e.g., \citealt{vallini2021,vallini2024,algera2024}). 

We moreover compare to local dwarf galaxies from the \textit{Herschel} Dwarf Galaxy Survey (DGS; \citealt{madden2013}) for which similar measurements are available. The high-redshift galaxies and local dwarfs span similar metallicities and ionization parameters, although on average the $z>6$ galaxies are more bursty and have higher \oiiitocii{} ratios. We quantitatively compare the dependence of \oiiitocii{} on metallicity, ionization parameter and burstiness for both samples, from which we conclude the following:

\begin{enumerate}
    \item An enhanced burstiness is the main driver of the high \oiiitocii{} ratios seen in both high-redshift galaxies and local dwarfs.

    \item The high \oiiitocii{} ratios are the result of \cii{} being suppressed in bursty galaxies, while the luminosity of \oiii{} per unit SFR is not strongly affected.

    \item Less significant (anti-)correlations between \oiiitocii{} and metallicity or ionization parameter are also obtained for the high-redshift and dwarf galaxy samples, but are likely a secondary manifestation of a primary (anti-)correlation of \oiiitocii{} with burstiness.
    
\end{enumerate}

Given that the typical high-redshift population identified by \textit{JWST} appears metal-poor with extreme ionizing conditions and bursty star formation, \oiii{} is likely to remain a powerful tool to study the earliest galaxies (c.f., \citealt{carniani2024_oiii,schouws2024,zavala2024_alma,witstok2025}; Algera et al.\ in preparation). At the same time, detecting \cii{} at similarly early epochs is likely to be challenging, as already hinted at by the lack of success in robustly detecting the line above $z\gtrsim8.5$ (c.f., \citealt{laporte2019,fujimoto2024_s4590,schouws2025}). 

We are now finally at a stage where we are beginning to unravel the ISM conditions of galaxies all the way out to -- and possibly soon beyond -- $z\approx14$ with far-infrared fine structure diagnostics. However, our comparison samples are by and large limited to galaxies at $z\approx0$, with very little available information across the nearly 13 Gyr of cosmic time between $z\sim0-6$. Planned far-infrared missions such as the PRobe far-Infrared Mission for Astrophysics (\textit{PRIMA}; \citealt{moullet2023}) are urgently required to fill in this gap, and paint a comprehensive evolutionary picture of the ISM conditions in the early Universe through Cosmic Noon to the those observed in the galaxy population around us today.

\section*{Acknowledgments}

LR gratefully acknowledges funding from the DFG through an Emmy Noether Research Group (grant number CH2137/1-1). MA is supported by FONDECYT grant number 1252054, and gratefully acknowledges support from ANID Basal Project FB210003 and ANID MILENIO NCN2024\_112. RB acknowledges support from an STFC Ernest Rutherford Fellowship [grant number ST/T003596/1]. C.-C.C. acknowledges support from the National Science and Technology Council of Taiwan (NSTC 111-2112-M-001-045-MY3 and 114-2628-M-001-006-MY4), as well as Academia Sinica through the Career Development Award (AS-CDA-112-M02). JW gratefully acknowledges support from the Cosmic Dawn Center through the DAWN Fellowship. The Cosmic Dawn Center (DAWN) is funded by the Danish National Research Foundation under grant No. 140.

This paper makes use of the following ALMA data: \\
ADS/JAO.ALMA\#2017.1.01217.S, ADS/JAO.ALMA\#2019.1.01634.L, ADS/JAO.ALMA\#2021.1.00318.S, ADS/JAO.ALMA\#2021.1.01297.S, ADS/JAO.ALMA\#2022.1.01324.S, ADS/JAO.ALMA\#2022.1.01384.S, ADS/JAO.ALMA\#2024.1.00406.S.

ALMA is a partnership of ESO (representing its member states), NSF (USA) and NINS (Japan), together with NRC (Canada), MOST and ASIAA (Taiwan), and KASI (Republic of Korea), in cooperation with the Republic of Chile. The Joint ALMA Observatory is operated by ESO, AUI/NRAO and NAOJ.

\bibliographystyle{mn2e}
\bibliography{main}

\appendix

\section{Tapered \oiii{} images and flux measurements}
\label{app:tapering}

\begin{figure*}
    \centering
    \includegraphics[width=1.0\linewidth]{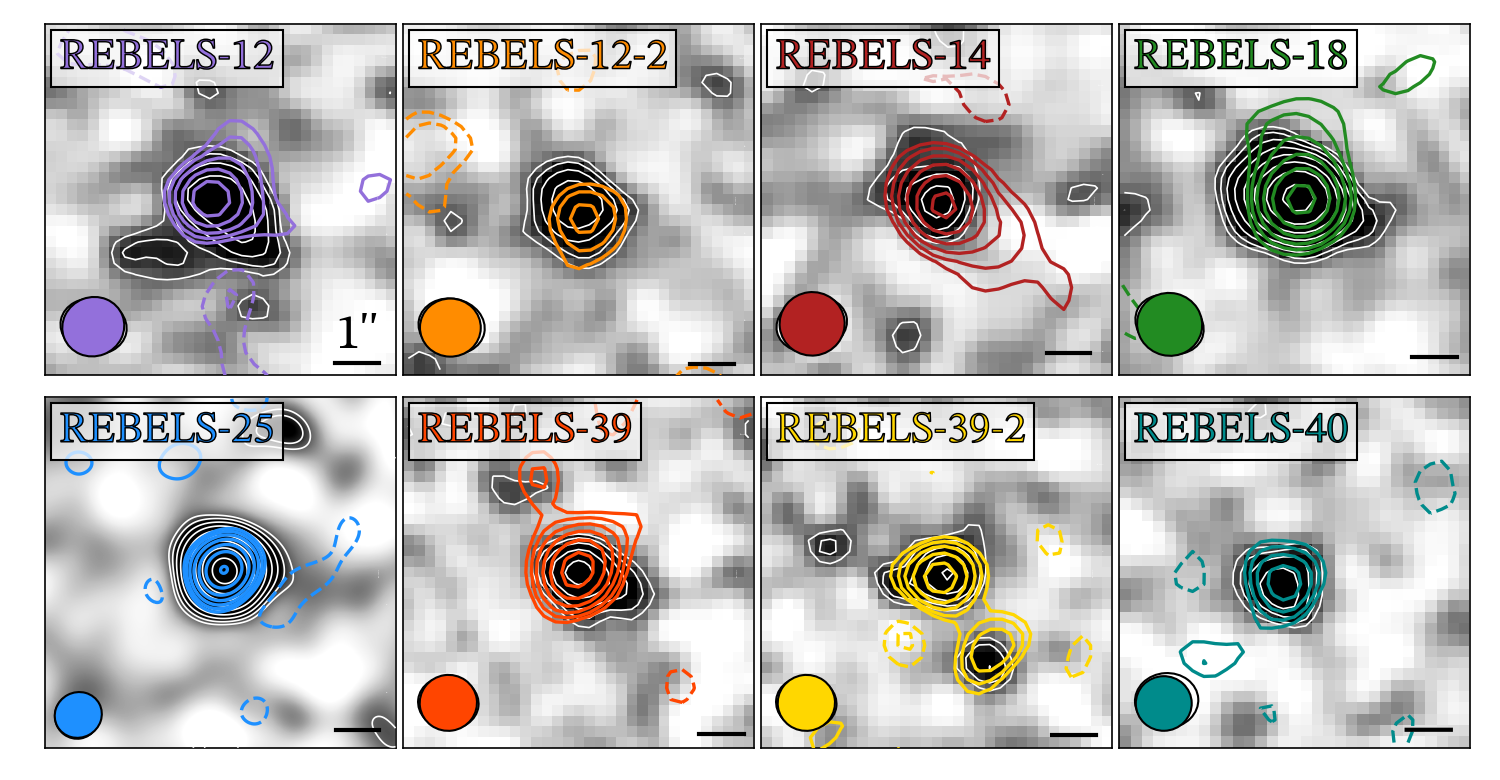}
    \caption{Same as Figure \ref{fig:CII_OIII_comparison}, this time with the \oiii{} data tapered to (approximately) the \cii{} resolution (except for REBELS-25 where we show the same $1''$ tapered data in both figures). While the line morphologies may differ in some cases -- see for example the extended \oiii{} tail in REBELS-14 -- these elongated features are generally low signal-to-noise, and the bulk of the \cii{} and \oiii{} emission originates from similar regions.}
    \label{fig:tapered_CII_OIII_comparison}
\end{figure*}

As is clear from Figure \ref{fig:CII_OIII_comparison}, the native resolution of the \oiiil{} data presented in this work ($\sim0.6''-0.8''$) is higher than that of the \cii{} data, which was originally taken as part of the REBELS survey at $1.3-1.5''$ resolution \citep{bouwens2022}. For REBELS-25, high-resolution ($\sim0.17''$) observations of both emission lines are available, although in this work we focus on the $1''$ tapered images provided by Rowland et al.\ (in preparation), which yield more reliable flux density measurements for our unresolved analysis. For the other seven sources, our fiducial flux measurements make use of the naturally-weighted \oiii{} datacubes (Table \ref{tab:data}). Given the coarser resolution of the \cii{} observations, we repeat the flux measurements in resolution-matched data to ensure that \textit{i)} no faint, extended flux is lost in our fiducial analysis, and \textit{ii)} our naturally-weighted fluxes are not boosted by the \citet{jorsater1995} effect.

For this purpose, we taper the \oiii{} datacubes to match the resolution of the \cii{} observations. The tapered moment-0 maps, extracted in the same manner as for the naturally-weighted data (Section \ref{sec:fluxMeasurements}, and discussed further below) are shown in Figure \ref{fig:tapered_CII_OIII_comparison}. The \cii{} and \oiii{} morphologies of both lines are broadly consistent, although some differences are apparent. For example, REBELS-14 shows an extended feature towards the north-west which is not seen in the \cii{} data. This feature corresponds to a $\sim4\sigma$ ``clump'' also seen in the naturally-weighted moment-0 map (c.f., Figure \ref{fig:CII_OIII_comparison}), where it is found to be spatially offset although possibly connected to REBELS-14. Its nature is currently unclear, and deeper observations area needed to assess if it could be a faint, \oiii{}-emitting companion galaxy or an ionized outflow. Broadly speaking, however, the bulk of the \cii{} and \oiii{} emission of our targets show consistent morphologies and centroids, supporting our unresolved analysis of their \oiiitocii{} ratios. \\

\begin{figure}
    \centering
    \includegraphics[width=0.49\textwidth]{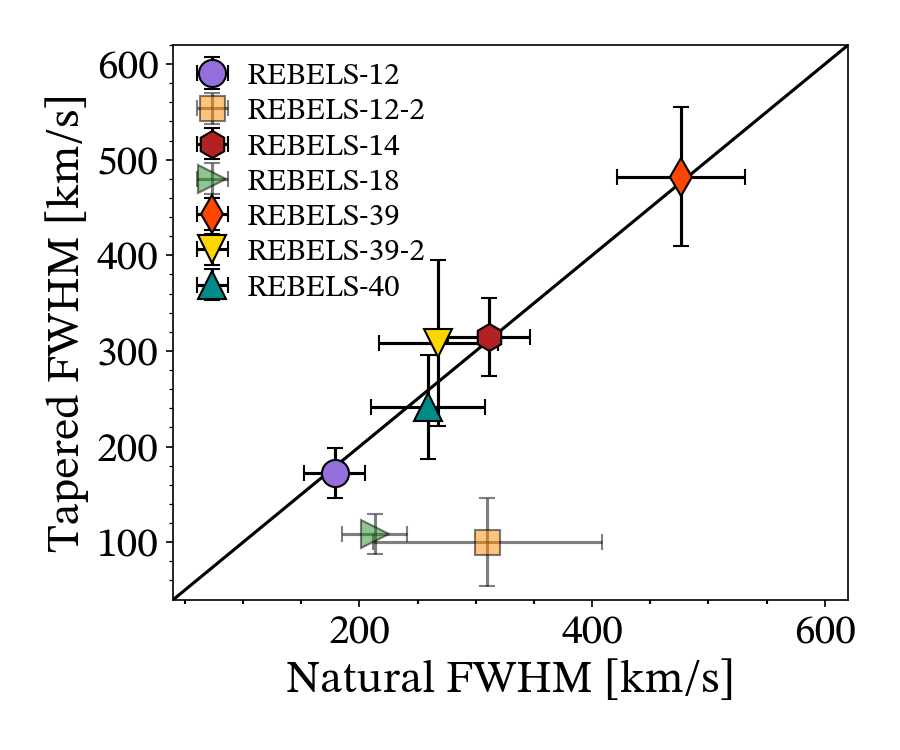}
    \includegraphics[width=0.49\textwidth]{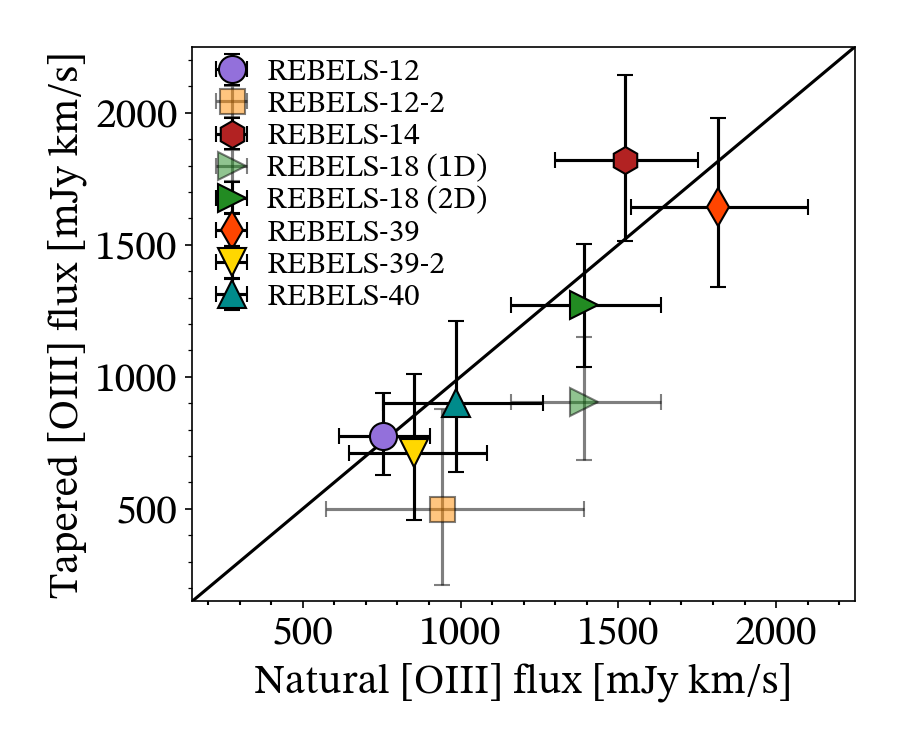}
    \caption{\textit{Left:} the FWHM of the \oiii{} lines extracted from the tapered cubes (with a typical resolution of $1.3'' - 1.5''$) is plotted against the FWHM obtained from the fiducial naturally-weighted cubes. The line widths agree for 5/7 sources, but the tapered extractions yield narrower FWHMs for REBELS-12-2 and REBELS-18. \textit{Right:} a comparison of the tapered and natural \oiii{} flux densities. As a consequence of their lower FWHM, the tapered flux densities of REBELS-12-2 and REBELS-18 are lower than the fiducial measurements. For REBELS-12-2, this is due to limited S/N in the tapered cube. For REBELS-18, we demonstrate that a consistent line flux is obtained from the tapered map if we fit its moment-0 map with a 2D Gaussian (see text). Overall, this suggests our fiducial flux measurements in the $0.6'' - 0.8''$ \oiii{} cubes are robust.}
    \label{fig:oiiiFluxesTapered}
    \vspace*{0.5cm}
\end{figure}

We re-extract the emission lines fluxes of our targets from the tapered cubes following the approach in Section \ref{sec:fluxMeasurements}. We compare the tapered \oiii{} FWHMs and fluxes to the fiducial ones in Figure \ref{fig:oiiiFluxesTapered}. For 5/7 sources, we find that the naturally-weighted and tapered \oiii{} fluxes are in good agreement. For the remaining two sources, REBELS-12-2 and REBELS-18, we however recover a flux density from the tapered maps that is $\sim2\times$ lower (albeit consistent within the uncertainties for REBELS-12-2). As is clear from the left panel of Figure \ref{fig:oiiiFluxesTapered}, this is due to these sources being assigned a narrower FWHM when the fluxes are extracted from the tapered maps. For REBELS-12-2 this is readily explainable; not only is its \oiii{} detection the lowest S/N one among our sample ($6.2\,\sigma$ in the naturally-weighted data), it also has the highest native resolution ($0.60''\times0.44''$; Table \ref{tab:data}), which means that tapering the cube to $1.3''$ to match the \cii{} beam comes at the cost of significant sensitivity. Consequently, the wings of the line fall well below the noise, and the resulting fit yields a narrower FWHM and a lower line flux (which is directly proportional to the FWHM). We note that the \oiii{} FWHM obtained from extracting the line from the naturally-weighted map moreover agrees with the width of the \cii{} line (Figure \ref{fig:CII_OIII_comparison}), which lends further credibility to the robustness of the fiducial line flux measurement.

For REBELS-18, the situation is similar: extracting the line from the tapered cube yields a $\sim2\times$ narrower FWHM than in the fiducial fit. In this case, however, the naturally-weighted data yield a high-S/N detection of the \oiii{} line, and even in the tapered cube REBELS-18 is detected at high significance. As such, it is less clear-cut whether the line flux extracted from the naturally-weighted image is overestimated, or whether the increased noise in the tapered maps suppresses the flux as in the case of REBELS-12-2. To investigate this, we re-measure the line flux of REBELS-18 by collapsing the tapered \oiii{} datacube around the expected line center within a velocity width of $\pm440\,\mathrm{km/s}$, equal to twice the FWHM of the \cii{} line (Table \ref{tab:lineFluxes}).\footnote{In this case, we use a datacube from which the continuum emission has been removed through {\sc{uvcontsub}}.} In the case of a Gaussian line profile with the same FWHM as the \cii{} line, this should capture virtually all of the line flux (namely 98\%, or even more if the true \oiii{} FWHM is narrower). We fit the resulting detection in the moment-0 map with a 2D Gaussian, and measure a flux of $S_\text{\oiii{}} = 1249 \pm 230\,\mathrm{km/s}$. This agrees well with our fiducial measurement obtained from the naturally-weighted cube of $S_\text{\oiii{}} = 1391_{-231}^{+250}\,\mathrm{mJy\,km/s}$, while it exceeds that obtained from 1D Gaussian fit to the flux extracted from the tapered map ($S_\text{\oiii{}} = 905_{-220}^{+244}\,\mathrm{mJy\,km/s}$).

Altogether, this exercise suggests that our fiducial fluxes, obtained from the naturally-weighted \oiii{} cubes, are reliable. However, it also underscores that low-level flux in the emission line wings can be missed, and that it is important to attempt multiple extraction methods and compare them for the most robust line flux measurements.

\section{Convolved NIRSpec \oiiiopt{} narrow-band images}
\label{app:jwst_convolved}

\begin{figure*}
    \centering
    \includegraphics[width=0.95\textwidth]{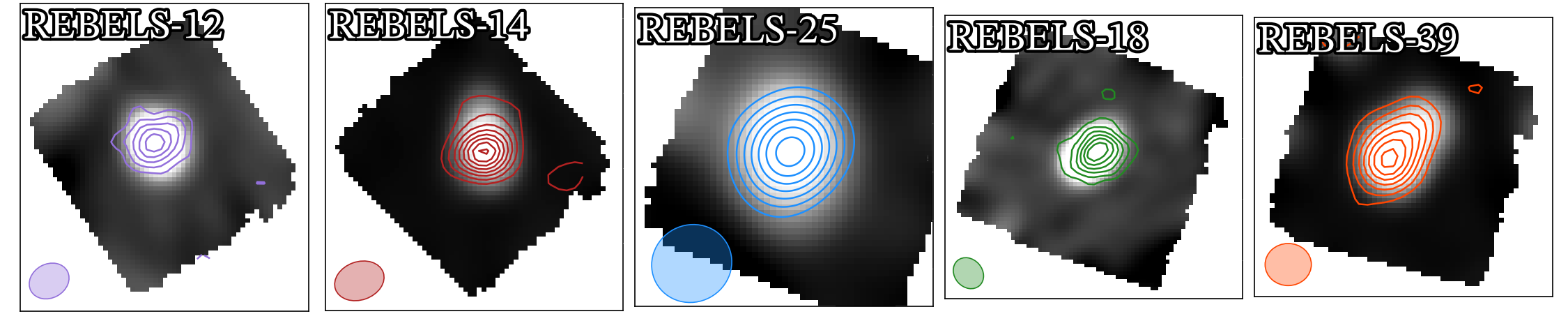}
    \caption{ALMA \oiii{} emission line maps (contours) on top of \textit{JWST}/NIRSpec \oiiiopt{} narrow-band images convolved to the same resolution for our five REBELS-IFU targets (greyscale background; c.f., Figure \ref{fig:momentZeroJWST} where the IFU data are shown at their native resolution). The ALMA contours start at $3\sigma$, and increase in steps of $2\sigma$. Overall, the rest-optical and far-infrared \oiiinowave{} emission trace each other well on scales of several kpc, motivating our focus on galaxy-integrated line ratios.}
    \label{fig:momentZeroConvolvedJWST}
    \vspace*{0.3cm}
\end{figure*}

In Figure \ref{fig:momentZeroConvolvedJWST}, we convolve the \textit{JWST}/NIRSpec \oiiiopt{} narrow-band images to the naturally-weighted \oiii{} resolution. For the NIRSpec point-spread function we adopt a simple circular 2D Gaussian following Equations 3 and 4 in \citet{deugenio2024}. This is appropriate given that the large difference in resolution between the \textit{JWST} and ALMA data ensures the size of convolution kernel depends mostly on the beam size of the latter, and not significantly on the exact NIRSpec PSF. We find that the \oiii{} and \oiiiopt{} profiles trace one another well, as we previously concluded based on the non-convolved IFU data. This good spatial correspondence reinforces the validity of our galaxy-integrated comparison of the \textit{JWST} and ALMA emission.

\section{Literature compilation}
\label{app:literatureCompilation}

A compilation of the $z>6$ sample with \oiii{} observations and ancillary \textit{JWST}/NIRSpec data used in this work is provided in Table \ref{tab:literatureComplation}. While GS-z11-0 was also recently detected in \oiii{} by \citet{witstok2025}, no rest-optical emission lines are confidently detected in its NIRSpec spectrum \citep{hainline2024}, and we therefore do not include it in our analysis. Furthermore, some of the literature works referenced in Table \ref{tab:literatureComplation} provide only the individual \oiiiopt{} and \oii{} line fluxes rather than a dust-corrected $O32$ ratio. In these cases, we use the measured Balmer decrement from these studies in combination with a \citet{calzetti2000} attenuation law to obtain a $O32$ measurement corrected for dust.

\begin{table*}
    \def\arraystretch{1.5}
    \centering
    \caption{Compilation of $z>6$ galaxies with ALMA \oiii{} observations and ancillary \textit{JWST}/NIRSpec data.}
    \label{tab:literatureComplation}
    \begin{threeparttable}
    \begin{tabular}{lccccccc}
    
    \hline\hline 
    ID & Redshift & $\log(\text{\oiiitocii{}})$ & $12+\log(\mathrm{O/H})$ & $\log(O32)$ & $\log\,$\EW{}\,[\AA] & References \\
    \hline
    (1) & (2) & (3) & (4) & (5) & (6) & (7) \\
    \hline 
GS-z14-0 & $14.18$ & $>0.54$ & $7.89\pm0.17$ & - & - & Car24ab, S24, S25 \\
GHZ2 & $12.33$ & - & $7.47\pm0.10$ & $1.26\pm0.19$ & - & Cas24, Cal24, Z24, Z25 \\
GS-z11-0$^\dagger$ & 11.12 & - & $\sim8.17$ & - & - & CL23, H24, W25 \\
MACS1149-JD1 & $9.11$ & $0.79_{-0.14}^{+0.15}$ & $7.82\pm0.07$ & $1.54\pm0.02$ & $3.19\pm0.02$ & H18, L19, C20, S23 \\
ID4590 & $8.50$ & $>0.71$ & $7.26\pm0.18$ & $>1.17$ & - & N23, F24 \\
MACS0416-Y1 & $8.31$ & $0.93_{-0.14}^{+0.12}$ & $7.76\pm0.10$ & $0.63\pm0.04$ & $3.33\pm0.03$ & T19, B20, Har24, M24 \\
REBELS-18 & $7.67$ & $0.47_{-0.09}^{+0.09}$ & $8.50\pm0.13$ & $0.35\pm0.03$ & $2.56\pm0.02$ & R25, this work \\
REBELS-12 & $7.35$ & $0.49_{-0.18}^{+0.20}$ & $8.23\pm0.20$ & $0.68\pm0.05$ & $2.81\pm0.02$ & A24, R25, this work \\
REBELS-25 & $7.31$ & $0.26_{-0.08}^{+0.08}$ & $8.62\pm0.17$ & $0.46\pm0.06$ & $2.91\pm0.04$ & A24, R25, this work \\
SXDF-NB1006-2 & $7.21$ & $1.32_{-0.22}^{+0.33}$ & $7.99\pm0.10$ & $1.11\pm0.22$ & $3.12\pm0.02$ & I16, R23, H25 \\
B14-65666 & $7.15$ & $0.49_{-0.07}^{+0.08}$ & $8.18\pm0.11$ & $0.51\pm0.07$ & - & H19, J24, Su25 \\
REBELS-14 & $7.08$ & $0.98_{-0.17}^{+0.20}$ & $7.90\pm0.12$ & $0.81\pm0.04$ & $3.18\pm0.01$ & R25, this work \\
COS-3018555981 & $6.85$ & $0.91_{-0.09}^{+0.08}$ & $8.05\pm0.15$ & $0.52\pm0.02$ & $3.21\pm0.04$ & S18, W22, Sch24 \\
REBELS-39 & $6.84$ & $0.67_{-0.12}^{+0.13}$ & $8.02\pm0.29$ & $0.70\pm0.06$ & $3.18\pm0.01$ & R25, this work \\
COS-2987030247 & $6.81$ & $0.75_{-0.16}^{+0.16}$ & $7.54\pm0.15$ & $0.52\pm0.08$ & $3.21\pm0.03$ & S18, W22, M25 \\
J0217-0208 & $6.20$ & $0.78_{-0.13}^{+0.11}$ & $8.20\pm0.13$ & $0.51\pm0.20$ & $2.87\pm0.02$ & H20, H25 \\
J1211-0118 & $6.03$ & $0.53_{-0.07}^{+0.07}$ & $7.55\pm0.14$ & $0.24\pm0.25$ & $2.23\pm0.02$ & H20, H25 \\
\hline\hline 
\end{tabular}
\begin{tablenotes}
    \item $^\dagger$ The metallicity of GS-z11-0 is uncertain due to the lack of strong emission line detections \citep[e.g.,][]{witstok2025}, and we therefore do not include it in our analysis in Section \ref{sec:lineRatioMetallicity} \vspace*{0.05cm}
    \item (1) Source name; (2) Spectroscopic redshift; (3) \oiii{}-to-\cii{} line ratio; (4) Oxygen abundance, used as a proxy for metallicity; (5) Dust-corrected $O32$ ratio, used as a proxy for ionization parameter; (6) Rest-frame equivalent width of the \oiiidopt{} + H$\beta$ complex in Angstrom, used as a proxy of burstiness; (7) References for the ALMA and \textit{JWST} observations:  \citet[][I16]{inoue2016}; \citet[][H18]{hashimoto2018}; \citet[][S18]{smit2018}; \citet[][H19]{hashimoto2019}; \citet[][L19]{laporte2019}; \citet[][T19]{tamura2019}; \citet[][B20]{bakx2020}; \citet[][C20]{carniani2020}; \citet[][H20]{harikane2020}; \citet[][W22]{witstok2022}; \citet[][CL23]{curtis-lake2023}; \citet[][N23]{nakajima2023}; \citet[][R23]{ren2023}; \citet[][S23]{stiavelli2023}; 
    \citet[][A24]{algera2024}; \citet[][Cal24]{calabro2024}; \citet[][C24a]{carniani2024_oiii}; \citet[][C24b]{carniani2024}; \citet[][Cas24]{castellano2024}; \citet[][F24]{fujimoto2024_s4590}; \citet[][H24]{hainline2024}; \citet[][Har24]{harshan2024}; \citet[][J24]{jones2024}; \citet[][M24]{ma2024}; \citet[][Sch24]{scholtz2024}; \citet[][S24]{schouws2024}; \citet[][Z24]{zavala2024_alma}; \citet[][H25]{harikane2025}; \citet[][M25]{mawatari2025}; \citet[][R25]{rowland2025}; \citet[][S25]{schouws2025}; \citet[][Su25]{sugahara2025}; \citet[][W25]{witstok2025}; \citet[][Z25]{zavala2025_miri}
\end{tablenotes}
\end{threeparttable}
\end{table*}

\section{Multi-variate fit to the \oiiitocii{} ratios of high-$z$ and dwarf galaxies}
\label{app:corner}

\begin{figure}
    \centering
    \includegraphics[width=0.8\linewidth]{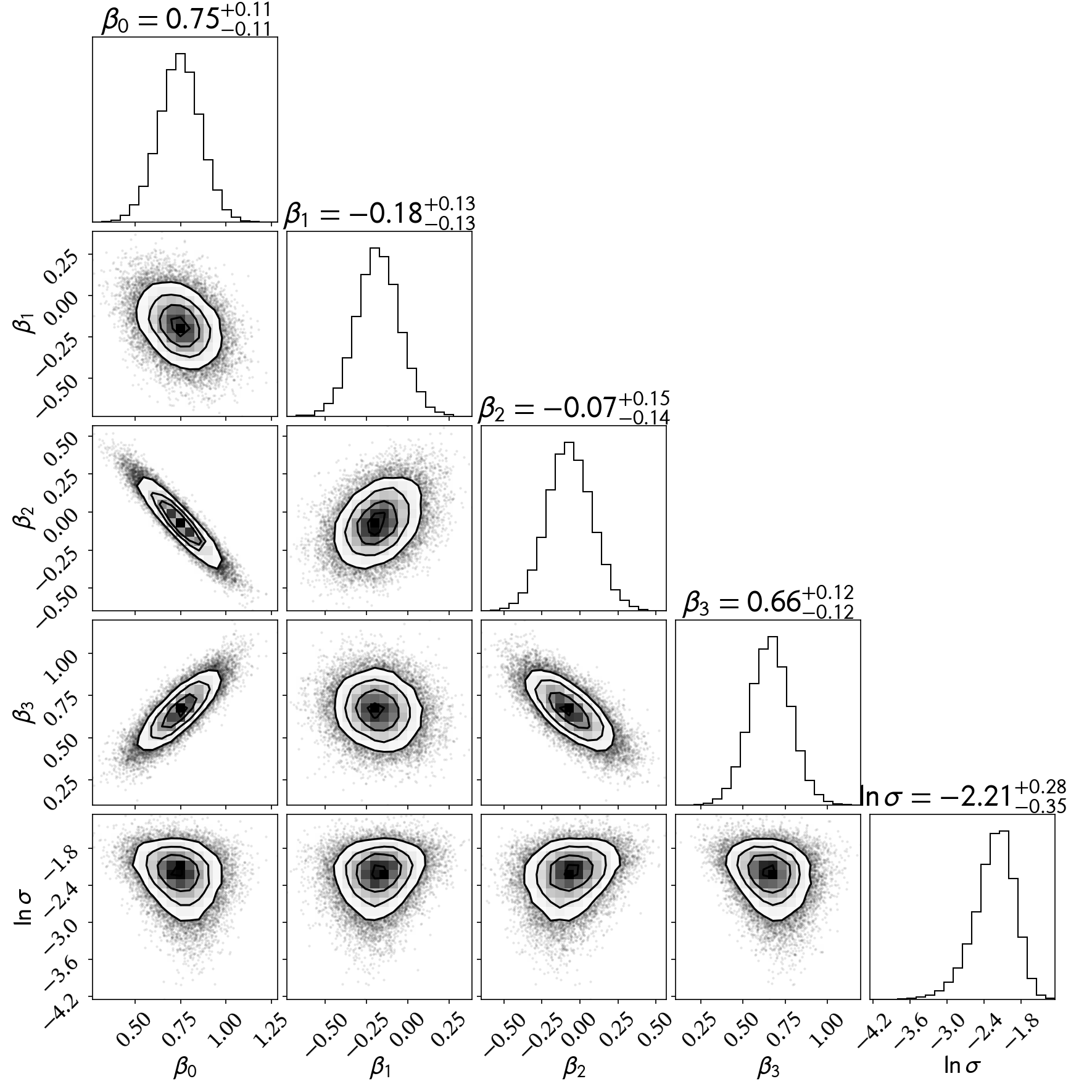}
    \caption{Corner plot of the trivariate fit to the \oiiitocii{} ratios of high-redshift ($z>6$) and dwarf galaxies (Equation \ref{eq:trivariate_fit}). The parameters $\beta_1, \beta_2, \beta_3$ represent the linear scaling with metallicity, ionization parameter and burstiness, respectively, while $\beta_0$ represents the intercept. We also include a scatter term $\ln\sigma$, which suggests an intrinsic scatter about the best fit of $\sigma \approx0.11\,\mathrm{dex}$.}
    \label{fig:corner}
\end{figure}

In Section \ref{sec:lineRatioMultivariateFit}, we perform an MCMC-based trivariate fit to the combined $z>6$ and DGS samples with measurements of their metallicity, $O32$, and \EW{}. We show the corresponding corner plot in Figure \ref{fig:corner}. While mild correlations are seen between the various fitting parameters, our conclusion that the elevated \oiiitocii{} ratios of high-redshift galaxies are driven primarily by their high burstiness is robust against these degeneracies.

\section{The impact of metallicity and ionization parameter on \oiii{}/SFR and \cii{}/SFR}
\label{app:oiii2sfr_cii2sfr}

\begin{figure*}
    \centering
    \includegraphics[width=1.0\textwidth]{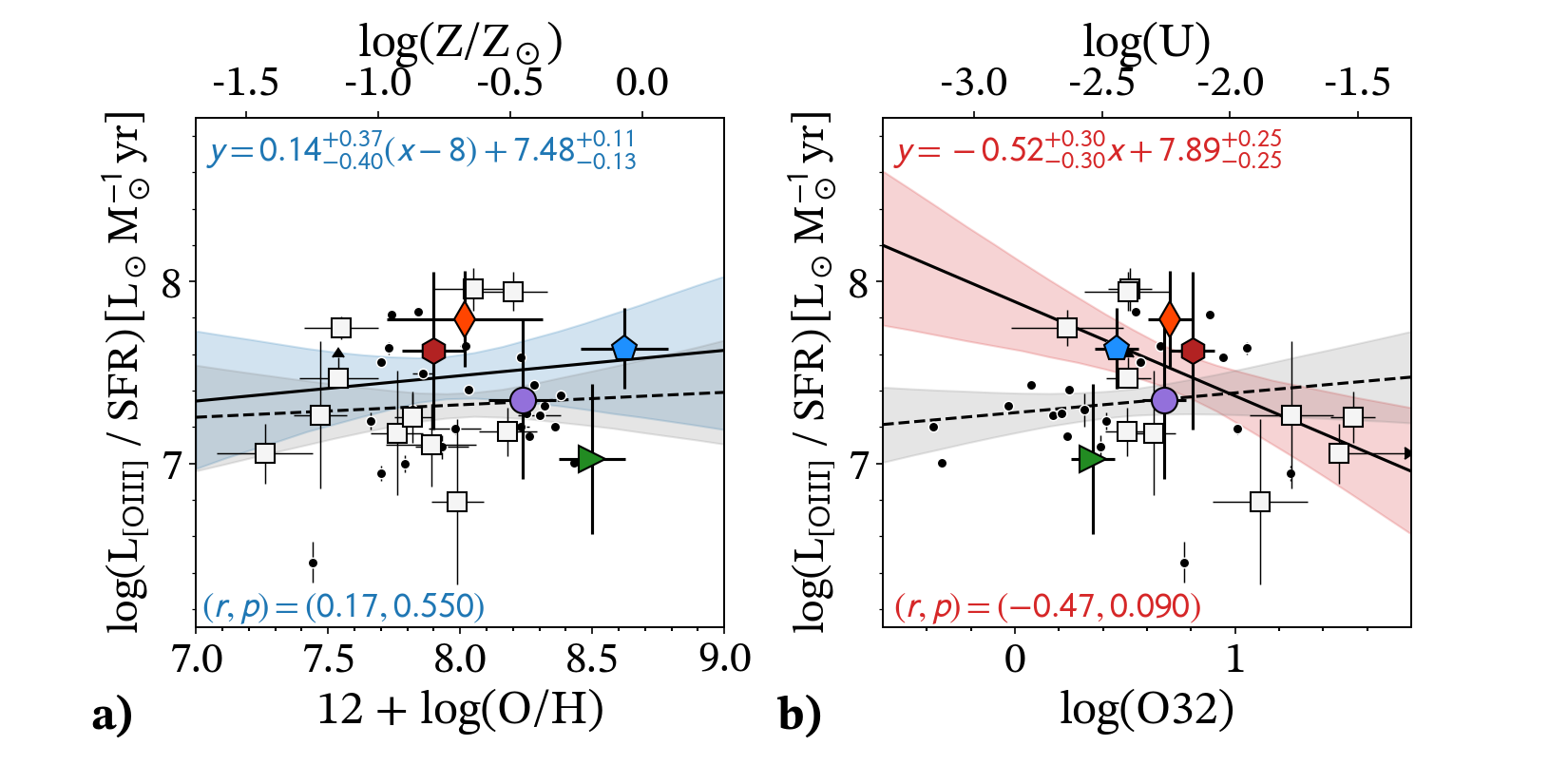}
    \caption{\textbf{a)} The ratio of the \oiii{} luminosity and SFR as a function of metallicity. \textbf{b)} the \oiii{}/SFR ratio as a function of ionization parameter. Symbols, lines and colors are as in Figure \ref{fig:oiii_cii_metallicity_ionization}. We do not see a strong correlation between the \oiii{}/SFR and either metallicity or ionization parameter.}
    \label{fig:oiii88sfr_metal_ionization}
\end{figure*}

\subsection{\oiii{}/SFR}
Analogous to our analysis in Section \ref{sec:burstiness_on_oiii_and_cii}, where we focus on the burstiness, we now plot the ratio of \oiii{}/SFR for both the $z>6$ and DGS samples against metallicity and ionization parameter (Figure \ref{fig:oiii88sfr_metal_ionization}). 

In Figure \ref{fig:oiii88sfr_metal_ionization}a, we find no strong evidence for a correlation between \oiii{}/SFR and metallicity, with a linear fit returning a slope that is consistent with zero within $< 1\sigma$ (Table \ref{tab:correlations}). A similar fit for the DGS sample yields consistent results, which thus suggests that the \oiii{} luminosity does not strongly vary with metallicity at fixed SFR. 

We next investigate $L_\text{\oiii{}} / \mathrm{SFR}$ as a function of ionization parameter, traced via $O32$, in Figure \ref{fig:oiii88sfr_metal_ionization}b. No strong correlation is seen here either, with a linear fit to both the high-redshift and DGS samples returning a slope that is consistent with no correlation within $<1.7\sigma$ and $< 1\sigma$, respectively.

\begin{figure*}
    \centering
    \includegraphics[width=1.0\textwidth]{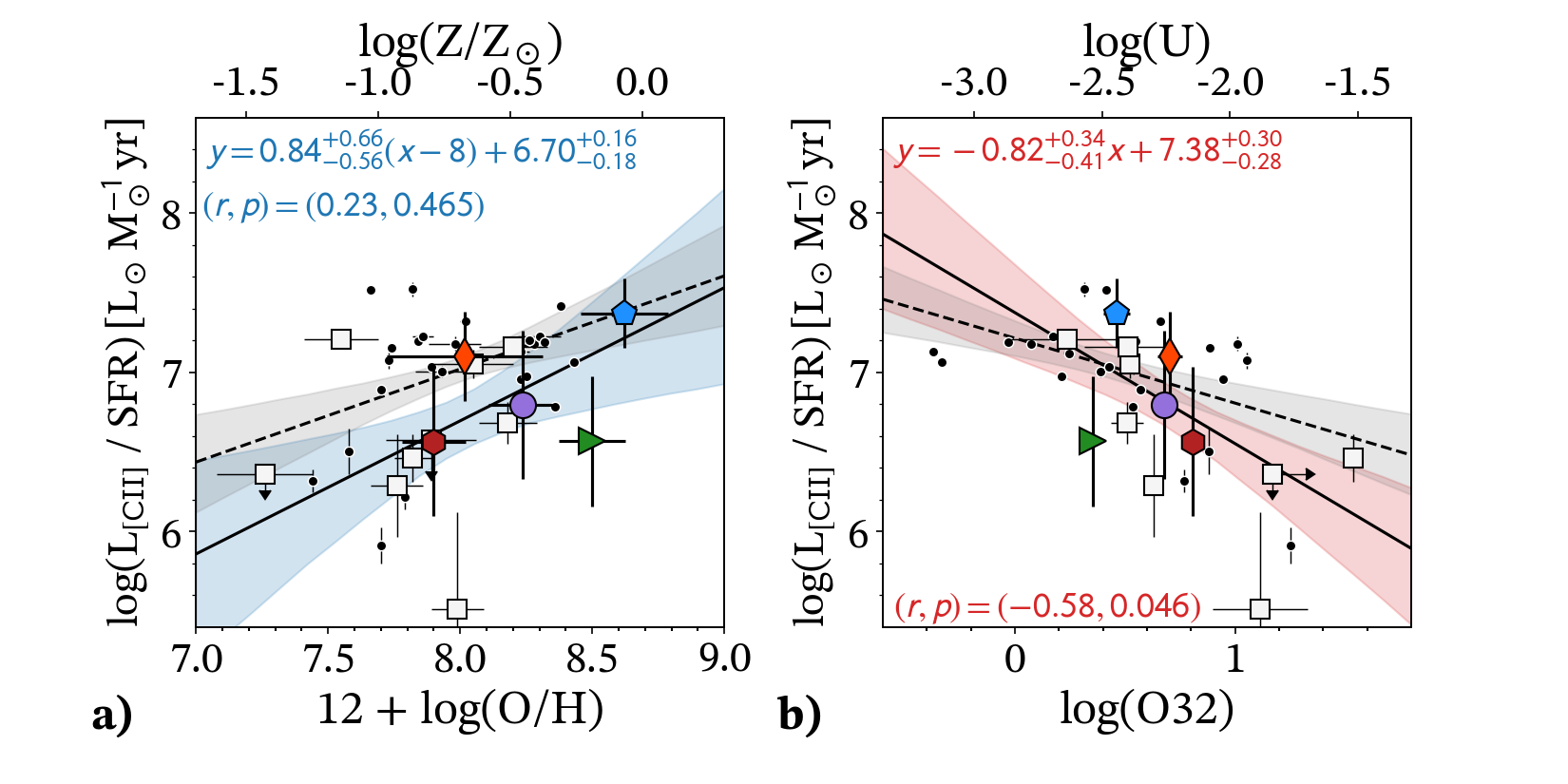}
    \caption{\textbf{a)} The ratio of the \cii{} luminosity and SFR as a function of metallicity. \textbf{b)} the \cii{}/SFR ratio as a function of ionization parameter. Symbols are as in Figure \ref{fig:oiii_cii_metallicity_ionization}. We find a statistically significant correlation between \cii{}/SFR and ionization parameter, with the \cii{} luminosity per unit SFR being suppressed at high $U$ as predicted by several theoretical works (e.g., \citealt{vallini2015,ferrara2019,harikane2020,sugahara2022,liang2024}). However, this is likely a manifestation of a primary anti-correlation with burstiness (Section \ref{sec:burstiness_on_oiii_and_cii}).}
    \label{fig:cii158sfr_metal_ionization}
\end{figure*}

\subsection{\cii{}/SFR}
We plot the \cii{}/SFR ratio against metallicity and ionization parameter in Figure \ref{fig:cii158sfr_metal_ionization}. We first investigate the relation between $L_\text{\cii{}}/\mathrm{SFR}$ and oxygen abundance. A correlation between these quantities has been predicted by several theoretical works, which suggests that \cii{} emission is suppressed at low metallicities (e.g., \citealt{vallini2015,vallini2025,ferrara2019,liang2024}). We do not find any evidence for a relation between $L_\text{\cii{}}/\mathrm{SFR}$ and metallicity for the high-redshift galaxy sample in Figure \ref{fig:cii158sfr_metal_ionization}a. However, for the DGS galaxies we find evidence for a modest correlation with a $p$-value of $0.048$ and slope of $\alpha_\mathrm{DGS} = 0.58 \pm 0.31$. However, the scatter about this relation is large ($\sigma = 0.36_{-0.05}^{+0.07}$), in agreement with earlier findings by \citet{cormier2015}. 

In Figure \ref{fig:cii158sfr_metal_ionization}b we present the \cii{}/SFR ratio as a function of ionization parameter. For the $z>6$ sample, we see a moderate anti-correlation with a slope of $\alpha_{z>6} = -0.82_{-0.41}^{+0.34}$, different from zero at the $2.4\sigma$ level. The corresponding $p$-value of $0.046$ suggests the anti-correlation is statistically significant (Table \ref{tab:correlations}).

For the DGS galaxies we also find evidence for an anti-correlation between $L_\text{\cii{}}/\mathrm{SFR}$ and $\log U$. A linear fit yields a slope of $\alpha_\mathrm{DGS} = -0.41_{-0.19}^{+0.18}$, inconsistent with zero at the $2.3\sigma$ level. The $p$-value of $0.024$ moreover suggests it is statistically significant.

A relation between $L_\text{\cii{}}$/SFR and ionization parameter is expected based on the photo-ionization modeling from \citet{harikane2020}. They find that \cii{} emission is suppressed when exposed to strong radiation fields due to C$^+$ being further ionized into C$^{++}$ (see also \citealt{moriwaki2018,pallottini2019,sugahara2022}). Our results thus support these predictions, although suggest that an enhanced burstiness is likely primarily responsible for driving up the ionization parameter.

\end{document}